\documentclass{article}
\usepackage{fullpage}
\usepackage{amssymb}
\usepackage{graphicx}
\usepackage{srcltx}

\def\ds{\displaystyle}

\newcommand{\bm}[1]{\mbox{\boldmath{$\rm #1$}}}

\def\roughly#1{\mathrel{\raise.3ex\hbox
{$#1$\kern-.75em\lower1ex\hbox{$\sim$}}}}

\begin{document}

\begin{flushright}
{\large DESY 07-097\\}
\end{flushright}

\vspace{2.0truecm}
\begin{center}
{\Large \bf Probing New Physics in the Neutrinoless Double Beta Decay
Using Electron Angular Correlation}
\end{center}

\vspace{0.9truecm}
\begin{center}
{\large \bf A.~Ali} \footnote{e-mail: ahmed.ali@desy.de}\\[0.1cm]
{\it Deutsches Elektronen-Synchrotron, DESY, 22607 Hamburg,
Germany}\\[0.3cm] 
{\large \bf A.V.~Borisov}\footnote{e-mail:
borisov@phys.msu.ru}
\\[0.1cm] {\it Faculty of Physics, Moscow State
University, 119991 Moscow, Russia}\\[0.3cm]
{\large \bf D.V.~Zhuridov}\footnote{e-mail:
	dmitry.zhuridov@uwr.edu.pl}\\[0.1cm] {\it Faculty of Physics, Moscow State
	University, 119991 Moscow, Russia}\\[0.1cm] {\it Faculty of Physics and Astronomy, University of Wroc{\l}aw, 50-204 Wroc{\l}aw, Poland}
\end{center}
\date{}

\vspace{2.0truecm}

\begin{abstract}
The angular correlation of the electrons emitted in the
neutrinoless double beta decay ($0\nu2\beta$) is presented
using a general Lorentz invariant effective Lagrangian for the leptonic
and hadronic charged weak currents. We show that the
 coefficient $K$ in the angular correlation $d\Gamma/d\cos \theta
\propto (1-K\cos \theta)$ is essentially independent of the
nuclear matrix element models and present its numerical values for
the five nuclei of interest ($^{76}\mbox{Ge}$, $^{82}\mbox{Se}$,
 $^{100}\mbox{Mo}$, $^{130}\mbox{Te}$, and $^{136}\mbox{Xe}$), assuming that
the $0\nu2\beta$-decays in these nuclei are induced solely by a
 light Majorana neutrino,  $\nu_M$. This coefficient varies between
$K=0.81$ (for the $^{76}\mbox{Ge}$ nucleus) and  $K=0.88$ (for the
$^{82}\mbox{Se}$ and  $^{100}\mbox{Mo}$ nuclei), calculated taking
into account the effects from the nucleon recoil, the $S$ and
$P$-waves for the outgoing electrons and the
 electron mass. Deviation of $K$ from its values derived here  would
indicate the presence of New Physics (NP) in addition to a light
Majorana neutrino, and we work out the angular coefficients in
several $\nu_M + \mbox{NP}$ scenarios for the $^{76}\mbox{Ge}$
nucleus. As an illustration of the correlations among the
 $0\nu2\beta$ observables (half-life $T_{1/2}$,  the coefficient $K$,
and the effective Majorana neutrino mass
 $|\langle m\rangle|$) and the parameters of the underlying NP model,
 we analyze the left-right symmetric models,
 taking into account current phenomenological bounds on the right-handed
$W_R$-boson mass and the left-right mixing parameter $\zeta$.
\end{abstract}

\vfill
\newpage
\section{Introduction}
It is now established beyond any doubt that the observed neutrinos
have tiny but non-zero masses and they mix with each other, with
both of these features following from the observation of the
atmospheric and solar neutrino oscillations and from the long
baseline neutrino oscillation  experiments~\cite{PDG}.
Theoretically, it is largely anticipated that the neutrinos are
Majorana particles.  Experimental evidence for the neutrinoless
double beta decay ($0\nu2\beta$) would deliver a conclusive
confirmation of the Majorana nature of neutrinos, establishing the
existence of physics beyond the standard model. This is
the overriding interest in carrying out these experiments and in
the related phenomenology~\cite{Vogel:2006sq}.

 We recall that  $0\nu2\beta$-decays are forbidden in the
standard model (SM) by lepton number (LN) conservation, which is a
consequence of the renormalizability of the SM. However, being the
low energy limit of a more general theory, an extended version of
the SM could contain nonrenormalizable terms (tiny to be
compatible with experiments), in particular, terms that violate LN
and allow the $0\nu2\beta$ decay. Probable mechanisms of LN
violation may include exchanges by: Majorana neutrinos $\nu_M$s
~\cite{ZK,Shchep,Doi} (the preferred mechanism after the
observation of neutrino oscillations \cite{PDG}), SUSY
particles~\cite{Mohapatra86,Vergados87,SUSY1,Babu95,SUSY,Faessler97},
scalar
 bilinears (SBs)~\cite{BL}, e.g. doubly charged dileptons (the component
$\xi^{--}$ of the $SU(2)_L$ triplet Higgs scalar etc.),
leptoquarks (LQs)~\cite{LQ}, right-handed $W_R$ bosons~\cite{Doi,HKP}
etc. From these particles light $\nu$s are much
lighter than the electron and others are much heavier than the
proton. Therefore, there are two possible classes of mechanisms for
the $0\nu2\beta$ decay. With the light $\nu$s in the intermediate
state the mechanism is called long range and otherwise it is referred to
as the short range mechanism. For both these classes, the separation of
the lepton physics from the hadron physics takes place~\cite{Vergados},
 which simplifies calculations.  According to the Schechter--Valle theorem
\cite{Valle}, any mechanism inducing the $0\nu2\beta$ decay produces an
effective Majorana mass for the neutrino, which must therefore
contribute to this decay. These various contributions
will have to be disentangled to extract information from the
$0\nu2\beta$ decay on the characteristics of the sources of LN
violation, in particular, on the neutrino masses and mixing.
Measurements of the neutrinoless double beta decay in different nuclei
will help in determining the underlying physics
 mechanism~\cite{Deppisch:2006hb,Gehman:2007qg}.

Our aim in this paper is to examine the possibility to discriminate
among the various possible mechanisms  contributing to the
$0\nu2\beta$-decays using the information on the angular correlation
of the final electrons in the process $N_i(A,Z) \to N_f(A,Z+2) +e^- + e^-$.
 A preliminary study along these
lines was published by us in 2006~\cite{Ali:2006iu}, with admittedly
 simplified treatment neglecting  the nucleon recoil and
the $P$-wave effects in the outgoing electron wave function. We rectify
these shortcomings and provide in this paper a detailed account of the
improved treatment. Restricting ourselves to
the long-range mechanism,  treating the  electrons
relativistically but with non-relativistic nucleons,
 we derive the angular correlation between the
electrons using  the general Lorentz invariant effective
Lagrangian involving the leptonic and hadronic charged weak
currents. Generally, this angular correlation can be expressed as
$d \Gamma /d \cos \theta \sim 1-K\cos \theta$, where $\theta$ is
the angle between the electron momenta in the rest frame of the
parent nucleus. Expressing $K={\mathcal B}/{\mathcal A}$, with
$-1<K< 1$, we derive the analytic expressions for ${\mathcal A}$
and ${\mathcal B}$ for the effective Lagrangian characterized by
the coefficients $\epsilon_{\alpha i}^\beta$ encoding the
standard, ($V-A) \otimes (V-A)$, and new physics contributions
(see Eq.~(1)). Essential steps of these derivations are presented
in section 2.  The analytic expressions derived here confirm the
earlier detailed derivations by Doi et al.~\cite{Doi}, and we
specify where the treatment presented here transcends the earlier
work.
 Specific cases are relegated to Appendix A (for the decays
involving scalar nonstandard terms), Appendix B (for the vector
nonstandard terms), and Appendix C (for the tensor nonstandard
terms). We hope to return to the discussion of including the short-range
mechanism, neglected in this paper, in future work.

Numerical analysis of the electron angular correlation is presented in
section 3, and the coefficient $K$ for the various underlying mechanisms  in
$0\nu2\beta$-decays are worked out. In particular,
 numerical values of $K$ for the five
nuclei of current experimental interest: $^{76}\mbox{Ge}$,
$^{82}\mbox{Se}$,
 $^{100}\mbox{Mo}$, $^{130}\mbox{Te}$, and $^{136}\mbox{Xe}$ are
presented for the light Majorana neutrino  $\nu_M$ case. Their
values range from $K=0.81$ (for the $^{76}\mbox{Ge}$ nucleus) and
$K=0.88$ (for the $^{82}\mbox{Se}$ and  $^{100}\mbox{Mo}$ nuclei).
To study the uncertainty in
 the nuclear matrix elements,  we have employed
the so-called QRPA model with and without the p-n pairing for
 the $^{76}\mbox{Ge}$ nucleus~\cite{Pantis}, and a more modern
QRPA model, fixing the particle-particle pairing
 strength~\cite{Kortelainen:2007rh}. While the uncertainty due to
 the nuclear matrix element model is quite
 marked for  $T_{1/2}$ in some cases, we show that it is
 rather modest for $K$, not exceeding 10\% for the models discussed here.
 For the  $\nu_M + \mbox{NP}$ scenarios, we remark that the nonstandard
coefficients $\epsilon^{V-A}_{V \mp A}$, $\epsilon^{T_L}_{T_R}$, and
$\epsilon^{T_R}_{T_L}$ do not change the value of the angular coefficient
 $K$. The contribution of the scalar nonstandard term from the
$\epsilon^{S+P}_{S\mp P}$ coefficients is found to be numerically small.
So, what concerns the angular correlation, we have essentially three distinct
 scenarios: (i) Standard ($\nu_M$), (ii) R-parity violating
 SUSY ($\nu_M + \epsilon_{T_R}^{T_R}$), and (iii) left-right-symmetric models
($\nu_M + \epsilon_{V+A}^{V\mp A}$).
 Numerical analysis of the
coefficient $K$ in the extended $\nu_M + \mbox{NP}$ scenario is
carried out for the decay of the  $^{76}\mbox{Ge}$ nucleus using
the nuclear matrix element model already specified.

 We take a closer look at the underlying physics behind the coefficients
$\epsilon^{V\mp A}_{V \mp A}$ in section 4. These coefficients appear in the
 context of the
left-right symmetric models which are  theoretically well motivated
 \cite{LR}. Also, the corresponding
nuclear matrix elements are available in the literature. Making
use of them, we work out the correlations among the angular
coefficient $K$, the half-life $T_{1/2}$ and either the mass of
the right-handed $W_R$ boson, $m_{W_R}$, or the $W$ boson's mixing
angle $\zeta$, taking into account the current bounds on the
various parameters. Results are presented in Figs.~1 -- 4.  The
differential distribution $d\Gamma/d\cos \theta$ for the
$0\nu2\beta$ decay of the $^{76}\mbox{Ge}$ nucleus is shown in
Fig.~5 for some representative values of $|\langle m \rangle|$ for
$m_{W_R}=1,~1.5$ TeV and for an infinitely heavy $m_{W_R}$. It is
seen that the effect of the right-handed $W_R$-boson is more
marked in the angular correlation for smaller values of $|\langle
m \rangle|$.
\vspace{-0.25cm}

\section{Angular correlation for the long range mechanism of $0\nu2\beta$ decay}
\subsection{General effective Lagrangian}
\label{section_Lagrangian}
For the decay mediated by light $\nu_M$s, the most general
effective Lagrangian is the Lorentz invariant combination of the
leptonic $j_\alpha$ and the hadronic $J_\alpha$ currents of
definite tensor structure and chirality \cite{Limits,Gamov}
\begin{equation}\label{L}
    {\mathcal L}=\frac{G_FV_{ud}}{\sqrt{2}}[(U_{ei}+\epsilon^{
    V-A}_{V-A,i})j_{V-A}^{\mu
    i}J^+_{V-A,\mu}+\sum\limits_{\alpha,\beta}\!^{^\prime}
\epsilon^\beta_{\alpha i}j^i_\beta J^+_\alpha+{\rm H.c.}]~,
\end{equation}
where the hadronic and leptonic currents are defined as:
$J^+_\alpha=\bar{u} O_\alpha d$ and $j^i_\beta =\bar{e} O_\beta
\nu_i$; the leptonic currents contain neutrino mass eigenstates
and the index $i$ runs over the light eigenstates. Here and
thereafter, a summation over the repeated indices is assumed;
$\alpha$,\,$\beta$=$V\!\mp\!A$,\,$S\!\mp\!P$,\,$T_{L,R}$
($O_{T_\rho}=2\sigma^{\mu\nu}P_\rho$,
$\sigma^{\mu\nu}=\frac{i}{2}\left[\gamma^\mu,\gamma^\nu\right]$,
$P_\rho=(1\mp \gamma_5)/2$ is the projector, $\rho=L,\,R$); the
prime indicates the summation over all the Lorentz invariant
contributions, except for $\alpha=\beta=V-A$, $U_{ei}$ is the PMNS
mixing matrix~\cite{PMNS} and $V_{ud}$ is the CKM matrix element
\cite{PDG}. Note that in Eq.~(\ref{L}) the currents have been
scaled relative to the strength of the usual $V-A$ interaction
with $G_F$ being the Fermi coupling constant. The coefficients
$\epsilon_{\alpha i}^\beta$ encode new physics, parametrizing
deviations of the Lagrangian from the standard $V-A$
current-current form and mixing of the non-SM neutrinos.


\vspace*{3mm}
In discussing the extension of the SM for the $0\nu2\beta$ decay,
 Ref.~\cite{Doi} considered explicitly only nonstandard terms with
\begin{equation}\label{Doi}
    \epsilon^{V-A}_{V+A,i}=\kappa \frac{g^\prime_V}{g_V}U^\prime_{ei},\quad
\epsilon^{V+A}_{V-A,i}=\eta V^\prime_{ei},\quad
\epsilon^{V+A}_{V+A,i}=\lambda \frac{g^\prime_V}{g_V}V_{ei}~.
\end{equation}
Implicitly, also the contributions encoded by the coefficients
 $\epsilon^{V-A}_{V-A,i}$ are discussed arising from the non-SM contribution
 to $U_{ei}$ in
$SU(2)_L\times SU(2)_R\times U(1)$ models with mirror leptons (see
Ref.~\cite{Doi}, Eq.~(A.2.17)). Here $V$, $U^\prime$ and
$V^\prime$ are the $3\times3$ blocks of mixing matrices for non-SM
neutrinos, e.g., for the usual $SU(2)_L\times SU(2)_R\times U(1)$
model $V$ describes the lepton mixing for neutrinos from
right-handed lepton doublets; for $SU(2)_L\times SU(2)_R\times
U(1)$ model with mirror leptons \cite{Pati} $U^\prime$
($V^\prime$) describes the lepton mixing for mirror
left(right)-handed neutrinos \cite{Doi} etc. The form factors
$g_{V}$ and $g^\prime_V$ are expressed through the mixing angles
for left- and right-handed quarks. Thus,  $g_V=\cos
\theta_C=V_{ud}$ and $g_V^\prime = e^{i \delta} \cos
\theta_C^\prime$, with $\theta_C$ being the Cabibbo angle,
$\theta_C^\prime$ is its right-handed mixing analogoue, and the CP
violating phase $\delta $ arises in these models due to both the
mixing of right-handed quarks and the mixing of left- and
right-handed gauge bosons (see Ref. \cite{Doi}, Eq. (3.1.11)). The
parameters $\kappa$, $\eta$, and $\lambda$ characterize the
strength of nonstandard effects. Below, we give some illustrative
examples relating the coefficients $\epsilon^{V-A}_{V-A,i}$,
$\epsilon^{V+A}_{V\pm A,i}$ and the particle masses, couplings and
the mixing parameters in the underlying theoretical models.

\vspace*{3mm} In the R-parity-violating (RPV) SUSY accompanying
the neutrino exchange
mechanism~\cite{Mohapatra86,Vergados87,SUSY1,Babu95,SUSY,Faessler97},
SUSY particles (sleptons, squarks) are present in one of the two
effective 4-fermion vertices. (The other vertex contains the usual
$W_L$ boson.) The nonzero parameters are
\begin{eqnarray}\label{SUSY}
&\ds   \epsilon^{V-A}_{V-A,i}=\frac{1}{2}\eta^{n1}_{(q)RR}U_{ni},\
\
   \epsilon^{S-P}_{S+P,i}=2\eta^{n1}_{(l)LL}U_{ni},\nonumber \\
&\ds   \epsilon^{S+P}_{S+P,i}=-\frac{1}{4}\left(\eta^{n1}_
   {(q)LR}-4\eta^{n1}_{(l)LR}\right)U^*_{ni},\
   \epsilon^{T_R}_{T_R,i}=\frac{1}{8}\eta^{n1}_{(q)LR}U^*_{ni},
\end{eqnarray}
where the index $n$ runs over $e$, $\mu$, $\tau$ (1, 2, 3), and
the RPV Minimal Supersymmetric Model (MSSM) parameters $\eta$s
depend on the couplings of the RPV MSSM superpotential, the masses
of the squarks and the sleptons, the mixings among the squarks and
among the sleptons. Concentrating on the dominant contributions
$\epsilon^{S+P}_{S+P,i}$ and $\epsilon^{T_R}_{T_R,i}$ (as the
others are helicity-suppressed), one can express $\eta^{n1}_{(q)
LR}$ and $\eta^{n1}_{(l) LR}$ as follows~\cite{SUSY}
\begin{eqnarray}\label{SUSY-Etas}
&\ds \eta^{n1}_{(q)LR} =
 \sum_k \frac{\lambda^\prime_{11k} \lambda^\prime_{nk1}}{2\sqrt{2}G_F}
  \sin 2 \theta^d_{(k)} \left(\frac{1}{m^2_{\tilde{d}_1(k)}}
  -  \frac{1}{m^2_{\tilde{d}_2(k)}}   \right),\nonumber \\
&\ds  \eta^{n1}_{(l)LR} =
  \sum_k \frac{\lambda^\prime_{k11} \lambda_{n1k}}{2\sqrt{2}G_F}
  \sin 2 \theta^e_{(k)} \left(\frac{1}{m^2_{\tilde{e}_1(k)}}
  -  \frac{1}{m^2_{\tilde{e}_2(k)}}   \right),
\end{eqnarray}
where $k$ is the generation index, $\theta^d_{(k)}$ and
$\theta^e_{(k)}$ are the squark and slepton mixing angles,
respectively, $ m_{\tilde{f}_1}$ and $ m_{\tilde{f}_2}$ are the
sfermion mass eigenvalues, and $\lambda_{ijk}$ and
$\lambda^\prime_{ijk}$ are the RPV-couplings in the
superpotential.

For the mechanism with LQs in one of the effective vertices
\cite{LQ}, the nonzero coefficients are
\begin{eqnarray}\label{LQ1}
&\ds
\epsilon^{S+P}_{S-P}=-\frac{\sqrt{2}}{4G_F}\frac{\epsilon_V}{M_V^2},\
   \ \epsilon^{S+P}_{S+P}=-\frac{\sqrt{2}}{4G_F}\frac{\epsilon_S}{M_S^2},
   \nonumber \\
&\ds
\epsilon^{V+A}_{V-A}=-\frac{1}{2G_F}\left(\frac{\alpha_S^{(L)}}{M_S^2}
+\frac{\alpha_V^{(L)}}{M_V^2}\right),\ \
\epsilon^{V+A}_{V+A}=-\frac{\sqrt{2}}{4G_F}\left(\frac{\alpha_S^{(R)}}{M_S^2}
+\frac{\alpha_V^{(R)}}{M_V^2}\right),
\end{eqnarray}
where
\begin{equation}\label{eps}
\epsilon^\beta_\alpha=U_{ei}\epsilon^\beta_{\alpha i},
\end{equation}
the parameters $\epsilon_{S(V)}$, $\alpha_{S(V)}^{(L)}$, and
$\alpha_{S(V)}^{(R)}$ depend on the couplings of the
renormalizable LQ-quark-lepton interactions consistent with the SM
gauge symmetry, the mixing parameters and the common mass scale
$M_{S(V)}$ of the scalar (vector) LQs~\cite{BRW87}.

The nonzero $\epsilon_\alpha^\beta$ for the discussed models are
collected in Table 1. \vspace*{2mm}

\begin{center}
\noindent Table 1: Nonzero coefficients
$\epsilon_\alpha^\beta$ for various models.

\label{Tab1}
\begin{tabular}{|c|c|}
  \hline
   Model & Nonzero $\epsilon$s \\
  \hline
   with $W_R$s & $\epsilon_{V+A}^{V-A}$, $\epsilon_{V\mp A}^{V+A}$  \\
  \hline
   RPV SUSY & $\epsilon_{S+P}^{S\mp P}$, $\epsilon_{V-A}^{V-A}$, $\epsilon_{T_R}^{T_R}$  \\
  \hline
   with LQs & $\epsilon_{S\mp P}^{S+P}$, $\epsilon_{V\mp A}^{V+A}$  \\
  \hline
\end{tabular}
\end{center}

The upper bounds on some of the $\epsilon^\beta_\alpha$ parameters
(\ref{eps}) from the Heidelberg--Moscow experiment were derived in
Ref. \cite{0002109} using the $S$-wave approximation for the
electrons, considering nucleon recoil terms and only one nonzero
parameter $\epsilon^\beta_{\alpha i}$ in the Lagrangian (\ref{L})
at a time.

The coefficients $\epsilon_{\alpha i}^{\beta}$ entering the
Lagrangian (\ref{L}) can be expressed as
\begin{equation}
\label{hat} \epsilon _{\alpha i}^\beta   = \hat \epsilon _\alpha
^\beta U^{(\alpha,\beta)}_{ei},
\end{equation}
where $U^{(\alpha,\beta)}_{ei}$ are mixing parameters for non-SM
neutrinos (see, e.g., Eq. (\ref{Doi})). As this Lagrangian
describes also ordinary $\beta$-decays (without LN violation), the
coefficients $\hat \epsilon _\alpha ^\beta$ are constrained by the
existing data on precision measurements in allowed nuclear beta
decays, including neutron decay \cite{sev}. For example, from
these data we obtain the conservative bound
\begin{equation}
\label{RR}
\left|{\hat \epsilon _{V + A}^{V + A}}\right| < 7
\times 10^{- 2}.
\end{equation}
From Eqs. (\ref{eps}), (\ref{hat}), (\ref{RR}) and the bound
$\left|\epsilon_{V+A}^{V+A}\right|<7.9\times 10^{-7}$ (see section
3.2) we can assume that the nonstandard mixing is small:
\begin{equation}
\label{nsm} \left| {U_{ei} V_{ei} } \right|\lesssim 10^{- 5},
\quad V_{ei}=U^{(V+A,V+A)}_{ei}.
\end{equation}

\subsection{Methods and approximations}

We have calculated the leading order in the Fermi constant taking into
account the
leading contribution of the parameters $\epsilon_\alpha^\beta$ to
the decay matrix elements using the approximation of the relativistic electrons
and non-relativistic nucleons.  The wavefunction of an electron
with the asymptotic momentum ${\bf p}$ and the spin projection $s$ can be
expanded in terms of spherical waves as \cite{Doi,Rose}
\begin{eqnarray}
   e_{{\bf p}s}({\bf r}) = e_{{\bf p}s}^{S_{1/2}}({\bf r})
   + e_{{\bf p}s}^{P_{1/2}}({\bf r}) + \dots
\end{eqnarray}
We take into account the $S_{1/2}$ and the $P_{1/2}$ waves for the
outgoing electrons:
\begin{eqnarray}\label{spherical waves}
   e_{{\bf p}s}^{S_{1/2}}({\bf r}) = \left(%
\begin{array}{c}
  \tilde g_{-1}\chi_s \\
  \tilde f_1 \bm\sigma\cdot\hat{\bf p} \chi_s \\
\end{array}%
\right), \\
   e_{{\bf p}s}^{P_{1/2}}({\bf r}) = i\left(%
\begin{array}{c}
  \tilde g_{1}\bm\sigma\cdot\hat{\bf r}\bm\sigma\cdot\hat{\bf p}\chi_s \\
  -\tilde f_{-1}\bm\sigma\cdot\hat{\bf r}\chi_s \\
\end{array}%
\right),
\end{eqnarray}
with $\hat{\bf r}={\bf r}/r$, $\hat{\bf p}={\bf p}/p$ and the two
component spinor $\chi_s$. We use the approximate radial wave
functions \cite{Doi}
\begin{eqnarray}
  && \left(%
\begin{array}{c}
  \tilde g_{-1} \\
  \tilde f_1  \\
\end{array}%
\right) = \tilde A_{\mp1}\left[ 1-\frac{1}{6}(\bar pr)^2\right], \\
  && (\bar pr)^2 = \left(\frac{3}{2}\alpha Z\right)^2\left(\frac{r}{R}\right)^2
  + 3\alpha Z\frac{r}{R}\varepsilon r + (pr)^2, \\
  && \left(%
\begin{array}{c}
  \tilde g_{1} \\
  \tilde f_{-1} \\
\end{array}%
\right) = \pm \tilde A_{\mp1}\xi_\pm(\varepsilon)\frac{r}{R},
\quad \xi_\pm=\frac{1}{2}\alpha Z+\frac{1}{3}(\varepsilon\pm
m_e)R,
\end{eqnarray}
including the finite de Broglie wave length correction (FBWC) for
the $S_{1/2}$ wave. Here $R$ is the nuclear radius, $\varepsilon$
is the electron energy and $\alpha$ is the fine structure
constant. For the normalization constants $\tilde A_{\pm1}$ we use
the approximate Eq. (\ref{tildA_pm}) (see below).

The nucleon matrix elements of the color singlet quark currents
are~\cite{SUSY1,Adler,Tomoda,Vergados2002}
\begin{eqnarray}\label{C:scalar}
    &\langle P(k^\prime)| \bar u(1\mp \gamma_5)d| N(k) \rangle =
    \bar\psi(k^\prime)\left[F^{(3)}_S(q^2) \mp F_P^{(3)}(q^2)\gamma_5\right]
    \tau_+ \psi(k),\\
 \label{C:vector}
   & \langle P(k^\prime)| \bar u \gamma^\mu (1\mp \gamma_5)d| N(k) \rangle =
    \bar\psi(k^\prime)\left[g_V(q^2)\gamma^\mu\mp
    g_A(q^2)\gamma^\mu\gamma_5-ig_M(q^2)\frac{\sigma^{\mu\nu}q_\nu}{2m_p}
    \pm g_P(q^2)\gamma_5q^\mu\right] \tau_+ \psi(k),\\
 \label{C:tenzor}
  &  \langle P(k^\prime)| \bar u\sigma^{\mu\nu} (1\mp \gamma_5)d| N(k) \rangle =
    \bar\psi(k^\prime)\left[J^{\mu\nu}\mp
    \frac{i}{2}\epsilon^{\mu\nu\rho\sigma}J_{\rho\sigma}\right]
    \tau_+\psi(k),
 \end{eqnarray}
 \begin{eqnarray}
    J^{\mu\nu} = T_1^{(3)}(q^2)\sigma^{\mu\nu} +
    \frac{iT_2^{(3)}}{m_p}(\gamma^\mu q^\nu-\gamma^\nu q^\mu) +
    \frac{T_3^{(3)}}{m_p^2}(\sigma^{\mu\rho}q_\rho q^\nu-\sigma^{\nu\rho}q_\rho
    q^\mu),
\end{eqnarray}
where
\begin{equation}
    \psi=\left(%
\begin{array}{c}
  P \\
  N \\
\end{array}%
\right)
\end{equation}
is a nucleon isodoublet.

The non-relativistic structure of the nucleon currents in the
impulse approximation is derived using Refs \cite{Tomoda,Ericson},
see Appendices A, B, and C. We have calculated the nucleon recoil
terms including the recoil terms due to the pseudoscalar form
factor.

%
%
\subsection{Electron angular correlation}

Taking into account the dominant terms introduced in the
Appendices A, B, and C in the closure approximation \cite{Doi} we
obtain the differential width in $\cos \theta$ for the
$0^+\!(A,Z)\rightarrow\!0^+\!(A,Z+2) e^- e^-$ transitions:
\begin{eqnarray}\label{dG}
\frac{d\Gamma}{d\cos\theta} = \frac{\ln2}{2}|M_{\rm GT}|^2{\mathcal
A}(1-K\cos\theta),
\end{eqnarray}
where $\theta$ is the angle  between the electron momenta in the
rest frame of the parent nucleus and the angular correlation
coefficient is
\begin{equation}\label{K}
K=\frac{{\mathcal B}}{{\mathcal A}}~,\quad -1<K< 1.
\end{equation}
The Gamow--Teller nuclear matrix element $M_{\rm GT}$ is defined in
 Eq.~(\ref{M_GT}) below.

The expressions for $\mathcal{A}$ and $\mathcal{B}$ for different
choices of $\epsilon_\alpha^\beta$, with only one coefficient
considered  at a time, are shown in Tables 2 and 3.


\begin{table}
\begin{center}
\noindent Table 2: Expressions for $\mathcal{A}$ in Eqs.
(\ref{dG}) and (\ref{K}) for the stated choice of
$\epsilon_\alpha^\beta$. \nopagebreak \label{Tab2}
\vspace*{3mm}
\begin{tabular}{|c|c|}
  \hline
   $\epsilon$ & $\mathcal{A}$ \\
  \hline
  $\epsilon^{V-A}_{V-A}$ & ${\mathcal A}_0 + 4C_1|\mu||\mu_{V-A}^{V-A}|c_{02}
+ 4C_1|\mu_{V-A}^{V-A}|^2$ \\
  \hline
  $\epsilon^{V-A}_{V+A}$ & ${\mathcal A}_0 +
4C_0|\mu||\mu_{V+A}^{V-A}|c_{01} +
4C_{1+}|\mu_{V+A}^{V-A}|^2$ \\
  \hline
  $\epsilon^{V+A}_{V-A}$ & ${\mathcal A}_0 + C_3|\mu|
|\epsilon^{V+A}_{V-A}|c_2 + C_5|\epsilon^{V+A}_{V-A}|^2$ \\
  \hline
  $\epsilon^{V+A}_{V+A}$ & ${\mathcal A}_0 + C_2|\mu|
|\epsilon^{V+A}_{V+A}|c_1 + C_4|\epsilon^{V+A}_{V+A}|^2$ \\
  \hline
  $\epsilon^{S-P}_{S-P}$ & ${\mathcal A}_0 + 4C_0^{SP}|\mu||\mu_{S-P}^{S-P}|c_{04}
+ 4C_1^{SP}|\mu_{S-P}^{S-P}|^2$ \\
  \hline
  $\epsilon^{S-P}_{S+P}$ & ${\mathcal A}_0 +
4C_0^{SP}|\mu||\mu_{S+P}^{S-P}|c_{03} +
4C_1^{SP}|\mu_{S+P}^{S-P}|^2$ \\
  \hline
  $\epsilon^{S+P}_{S-P}$ & ${\mathcal A}_0 + 4C_2^{SP}|\mu|
|\epsilon^{S+P}_{S-P}|c_{4} + 4C_3^{SP}|\epsilon^{S+P}_{S-P}|^2$ \\
  \hline
  $\epsilon^{S+P}_{S+P}$ & ${\mathcal A}_0 + 4C_{2+}^{SP}|\mu|
|\epsilon^{S+P}_{S+P}|c_{3} + 4C_{3+}^{SP}|\epsilon^{S+P}_{S+P}|^2$ \\
  \hline
  $\epsilon^{T_L}_{T_L}$ & ${\mathcal A}_0 + 4C_0^T|\mu||\mu_{T_L}^{T_L}|c_{06}
+ 4C_1^T|\mu_{T_L}^{T_L}|^2$ \\
  \hline
  $\epsilon^{T_L}_{T_R}$, $\epsilon^{T_R}_{T_L}$ & ${\mathcal A}_0$ \\
  \hline
  $\epsilon^{T_R}_{T_R}$ & ${\mathcal A}_0 + C_2^T|\mu|
|\epsilon^{T_R}_{T_R}|c_5 + C_3^T|\epsilon^{T_R}_{T_R}|^2$ \\
  \hline
\end{tabular}
\end{center}
\begin{center}
\noindent Table 3: Expressions for
$\mathcal{B}$ in Eq. (\ref{K}) for the stated choice of
$\epsilon_\alpha^\beta$.  \nopagebreak \label{Tab3}
\vspace*{3mm}
\begin{tabular}{|c|c|}
  \hline
  $\epsilon$ & $\mathcal{B}$ \\
  \hline
  $\epsilon^{V-A}_{V-A}$ & ${\mathcal B}_0 + 4D_1|\mu||\mu_{V-A}^{V-A}|c_{02}
+ 4D_1|\mu_{V-A}^{V-A}|^2$ \\
  \hline
  $\epsilon^{V-A}_{V+A}$ & ${\mathcal B}_0 +
4D_0|\mu||\mu_{V+A}^{V-A}|c_{01} +
4D_{1+}|\mu_{V+A}^{V-A}|^2$ \\
  \hline
  $\epsilon^{V+A}_{V-A}$ & ${\mathcal B}_0 + |\mu||\epsilon^{V+A}_{V-A}|
(D_{3}c_2+D_{3-}s_2) + D_5|\epsilon^{V+A}_{V-A}|^2$ \\
  \hline
  $\epsilon^{V+A}_{V+A}$ & ${\mathcal B}_0 + |\mu||\epsilon^{V+A}_{V+A}|
(D_{2}c_1+D_{2-}s_1) + D_4|\epsilon^{V+A}_{V+A}|^2$ \\
  \hline
  $\epsilon^{S-P}_{S-P}$ & ${\mathcal B}_0 + 4D_{0-}^{SP}|\mu||\mu_{S-P}^{S-P}|s_{04}
+ 4D_1^{SP}|\mu_{S-P}^{S-P}|^2$ \\
  \hline
  $\epsilon^{S-P}_{S+P}$ & ${\mathcal B}_0 +
4D_{0-}^{SP}|\mu||\mu_{S+P}^{S-P}|s_{03} +
4D_1^{SP}|\mu_{S+P}^{S-P}|^2$ \\
  \hline
  $\epsilon^{S+P}_{S-P}$ & ${\mathcal B}_0 + 4|\mu||\epsilon^{S+P}_{S- P}|
(D_{2}^{SP}c_{4}+D_{2-}^{SP}s_{4}) + 4D_3^{SP}|\epsilon^{S+P}_{S- P}|^2$ \\
  \hline
  $\epsilon^{S+P}_{S+P}$ & ${\mathcal B}_0 + 4|\mu||\epsilon^{S+P}_{S+ P}|
(D_{2+}^{SP}c_{3}+D_{2-}^{SP}s_{3}) + 4D_{3+}^{SP}|\epsilon^{S+P}_{S+P}|^2$ \\
  \hline
  $\epsilon^{T_L}_{T_L}$ & ${\mathcal B}_0 + 4D_{0-}^T|\mu||\mu_{T_L}^{T_L}|s_{06}
+ 4D_1^T|\mu_{T_L}^{T_L}|^2$ \\
  \hline
  $\epsilon^{T_L}_{T_R}$, $\epsilon^{T_R}_{T_L}$ & ${\mathcal B}_0$ \\
  \hline
  $\epsilon^{T_R}_{T_R}$ & ${\mathcal B}_0 + D_2^T|\mu||\epsilon^{T_R}_{T_R}|
c_5 + D_3^T|\epsilon^{T_R}_{T_R}|^2$ \\
  \hline
\end{tabular}
\end{center}
\end{table}
%
In these tables
\begin{eqnarray}
c_i=\cos\psi_i, \quad s_i=\sin\psi_i
\end{eqnarray}
and
\begin{eqnarray}
\mu=\langle m\rangle/m_e, \quad
\mu_\alpha^\beta=m_\alpha^\beta/m_e,
\end{eqnarray}
with the standard effective Majorana mass $\langle
m\rangle=\sum_iU_{ei}^2m_i$ and the nonstandard ones:
\begin{eqnarray}
m_{S\mp P}^{S-P}=\sum_iU_{ei}\epsilon_{S\mp P,i}^{S-P}m_i, \quad
  m_{V\mp A}^{V-A}=\sum_iU_{ei}\epsilon_{V\mp A,i}^{V-A}m_i, \quad
m_{T_{L,R}}^{T_{L}}=\sum_iU_{ei}\epsilon_{T_{L,R},i}^{T_L}m_i.
\end{eqnarray}
The quantities $\mathcal{A}$ and $\mathcal{B}$ for all zero
$\epsilon_\alpha^\beta$ are
\begin{eqnarray}
   {\mathcal A}_0=C_1|\mu|^2, \quad {\mathcal B}_0=D_1|\mu|^2
\end{eqnarray}
and the relative phases are
\begin{eqnarray}
  \psi_{01} &=& \arg(\mu \,\mu_{V+A}^{V-A*}), \quad
\psi_{02} = \arg(\mu \,\mu_{V-A}^{V-A*}),\nonumber \\
  \psi_1 &=& \arg(\mu\epsilon_{V+A}^{V+A*}), \quad
  \psi_2 = \arg(\mu\epsilon_{V-A}^{V+A*}), \nonumber\\
  \psi_{03} &=& \arg(\mu \,\mu_{S+P}^{S-P*}), \quad
  \psi_{04} = \arg(\mu \,\mu_{S-P}^{S-P*}), \nonumber\\
  \psi_{3} &=& \arg(\mu \,\epsilon_{S+P}^{S+P*}), \quad
  \psi_{4} = \arg(\mu \,\epsilon_{S-P}^{S+P*}), \nonumber\\
  \psi_{06} &=& \arg(\mu \,\mu_{T_L}^{T_L*}), \nonumber\\
  \psi_{5} &=& \arg(\mu \,\epsilon_{T_R}^{T_R*}), \quad
  \psi_{6} = \arg(\mu \,\epsilon_{T_L}^{T_R*}).
\end{eqnarray}

The coefficients $C_i$ and $C_i^{(SP,T)}$ in Table 2 are
\begin{eqnarray}
   C_0 &=& (\chi_F^2-1)A_{01}, \nonumber\\
   C_1 &=& (\chi_F-1)^2A_{01}, \quad  C_{1+} = (\chi_F+1)^2A_{01}, \nonumber\\
   C_2 &=& (\chi_F-1)(\chi_{2-}A_{03}-\chi_{1+}A_{04}), \nonumber\\
   C_3 &=&-(\chi_F-1)(\chi_{2+}A_{03}-\chi_{1-}A_{04}-
   \chi_P^\prime A_{05}+\chi_R^\prime A_{06}), \nonumber\\
   C_4 &=& \chi_{2-}^2A_{02}-\frac{2}{9}\chi_{1+}\chi_{2-}A_{03}+
   \frac{1}{9}\chi_{1+}^2A_{04},\nonumber \\
   C_5 &=& \chi_{2+}^2A_{02}-\frac{2}{9}\chi_{1-}\chi_{2+}A_{03}+
   \frac{1}{9}\chi_{1-}^2A_{04}+\chi_P^{\prime2}A_{08}-
   \chi_P^\prime\chi_R^\prime A_{07}+\chi_R^{\prime2}A_{09};
\end{eqnarray}
\begin{eqnarray}
   C_0^{SP} &=& -(\chi_F-1)\chi_F^{SP}A_{00}^{SP}, \nonumber\\
   C_1^{SP} &=& (\chi_F^{SP})^2A_{01}^{SP}, \nonumber\\
   %
   C_2^{SP} &=& -(\chi_F-1)\left(\chi_B^{\prime SP}+\chi_D^{\prime SP}\right)A_{05}^{SP} + (\chi_F-1)\chi_P^{\prime SP}A_{06}^{SP},\nonumber\\
   %
   C_{2+}^{SP} &=& (\chi_F-1)\left(\chi_B^{\prime SP}-\chi_D^{\prime SP}\right)A_{05}^{SP} + (\chi_F-1)\chi_P^{\prime SP}A_{06}^{SP},\nonumber\\
   C_3^{SP} &=& \left(\chi_B^{\prime SP}+\chi_D^{\prime SP}\right)^2A_{02}^{SP} - \left(\chi_B^{\prime SP}+\chi_D^{\prime SP}\right)\chi_P^{\prime SP}A_{03}^{SP} + \left(\chi_P^{\prime SP}\right)^2A_{04}^{SP},\nonumber\\
   %
   C_{3+}^{SP} &=& \left(\chi_B^{\prime SP}-\chi_D^{\prime SP}\right)^2A_{02}^{SP} + \left(\chi_B^{\prime SP}-\chi_D^{\prime SP}\right)\chi_P^{\prime SP}A_{03}^{SP} + \left(\chi_P^{\prime SP}\right)^2A_{04}^{SP};
\end{eqnarray}
\begin{eqnarray}
   C_0^T &=& \frac{T_1^{(3)}}{g_A}(\chi_F-1)A_{00}^T, \nonumber\\
   C_1^T &=& \left(\frac{T_1^{(3)}}{g_A}\right)^2A_{01}^T, \nonumber\\
   C_2^T &=& -(\chi_F-1)\left[(\chi_{RC_\sigma}^{T\prime}+
   \chi_{R}^{T\prime} +\chi_{RT_\sigma}^{T\prime}-\chi_{RT}^{T\prime})A_{01}
   + \left(\frac{1}{3}\chi_{\rm GT}^{T\prime}-
   2\chi_T^{T\prime}\right)A_{02}^T\right],\nonumber \\
   C_3^T &=&
   (\chi_{RC_\sigma}^{T\prime}+\chi_R^{T\prime}+
   \chi_{RT_\sigma}^{T\prime}-\chi_{RT}^{T\prime})^2A_{09}+
   \left(\frac{1}{3}\chi_{\rm GT}^{T\prime}-
   2\chi_T^{T\prime}\right)^2A_{03}^T~.
\end{eqnarray}
The coefficients $D_i$ and  $D_i^{(SP,T)}$ entering in Table 3
are:
\begin{eqnarray}
   D_0 &=&  (\chi_F^2-1)B_{01}, \nonumber\\
   D_{1} &=& (\chi_F-1)^2B_{01}, \quad  D_{1+} = (\chi_F+1)^2B_{01}, \nonumber\\
   D_{2-} &=& (\chi_F-1)\chi_{2-}B_{03-}, \quad D_{2} = -(\chi_F-1)\chi_{1+}B_{04}, \nonumber\\
   D_{3} &=& (\chi_F-1)(\chi_{2+}B_{03}-\chi_P^\prime B_{05}), \nonumber\\
   D_{3-} &=& -(\chi_F-1)(\chi_{1-}B_{04-}-\chi_P^\prime B_{05-}+\chi_R^\prime B_{06-}), \nonumber\\
   D_4 &=& -\chi_{2-}^2B_{02}+\frac{1}{9}\chi_{1+}^2B_{04},\nonumber \\
   D_5 &=& \chi_{2+}^2B_{02}-\frac{1}{9}\chi_{1-}^2B_{04}
- \chi_P^{\prime2}B_{08} + \chi_P^\prime\chi_R^\prime B_{07} -
\chi_R^{\prime2}B_{09};
\end{eqnarray}
\begin{eqnarray}
  D_{0-}^{SP} &=&  (\chi_F-1)\chi_F^{SP}B_{00-}^{SP}, \nonumber\\
  D_{1}^{SP} &=& -(\chi_F^{SP})^2B_{01}^{SP},\nonumber\\
  D_{2}^{SP} &=& -(\chi_F-1)(\chi_B^{\prime SP}+\chi_D^{\prime SP})B_{05}^{SP} + (\chi_F-1)\chi_P^{\prime SP}B_{06}^{SP}, \nonumber\\
  D_{2-}^{SP} &=& (\chi_F-1)\chi_P^{\prime SP} B_{02-}^{SP},\nonumber \\
  D_{2+}^{SP} &=& (\chi_F-1)(\chi_B^{\prime SP}-\chi_D^{\prime SP})B_{05}^{SP} + (\chi_F-1)\chi_P^{\prime SP}B_{06}^{SP}, \nonumber\\
  D_3^{SP} &=& (\chi_B^{\prime SP}+\chi_D^{\prime SP})^2 B_{02}^{SP} - (\chi_B^{\prime SP}+\chi_D^{\prime SP})\chi_P^{\prime SP} B_{03}^{SP} + (\chi_P^{\prime SP})^2 B_{04}^{SP},\nonumber\\
  D_{3+}^{SP} &=& (\chi_B^{\prime SP}-\chi_D^{\prime SP})^2 B_{02}^{SP} + (\chi_B^{\prime SP}-\chi_D^{\prime SP})\chi_P^{\prime SP} B_{03}^{SP} + (\chi_P^{\prime SP})^2 B_{04}^{SP};
\end{eqnarray}
\begin{eqnarray}
  D_{0-}^T &=&  \frac{T_1^{(3)}}{g_A}(\chi_F-1)B_{00-}^T, \nonumber\\
  D_{1}^T &=& -\left(\frac{T_1^{(3)}}{g_A}\right)^2B_{01}^T,\nonumber\\
  D_{2}^T &=& -(\chi_F-1)\left[(\chi_{RC_\sigma}^{T\prime}+
   \chi_{R}^{T\prime}+\chi_{RT_\sigma}^{T\prime}-\chi_{RT}^{T\prime})B_{01}
   + \left(\frac{1}{3}\chi_{\rm GT}^{T\prime}-
   2\chi_T^{T\prime}\right)B_{02}^T\right], \nonumber\\
  D_3^T &=& (\chi_{RC_\sigma}^{T\prime}+
  \chi_R^{T\prime}+\chi_{RT_\sigma}^{T\prime}-\chi_{RT}^{T\prime})^2B_{09}
  +\left(\frac{1}{3}\chi_{\rm GT}^{T\prime}-
   2\chi_T^{T\prime}\right)^2B_{03}^T,
\end{eqnarray}
where the integrated phase space factors are
\begin{equation}
\left( {\begin{array}{*{20}c}
   {A_{0k} },& {A_{0k}^{(SP,T)} }  \\
   {B_{0k} },& {B_{0k}^{(SP,T)} }  \\
\end{array}} \right) = \frac{1}{{\ln 2}}\frac{{a_{0\nu } }}{{\left( {m_e R} \right)^2 }}\int {\left( {\begin{array}{*{20}c}
   {a_{0k} },& {a_{0k}^{(SP,T)} }  \\
   {b_{0k} },& {b_{0k}^{(SP,T)} }  \\
\end{array}} \right)} d\Omega _{0\nu },
\end{equation}
with the phase space element $ d\Omega_{0\nu} $ defined as follows:
\begin{eqnarray}
\label{ps}
 d\Omega_{0\nu} = m_e^{-5} |{\bf p}_1| |{\bf p}_2| \varepsilon_1\varepsilon_2
\delta(\varepsilon_1+\varepsilon_2+E_f-E_i) d\varepsilon_1
d\varepsilon_2 d({\hat {\bf p}_1\cdot \hat {\bf p}_2})~.
\end{eqnarray}
The constant $a_{0\nu}$ and the kinematic factors $a_{0k}, a_{0k}^{(S,P,T)},
b_{0k}$ and $ b_{0k}^{(S,P,T)}$ entering above are defined as follows:
\begin{equation}
\label{a0nu}
a_{0\nu}=(G_Fg_A)^4|V_{ud}|^4m_e^9/(64\pi^5),
\end{equation}
\begin{eqnarray}
  && a_{01} = \alpha_++\beta_+, \quad
a_{02} = \left(\frac{\varepsilon_{21}}{m_e}\right)^2\beta_+, \quad
a_{03} = 2\frac{\varepsilon_{21}}{m_e}\beta_-, \quad a_{04} =
\frac{4}{9}\beta_+, \nonumber\\
  && a_{05} = \frac{4}{3}\left(\frac{\zeta\alpha_-}{m_eR}-2\alpha_+\right), \quad
a_{06} = \frac{8}{m_eR}\alpha_-, \quad a_{07} =
\frac{1}{3}\left(\frac{4}{m_eR}\right)^2(\zeta\alpha_+-2m_eR\alpha_-),\nonumber\\
  && a_{08} =
  \left(\frac{2}{3m_eR}\right)^2[\zeta^2\alpha_++4m_eR(m_eR\alpha_+-\zeta\alpha_-)],
  \quad a_{09} = \left(\frac{4}{m_eR}\right)^2\alpha_+;
\end{eqnarray}
\begin{eqnarray}
  && a_{00}^{SP} = \alpha_-, \quad a_{01}^{SP} = \alpha_+, \quad a_{02}^{SP}
  = 
  \frac{1}{(m_eR)^2} \left(\alpha_+ + \beta_+\right), \nonumber\\
&& a_{03}^{SP} = 
\frac{1}{3m_eR}\left[\frac{\zeta}{m_eR} \left(\alpha_+ + \beta_+\right) - 2\alpha_-\right], \nonumber\\
&& a_{04}^{SP} = \frac{1}{9}\left\{\left[\left(\frac{\zeta}{2m_eR}\right)^2+1\right]\alpha_+ + \left(\frac{\zeta}{2m_eR}\right)^2\beta_+ - \frac{\zeta}{m_eR}\alpha_- \right\}, \nonumber\\
&& a_{05}^{SP} = \frac{1}{m_eR} \left(\alpha_+ + \beta_+\right), \nonumber\\
&& a_{06}^{SP} = \frac{1}{6}\left[\frac{\zeta}{m_eR} \left(\alpha_+ + \beta_+\right) - 2\alpha_-\right] = \frac{m_eR}{2} \, a_{03}^{SP};
\end{eqnarray}
\begin{eqnarray}
  && a_{00}^T = 2\beta_-, \quad a_{01}^T = 16\alpha_+ = 16a_{01}^{SP}, \quad
  a_{02}^T = \frac{8\zeta\beta_+}{m_eR}, \nonumber\\
  && a_{03}^T = \left(\frac{8\zeta}{m_eR}\right)^2\beta_+;
\end{eqnarray}
\begin{eqnarray}
  && b_{01} = \gamma_+ + \delta_+, \quad
b_{02} = \left(\frac{\varepsilon_{21}}{m_e}\right)^2\delta_+,
\nonumber\\
  && b_{03} = 2\frac{\varepsilon_{21}}{m_e}\delta_+, \quad
b_{03-} = 2\frac{\varepsilon_{21}}{m_e}\delta_-, \nonumber\\
  && b_{04} = \frac{4}{9}\delta_+, \quad
b_{04-} = \frac{4}{9}\delta_-, \nonumber\\
  && b_{05} = \frac{8}{3}\gamma_+, \quad
b_{05-} = \frac{4}{3}\frac{\zeta\gamma_-}{m_eR},\nonumber \\
  && b_{06-} = \frac{8\gamma_-}{m_eR}, \quad
b_{07} = \frac{16}{3}\frac{\zeta\gamma_+}{(m_eR)^2}, \nonumber\\
  && b_{08} = \frac{16}{9}\left[\left(\frac{\zeta}{2m_eR}\right)^2-1\right]\gamma_+, \quad
b_{09} = \left(\frac{4}{m_eR}\right)^2\gamma_+;
\end{eqnarray}
\begin{eqnarray}
  && b_{00-}^{SP} = \gamma_-, \quad b_{01}^{SP} = \gamma_+=\frac{3}{8}b_{05}, \nonumber\\
  && b_{02}^{SP} = \frac{1}{(2m_eR)^2}(\gamma_++\delta_+), \quad
  b_{02-}^{SP} = \frac{1}{3}\gamma_- = \frac{1}{3} b_{00-}^{SP}, \nonumber\\
  && b_{03}^{SP} = \frac{\zeta}{6m_e^2R^2} (\gamma_++\delta_+), \nonumber\\
  && b_{04}^{SP} = \frac{1}{9} \left\{ \left[\left(\frac{\zeta}{2m_eR}\right)^2 - 1\right]\gamma_+ + \left(\frac{\zeta}{2m_eR}\right)^2\delta_+ \right\}, \nonumber\\
  && b_{05}^{SP} = \frac{1}{m_eR}(\gamma_++\delta_+), \quad
  	b_{06}^{SP} = \frac{\zeta}{6m_eR} (\gamma_++\delta_+);
\end{eqnarray}
\begin{eqnarray}
  && b_{00-}^T = 4\gamma_- = 4b_{00-}^{SP}, \quad b_{01}^T = 16\gamma_+=6b_{05}, \nonumber\\
  && b_{02}^T = \frac{8\zeta\delta_+}{m_eR}, \quad
  b_{03}^T = \left(\frac{8\zeta}{m_eR}\right)^2\delta_+,
\end{eqnarray}
where $\varepsilon_{21}=\varepsilon_2-\varepsilon_1$ is the difference in the
electron energy. The characteristic features of the
$P_{1/2}$-wave are expressed as
\begin{eqnarray}
   \zeta = 3\alpha Z+(\varepsilon_1+\varepsilon_2)R
\end{eqnarray}
and the Coulomb corrections appear as the following combinations
\begin{eqnarray}
&& \alpha_\pm = |\alpha_{-1-1}|^2 \pm |\alpha_{11}|^2, \quad
 \beta_\pm = |\alpha_{1-1}|^2 \pm |\alpha_{-11}|^2, \nonumber\\
&& \gamma_+ = 2\mbox{Re}(\alpha_{11}\alpha_{-1-1}^*), \quad
\gamma_- =2\mbox{Im}(\alpha_{11}\alpha_{-1-1}^*),\nonumber\\ &&
\delta_+ = 2\mbox{Re}(\alpha_{-11}\alpha_{1-1}^*), \quad \delta_-
= 2\mbox{Im}(\alpha_{-11}\alpha_{1-1}^*),
\end{eqnarray}
with $\alpha_{ij}=\tilde A_i(\varepsilon_2)\tilde
A_j(\varepsilon_1)$.

For the normalization constants in the approximation including terms up to
$(\alpha Z)^2$ \cite{Doi}
\begin{eqnarray}
 && \tilde A_{\pm1} = \sqrt{\frac{\varepsilon\mp m_e}
 {2\varepsilon} F_0(Z, \varepsilon)}, \label{tildeA} \label{tildA_pm} \nonumber\\
&& F_0 = \frac{4}{\Gamma^2(2\gamma_1+1)} (2pR)^{2(\gamma_1-1)}
 |\Gamma(\gamma_1+iy)|^2e^{\pi y}, \nonumber\\
&& \gamma_1 = \sqrt{1-(\alpha Z)^2}, \quad y = \alpha
Z\varepsilon/p, \label{gamma1}
\end{eqnarray}
we have
\begin{eqnarray}
   && \left(%
\begin{array}{c}
  \alpha_+ \\
  \beta_+ \\
\end{array}%
\right) = \frac{1}{2}(\varepsilon_1\varepsilon_2\pm m_e^2)C_{00},
\quad
\left(%
\begin{array}{c}
  \alpha_- \\
  \beta_- \\
\end{array}%
\right) = \frac{1}{2}(\varepsilon_2\pm \varepsilon_1)m_eC_{00}, \\
  && \gamma_+=\delta_+=\frac{1}{2} |{\bf p}_1||{\bf p}_2|C_{00},
\quad \gamma_-=\delta_-=0,
\label{gamma-}
\end{eqnarray}
where
\begin{eqnarray}
   C_{00}=\frac{F_0(Z, \varepsilon_2)F_0(Z,
   \varepsilon_1)}{\varepsilon_2\varepsilon_1}.
\end{eqnarray}
Note that using Eq. (\ref{gamma-}) the expressions for
$\mathcal{B}$ from Table 3 are reduced to the form shown in Table
4.

\begin{center} \vspace*{2mm} \noindent Table 4: Expressions
for $\mathcal{B}$ in Eq. (\ref{K}) for the stated choice of
$\epsilon_\alpha^\beta$ for the $\tilde A_{\pm1}$ from Eq.
(\ref{tildeA}).
\nopagebreak \label{Tab4} \vspace*{3mm}
\begin{tabular}{|c|c|}
  \hline
  $\epsilon$ & $\mathcal{B}$ \\
  \hline
  $\epsilon^{V-A}_{V-A}$ & ${\mathcal B}_0 + 4D_1|\mu||\mu_{V-A}^{V-A}|c_{02}
+ 4D_1|\mu_{V-A}^{V-A}|^2$ \\
  \hline
  $\epsilon^{V-A}_{V+A}$ & ${\mathcal B}_0 +
4D_0|\mu||\mu_{V+A}^{V-A}|c_{01} +
4D_{1+}|\mu_{V+A}^{V-A}|^2$ \\
  \hline
  $\epsilon^{V+A}_{V-A}$ & ${\mathcal B}_0 + D_{3}|\mu||\epsilon^{V+A}_{V-A}|
c_2 + D_5|\epsilon^{V+A}_{V-A}|^2$ \\
  \hline
  $\epsilon^{V+A}_{V+A}$ & ${\mathcal B}_0 + D_{2}|\mu||\epsilon^{V+A}_{V+A}|
c_1 + D_4|\epsilon^{V+A}_{V+A}|^2$ \\
  \hline
  $\epsilon^{S-P}_{S\mp P}$ & ${\mathcal B}_0 + 4D_1^{SP}|\mu_{S\mp P}^{S-P}|^2$ \\
  \hline
  $\epsilon^{S+P}_{S-P}$ & ${\mathcal B}_0 + 4D_{2}^{SP}|\mu||\epsilon^{S+P}_{S- P}|
c_{4} + 4D_3^{SP}|\epsilon^{S+P}_{S- P}|^2$ \\
  \hline
  $\epsilon^{S+P}_{S+P}$ & ${\mathcal B}_0 + 4D_{2+}^{SP}|\mu||\epsilon^{S+P}_{S+ P}|
c_{3} + 4D_{3+}^{SP}|\epsilon^{S+P}_{S+P}|^2$ \\
  \hline
  $\epsilon^{T_L}_{T_L}$ & ${\mathcal B}_0 + 4D_1^T|\mu_{T_L}^{T_L}|^2$ \\
  \hline
  $\epsilon^{T_L}_{T_R}$, $\epsilon^{T_R}_{T_L}$ & ${\mathcal B}_0$ \\
  \hline
  $\epsilon^{T_R}_{T_R}$ & ${\mathcal B}_0 + D_2^T|\mu||\epsilon^{T_R}_{T_R}|
c_5 + D_3^T|\epsilon^{T_R}_{T_R}|^2$ \\
  \hline
\end{tabular}
\end{center}
\vspace*{3mm}

In the definitions of $C_i$ and $D_i$ we use some combinations of
nuclear parameters (which are assumed to be real) similar to the ones in Ref.~\cite{Doi}. Thus,
\begin{eqnarray}
&\ds
\chi_{2\pm}=\chi_{\rm GT\omega}\pm\chi_{F\omega}-\frac{1}{9}\chi_{1\mp};\quad
\chi_{1\pm}=(\chi^\prime_{\rm GT}-6\chi_T^\prime) \pm
3\chi_F^\prime;\nonumber\\ &\ds
\chi_F=\left(\frac{g_V}{g_A}\right)^2\frac{M_F}{M_{\rm GT}};\quad
\chi_k=\frac{g_V}{g_A}\frac{M_k}{M_{\rm GT}}, \ k=P,\,R,\,RT;\nonumber
\\ &\ds \chi_k=\frac{M_k}{M_{\rm GT}}, \
k=T,\,GT,\,RC_\sigma,\,RT_\sigma; \nonumber\\ 
&\ds
\chi_F^{SP}=\frac{F_S^{(3)}}{g_V}\chi_F; \quad
\chi_P^{SP}=\frac{F_S^{(3)}}{g_V}\left(\frac{g_V}{g_A}\right)^2\frac{M_F}{M_{{\rm GT}}};\nonumber\\ 
&\ds 
\chi_B^{SP}=\frac{F_P^{(3)}}{g_A}\frac{M_B}{M_{{\rm GT}}};
\quad 
\chi_D^{SP}=\frac{F_S^{(3)}}{g_A}\frac{g_V}{g_A}\frac{M_D}{M_{{\rm GT}}};
\nonumber\\
&\ds \chi_k^T=\frac{T_1^{(3)}}{g_A}\chi_k, \
k=R,\,RT,\,RC_\sigma,\,RT_\sigma; \quad
\chi_k^T=\frac{T_1^{(3)}}{g_A}\frac{M_k^T}{M_{\rm GT}}, \ k=GT,\,T,
\end{eqnarray}
where the index $F$ refers to Fermi, $GT$ to Gamow--Teller, $T$ to
tensor, $P$ to the $P$-wave effect and $R$ to the recoil effect.
If $\chi$ has prime or the index $\omega$ than the same has the
according matrix element in the numerator. The nuclear matrix
elements defined below contain the operator
 $\tau^a_+=(\tau_1+i\tau_2)^a/2$ converting
the $a$-th neutron into the $a$-th proton, and the initial (final)
nuclear state are denoted by $|0^+_i\rangle$ ($\langle0^+_f|$)
\begin{eqnarray}
&&\ds   M_{F} = \sum\limits_N\langle0^+_f||\sum\limits_{a\neq
b}h_+(r_{ab}, E_N)
   \tau^a_+\tau^b_+||0^+_i\rangle, \\
&&\ds   M_{\rm GT} = \sum\limits_N\langle0^+_f||\sum\limits_{a\neq
    b}h_+(r_{ab},
    E_N){\bm\sigma}_a\cdot{\bm\sigma}_b\tau^a_+\tau^b_+||0^+_i\rangle,
    \label{M_GT}\\
&&\ds   M_{T} = \sum\limits_N\langle0^+_f||\sum\limits_{a\neq
b}h_+(r_{ab}, E_N)
   \left[({\bm\sigma}_a\cdot\hat{\bf r}_{ab})({\bm \sigma}_b\cdot\hat{\bf r}_{ab})-
\frac{1}{3}{\bm\sigma}_a\cdot{\bm\sigma}_b\right]\tau^a_+\tau^b_+||0^+_i\rangle, \\
&&\ds   M_{\rm GT}^\prime =
\sum\limits_N\langle0^+_f||\sum\limits_{a\neq
    b}h_+^\prime(r_{ab}, E_N){\bm\sigma}_a\cdot{\bm\sigma}_b\tau^a_+\tau^b_+||0^+_i\rangle,\\
&&\ds   M_{F}^\prime =
\sum\limits_N\langle0^+_f||\sum\limits_{a\neq b}h_+^\prime
(r_{ab}, E_N) \tau^a_+\tau^b_+||0^+_i\rangle, \\
&&\ds   M_{T}^\prime =
\sum\limits_N\langle0^+_f||\sum\limits_{a\neq b}h_+^\prime
(r_{ab}, E_N)\left[({\bm\sigma}_a\cdot\hat{\bf r}_{ab})({\bm
\sigma}_b\cdot\hat{\bf r}_{ab})-
\frac{1}{3}{\bm\sigma}_a\cdot{\bm\sigma}_b\right]\tau^a_+\tau^b_+||0^+_i\rangle, \\
&&\ds   M_{P}^\prime =
\sum\limits_N\langle0^+_f||\sum\limits_{a\neq b}h_+^\prime
(r_{ab}, E_N)\frac{ir_{+ab}}{2r_{ab}}
\left\{(\bm\sigma_a-\bm\sigma_b)\cdot
\left[\hat{\bf r}_{ab}\times {\bf \hat r}_{+ab}\right]\right\}\tau^a_+\tau^b_+||0^+_i\rangle, \\
&&\ds   M_{R}^\prime =
\sum\limits_N\langle0^+_f||\sum\limits_{a\neq b}h_+^\prime
(r_{ab}, E_N)\frac{R}{2r_{ab}} \hat{\bf
r}_{ab}\cdot({\bm\sigma}_a\times {\bf D}_b +
{\bf D}_a\times {\bm\sigma}_b)\tau^a_+\tau^b_+||0^+_i\rangle, \\
&&\ds   M_{\rm GT\omega} =
\sum\limits_N\langle0^+_f||\sum\limits_{a\neq
    b}h_{0\omega}(r_{ab}, E_N){\bm\sigma}_a\cdot{\bm\sigma}_b\tau^a_+\tau^b_+||0^+_i\rangle,\\
&&\ds   M_{F\omega} =
\sum\limits_N\langle0^+_f||\sum\limits_{a\neq
b}h_{0\omega}(r_{ab}, E_N)
   \tau^a_+\tau^b_+||0^+_i\rangle, \\
%
%
&&\ds   M_B^\prime = \sum\limits_N\langle0^+_f||\sum\limits_{a\neq
		b}\frac{iR}{2r}h_+^\prime(r_{ab}, E_N)
	\hat{\bf r}_{ab} ({\bm\sigma}_aB_b-B_a{\bm\sigma}_b) \tau^a_+\tau^b_+||0^+_i\rangle, \\
&&\ds   M_D^\prime = \sum\limits_N\langle0^+_f||\sum\limits_{a\neq
		b}\frac{iR}{2r}h_+^\prime(r_{ab}, E_N)
	\hat{\bf r}_{ab} ({\bm D}_a-{\bm D}_b) \tau^a_+\tau^b_+||0^+_i\rangle, \\
&&\ds   M_{\rm GT}^{T\prime} =
\sum\limits_N\langle0^+_f||\sum\limits_{a\neq b}h_+^\prime(r_{ab},
E_N)\frac{iR}{r}{\bm\sigma}_a\cdot{\bm\sigma}_b
\tau^a_+\tau^b_+||0^+_i\rangle,\\
&&\ds   M_{T}^{T\prime} =
\sum\limits_N\langle0^+_f||\sum\limits_{a\neq b}h_+^\prime
(r_{ab}, E_N)\frac{iR}{r}\left[({\bm\sigma}_a\cdot\hat{\bf
r}_{ab})({\bm \sigma}_b\cdot\hat{\bf r}_{ab})-
\frac{1}{3}{\bm\sigma}_a\cdot{\bm\sigma}_b\right]\tau^a_+\tau^b_+||0^+_i\rangle,
\\
&&\ds   M_{RT}^\prime =
\sum\limits_N\langle0^+_f||\sum\limits_{a\neq b}h_+^\prime
(r_{ab}, E_N)\frac{R}{2r}\hat{\bf
r}_{ab}\cdot ({\bf T}_a-{\bf T}_b) \tau^a_+\tau^b_+||0^+_i\rangle, \\
&&\ds   M_{RC_\sigma}^\prime =
\sum\limits_N\langle0^+_f||\sum\limits_{a\neq b}h_+^\prime
(r_{ab}, E_N)\frac{iR}{2r} (\hat{\bf r}_{ab}\cdot{\bm\sigma}_aC_b
- C_a\hat{\bf r}_{ab}\cdot{\bm\sigma}_b)\tau^a_+\tau^b_+||0^+_i\rangle, \\
&&\ds   M_{RT_\sigma}^\prime =
\sum\limits_N\langle0^+_f||\sum\limits_{a\neq b}h_+^\prime
(r_{ab}, E_N)\frac{iR}{2r} \hat{\bf
r}_{ab}\cdot({\bm\sigma}_a\times {\bf T}_b + {\bf T}_a\times
{\bm\sigma}_b)\tau^a_+\tau^b_+||0^+_i\rangle~.
\end{eqnarray}
In the above expressions, the neutrino potentials
 $h_{i}(r_{ab}, \langle E_N\rangle)$ are defined as follows:
\begin{eqnarray}
 && h_+(r_{ab},\langle E_N\rangle) = \frac{R}{4\pi^2}\int\frac{d{\bf k}}{\omega}
  \left(\frac{1}{\omega+A_1}+\frac{1}{\omega+A_2}\right)e^{i{\bf k\cdot r}}
  \simeq RH(r, \bar A), \\
 && h_0(r_{ab},\langle E_N\rangle) = \frac{1}{2\pi^2\varepsilon_{12}}
 \int\frac{d{\bf k}}{\omega} \left(\frac{1}{\omega+A_1} -
 \frac{1}{\omega+A_2}\right)e^{i{\bf k\cdot r}} \nonumber\\
 && \simeq 2H(r, \bar A)+r\frac{\partial}{\partial r}H(r, \bar A), \\
 && h_{0\omega}(r_{ab},\langle E_N\rangle) = h_+ - \bar ARh_0, \quad
  h_+^\prime(r_{ab},\langle E_N\rangle) = h_+ + \bar ARh_0, \\
 && h_R(r_{ab},\langle E_N\rangle) = -\frac{\bar A}{m_p}\left[
  \frac{2}{\pi}\left(\frac{R}{r}\right)^2 - \bar ARh_+ \right],
\end{eqnarray}
with
\begin{equation}
  H(r,\bar A) = \frac{1}{2\pi^2}\int\frac{d{\bf k}}{\omega}
  \frac{e^{i{\bf k\cdot r}}}{\omega+\bar A},
\end{equation}
\begin{equation}
  A_j=\varepsilon_j+\langle E_N\rangle-E_i, \ i=1,2; \quad
  \bar A = \langle E_N\rangle - (E_i+E_f)/2,
\end{equation}
where $r_{ab}$ is the distance between the nucleons $a$ and $b$,
and $\langle E_N\rangle$ is the average energy of the intermediate
nucleus $N$.

To derive the expressions for $\mathcal{A}$ and $\mathcal{B}$
shown in Tables 2 and 3 we have used the formulas:
\begin{eqnarray}
  && C_1^A\left(\epsilon_{S\mp P}^{S-P}\right) =
  2\,M_{\rm GT}\frac{m_{S\mp
  		P}^{S-P}}{m_e}
  \chi_F^{SP}, \quad C_1^A\left(\epsilon_{S\mp P}^{S+P}\right) = 0, \nonumber\\
  && C_2^A\left(\epsilon_{S\mp P}^{S-P}\right) = 0, \quad
  C_2^A\left(\epsilon_{S\mp P}^{S+P}\right) = 
  	2M_{\rm GT}\epsilon_{S\mp P}^{S+P}\chi_P^{\prime SP}, \quad C_5^A\left(\epsilon_{S\mp P}^{S\mp P}\right) = 0, \nonumber\\
  && C_{4R}^B\left(\epsilon_{S\mp P}^{S-P}\right) = 0, \quad
  C_{4R}^B\left(\epsilon_{S\mp P}^{S+P}\right) = 
  	M_{\rm GT}\epsilon_{S\mp P}^{S+P} (\mp\chi_B^{\prime SP} - \chi_D^{\prime SP});
\end{eqnarray}

\begin{eqnarray}
  && Z_1^X\left(\epsilon_{V-A}^{V-A}\right) = M_{\rm GT}\left(\mu
+ 2\mu_{V-A}^{V-A}\right)(\chi_F-1), \nonumber\\
  && Z_1^X\left(\epsilon_{V+A}^{V-A}\right) = M_{\rm GT} \left[\mu
(\chi_F-1)+2\mu_{V+A}^{V-A}(\chi_F+1)\right], \nonumber\\
  && Z_3^X\left(\epsilon_{V\mp A}^{V+A}\right) = \pm M_{\rm GT} \epsilon_{V\mp A}^{V+A}
(\chi_{\rm GT\omega}\pm\chi_{F\omega}),\nonumber \\
  && Z_4^X\left(\epsilon_{V\mp A}^{V+A}\right) = \mp \frac{1}{3} M_{\rm GT}
\epsilon_{V\mp A}^{V+A}\chi_{1\mp}, \nonumber\\
  && Z_6^Y\left(\epsilon_{V-A}^{V+A}\right)\frac{r}{r_+} = M_{\rm GT}
\epsilon_{V-A}^{V+A}\chi_P^\prime, \nonumber\\
  && Z_{4R}^Y\left(\epsilon_{V-A}^{V+A}\right) = M_{\rm GT}
\epsilon_{V-A}^{V+A}\chi_R^\prime;
\end{eqnarray}

\begin{eqnarray}
  && W_1^U\left(\epsilon_{T_L}^{T_L}\right) =
  -4M_{\rm GT}\mu_{T_L}^{T_L}\frac{T_1^{(3)}}{g_A}, \nonumber\\
  && W_{4R}^V\left(\epsilon_{T_R}^{T_R}\right) =
  -2M_{\rm GT}\epsilon_{T_R}^{T_R}\frac{T_1^{(3)}}{g_A} (\chi_{RC_\sigma}^\prime
  + \chi_{R}^\prime+\chi_{RT_\sigma}^{T\prime}-\chi_{RT}^{T\prime}), \nonumber\\
  && W_2^U\left(\epsilon_{T_R}^{T_R}\right)\frac{r}{r_+} =
  2iM_{\rm GT}\epsilon_{T_R}^{T_R} \frac{T_1^{(3)}}{g_A}
  \chi_{\rm GT}^\prime, \nonumber\\
  && W_7^U\left(\epsilon_{T_R}^{T_R}\right)\frac{r}{r_+} =
  -4iM_{\rm GT}\epsilon_{T_R}^{T_R}\frac{T_1^{(3)}}{g_A}\left(
  \frac{1}{3}\chi_{\rm GT}^\prime+2\chi_T^\prime \right).
\end{eqnarray}
For all other arguments $\epsilon_\alpha^\beta$ these nucleon
matrix elements have zero values, except for
\begin{eqnarray}
  && Z_1^X\left(\epsilon_{V\mp A}^{V-A}=0\right) =
  M_{\rm GT}\mu(\chi_F-1).
\end{eqnarray}

We have calculated the numerical values of the
 integrated kinematic factors $A_{0i}$,
$A_{0i}^{(SP,\,T)}$, $B_{0i}$, and $B_{0i}^{(SP,\,T)}$
for all the five nuclei of current experimental interest. We shall
use them in the results shown below in Table 6 for the angular
coefficient $K$. However, as we will focus in
this paper mainly on the $0\nu2\beta$ decay of the $^{76}{\mbox{Ge}}$
nucleus, we give the values of these factors for this nucleus
in Table 5, where we have used
\begin{eqnarray}
   Q=E_i-E_f-2m_e=2.039\ \mbox{MeV}~,
\end{eqnarray}
taken from Ref.~\cite{Zuber}, the scaling factor for the
neutrino potentials
\begin{eqnarray}
  R=r_0A^{1/3}, \quad r_0=1.1\ \mbox{fm},
\end{eqnarray}
and the values of $g_A=1.254$ and $|V_{ud}|=0.97377$~\cite{PDG}. 
The values of $A_{00}^T$ and $B_{03}$ are of the order of
$10^{-44}$ yr$^{-1}$. Hence these values are not given in Table 5
and the terms with $A_{00}^T$ and $B_{03}$ can be safely
neglected.

\vspace*{5mm}
\begin{center}
\noindent Table 5: The integrated kinematic $A$- and
$B$-factors $[\mbox{in}~10^{-15} \mbox{yr}^{-1}]$ for the $0^+\to0^+$
transition\\ of the $0\nu2\beta$ decay of $^{76}$Ge.
\label{Tab5}
\nopagebreak
\vspace*{3mm}

\begin{tabular}{|c|c||c|c|}
\hline $A_{01}$& 6.69           & $B_{01}$& 5.45 \\
  \hline
$A_{02}$& 1.09$\times 10$  & $B_{02}$& 8.95\\
  \hline
$A_{03}$& 3.76           & $B_{03}$ & --- \\
  \hline
$A_{04}$& 1.30           & $B_{04}$& 1.21   \\
  \hline
$A_{05}$& 2.08$\times 10^2$& $B_{05}$& 7.27\\
  \hline
$A_{06}$& 1.69$\times 10^3$& --- & --- \\
  \hline
$A_{07}$& 1.05$\times 10^5$& $B_{07}$& 7.72$\times 10^4$\\
  \hline
$A_{08}$& 6.59$\times 10^3$& $B_{08}$& 4.97$\times 10^3$\\
  \hline
$A_{09}$& 4.14$\times 10^5$& $B_{09}$& 3.00$\times 10^5$\\
  \hline
  \hline
$A_{00}^{SP}$ & 2.55     & --- & --- \\
  \hline
$A_{01}^{SP}$ & 3.77     & $B_{01}^{SP}$& 2.73 \\
  \hline
$A_{02}^{SP}$&$4.60\times 10^4$& $B_{02}^{SP}$& $1.40\times10^2$   \\
\hline
$A_{03}^{SP}$&$1.17\times 10^4$&$B_{03}^{SP}$& $7.23\times10$ \\
\hline
$A_{04}^{SP}$&$7.45\times 10^2$&$B_{04}^{SP}$& $9.30$ \\
\hline
$A_{05}^{SP}$&$5.55\times 10^2$&$B_{05}^{SP}$& $6.77$ \\
\hline
$A_{06}^{SP}$&$7.06\times 10$&$B_{06}^{SP}$& $5.82\times10$ \\
  \hline
  \hline
$A_{01}^T$ & 6.03$\times 10$   &  $B_{01}^T$& 4.36$\times 10$\\
  \hline
$A_{02}^T$& 1.50$\times 10^{3}$ & $B_{02}^T$& 1.40$\times 10^{3}$\\
  \hline
$A_{03}^T$& 7.67$\times 10^{5}$ & $B_{03}^T$& 7.16$\times 10^{5}$\\
  \hline
 \end{tabular}
\end{center}

We recall that the analytic expressions associated with the
coefficients $\epsilon_{V\mp A}^{V+A}$ given in this section and
the values of $A_{0i}$ from Table 5 confirm the results of Ref.
\cite{Doi}.  The analytic expressions associated with the
coefficients $\epsilon_{V\mp A}^{V-A}$, $\epsilon_{S\mp P}^{S\mp
P}$, $\epsilon_{T_{L,R}}^{T_{L,R}}$ and the values of
$A_{0i}^{(SP,T)}$, $B_{0i}$, $B_{0i}^{(SP,T)}$ from Table 5
transcend the earlier work.


\section{Analysis of the electron angular correlation}
\subsection{Qualitative analysis}

If the effects of all the interactions beyond the SM extended by
the $\nu_M$s, which we call the ``nonstandard" effects, are zero
(i.e., all $\epsilon^\beta_\alpha=0$), then $K = B_{01}/A_{01}$.
Its values are given in Table 6 for various decaying nuclei.
 We will concentrate on the case of $^{76}$Ge nucleus in the
following. In this case the correlation (\ref{dG}) is proportional
to $1-0.81\cos\theta$. (Note that in the limit of $m_e/(E_i -
E_f)\to 0$ we have $\alpha_++\beta_+ =\gamma_++\delta_+$ and
$K=1$.) Tables 2 and 4 show that the presence of the
``nonstandard" parameters $\epsilon_{V\mp A}^{V-A}$,
$\epsilon_{T_R}^{T_L}$ or $\epsilon_{T_L}^{T_R}$ does not change
the value of $K$ and therefore the form of the angular
correlation. The presence of any other parameter
$\epsilon_\alpha^\beta$ does change this correlation. Due to the large values of $A_{0i}^{SP}$, $B_{0i}^{SP}$, $A_{0i}^T$ and $B_{0i}^T$ for $i\geq2$ there are three additional nonstandard parameters that can change the form of the angular correlation, namely, $\epsilon_{S\mp P}^{S+P}$ and $\epsilon_{T_R}^{T_R}$.

\vspace*{3mm}
\begin{center}
 \noindent Table 6: The values of angular correlation
coefficient $K$ for various decaying nuclei for the SM extended by
the $\nu_M$s.
\label{Tab6}

\nopagebreak

\vspace*{3mm}

\begin{tabular}{|c|c|c|c|c|c|}
  \hline
         &$^{76}\mbox{Ge}$      & $^{82}\mbox{Se}$       &$^{100}\mbox{Mo}$&$^{130}\mbox{Te}$ &$^{136}\mbox{Xe}$ \\
  \hline
     $K$ & 0.81         &  0.88         &  0.88         &  0.85        &   0.84   \\
  \hline
\end{tabular}
\end{center}
\vspace*{3mm}
Using Table 1 and taking into account the fact that
$|\mu_\alpha^\beta|$ are suppressed in comparison with
$|\epsilon_\alpha^\beta|$ by the factor $m_i/m_e$ (the chiral
suppression), we find the coefficient $K$ and the set
$\{\epsilon\}$ of nonzero $\epsilon_\alpha^\beta$s that change the
 $1-0.81\cos\theta$ form of the correlation for the SM plus $\nu_M$s,
see Table 7 (the lower two entries). They correspond to the
following extensions of the SM:  $\nu_M$s plus RPV SUSY
\cite{SUSY}, $\nu_M$s plus right-handed currents (RC) (connected
with right-handed $W$ bosons \cite{Doi} or vector LQs~\cite{LQ}), and $\nu_M$s plus scalar LQs (SLQ)~\cite{LQ}.
Hence, the angular coefficient $K$ can signal the presence of these
NP interactions.

\vspace*{3mm}
\begin{center}
 \noindent Table 7: The angular correlation coefficient $K$
for various SM extensions for decays of $^{76}$Ge. \label{Tab7}

\nopagebreak

\vspace*{3mm}

\begin{tabular}{|c|c|c|}
  \hline
  SM extension & $\{\epsilon\}$ & $K$ \\
  \hline
  $\nu_M$ & --- & 0.81 \\
  \hline
  $\nu_M$+RPV SUSY & $\epsilon_{S+P}^{S+P}$, $\epsilon_{T_R}^{T_R}$ & $-1<K<1$ \\
  \hline
  $\nu_M+$RC & $\epsilon_{V\mp A}^{V+A}$ & $-1<K<1$ \\
  \hline
  $\nu_M+{\rm SLQ}$ & $\epsilon_{S\mp P}^{S+P}$ & $-1<K<1$ \\
  \hline
\end{tabular}
\end{center}
\vspace*{3mm}

We remark here that in our earlier analysis~\cite{Ali:2006iu} we
had neglected the $P$-wave and recoil effects, which is not a
good assumption. Our current study shows that these
 effects give significant contribution to
the terms with $\epsilon_{V-A}^{V+A}$, $\epsilon_{S\mp P}^{S+P}$ and $\epsilon_{T_R}^{T_R}$.
Hence, they have to be included in any realistic analysis of the
data, as and when it becomes available. Including them,
 not only the model called $\nu_M +$ RC but also the
models $\nu_M+{\rm RPV}$ and $\nu_M+{\rm SLQ}$ can essentially change the angular coefficient
$K$ from being 0.81 in the decay of the $^{76}\mbox{Ge}$ nucleus.
Left-right symmetric models belong to the class $\nu_M +$ RC and
we have studied these models in detail in section 4, where the
correlations among the parameters $K$, $T_{1/2}$ and either $m_
{W_R}$ or $\zeta$ are worked out for the case $|\langle
m\rangle|\neq0$, $\cos\psi_i=0$ considered in section 3.2.

Note that the decay half-life and angular correlation do not give
any bounds on the parameters $\epsilon_{T_R}^{T_L}$ and
$\epsilon_{T_L}^{T_R}$ because the according expressions for
$\mathcal{A}$ and $\mathcal{B}$ do not depend on them.

\subsection{Quantitative analysis}

Let us now consider some particular cases for the parameter space.
We will analyze only the terms with $\epsilon_{V\mp A}^{V\mp A}$
as the corresponding nuclear matrix elements have been workd out in the
literature. We use various types of QRPA model for the
$^{76}\mbox{Ge}$ nucleus~\cite{Pantis,Kortelainen:2007rh}
as a test case.

 Using the case of $|\langle m\rangle|=0$, which gives conservative upper
bounds on $|\mu_\alpha^\beta|$ and  $|\epsilon_\alpha^\beta|$, the
decay half-life is expressed from Eq. (\ref{dG}) as
\begin{equation}\label{T_1/2}
    T_{1/2} = \ln2/\Gamma = \left( |M_{\rm GT}|^2\mathcal{A}
    \right)^{-1}.
\end{equation}

From Eq. (\ref{T_1/2}), using Tables 2, 5 and the values of the
nuclear matrix elements reported in Refs.~\cite{Pantis,Kortelainen:2007rh},
we have the following expressions for the half-life [in yr] for various
choices of the parameters $|\mu_{V\mp A}^{V-A}|$ and
$|\epsilon_{V\mp A}^{V+A}|$, taking only one parameter at a time:
\begin{eqnarray}\label{pnQRPA}
    T_{1/2}=1.1(1.3)\times10^{12}|\mu_{V-A}^{V-A}|^{-2},
    \quad T_{1/2}=3.2(4.0)\times10^{12}|\mu_{V+A}^{V-A}|^{-2},
\end{eqnarray}
\begin{eqnarray}\label{QRPA1}
    T_{1/2}=4.0(21)\times10^{12}|\mu_{V-A}^{V-A}|^{-2},
    \quad T_{1/2}=4.5(6.8)\times10^{12}|\mu_{V+A}^{V-A}|^{-2},
\end{eqnarray}
\begin{eqnarray}\label{QRPA2}
    T_{1/2}=3.7(27)\times10^8|\epsilon_{V-A}^{V+A}|^{-2},
    \quad T_{1/2}=1.0(9.7)\times10^{13}|\epsilon_{V+A}^{V+A}|^{-2}.
\end{eqnarray}
Eq.~(\ref{pnQRPA}) corresponds to using the pnQRPA model with
particle-particle strength parameter $g_{pp}$=1.02(1.06)
\cite{Kortelainen:2007rh} and Eqs~(\ref{QRPA1})--(\ref{QRPA2})
correspond to using the QRPA model without (with) the p-n pairing
\cite{Pantis} (note that the definitions of the nuclear matrix
elements $\chi_P^\prime$ and $\chi_R$ in Ref. \cite{Pantis} differ
from $\chi_P^\prime$ and $\chi_R^\prime$ in Ref. \cite{Doi} by the
factors $1/2$ and $4/(m_eR)$, respectively). Comparing the
numerical results in these equations, we note that the dispersion
in the half-lifes is less marked for the coefficient $
|\mu_{V+A}^{V-A}|$. However, the half-lifes involving the
coefficients $ |\mu_{V-A}^{V-A}|$ and $|\epsilon_{V+A}^{V\pm A}|$
show a very strong nuclear matrix element dependence. For the QRPA
 model worked out in \cite{Pantis}, it is not clear to us if this is
due to a numerical artifact or the treatment of the isoscalar neutron-proton
pairing. An important, and related point, is how to fix correctly the
particle-particle strength of the nuclear Hamiltonian. Fixing the particle-particle
pairing parameter, and varying it as done in \cite{Kortelainen:2007rh},
leads to rather stable values for the half-life of $^{76}\mbox{Ge}$ nucleus.
Clearly, these issues remain to be further discussed and clarified.
A detailed discussion of these nuclear models will take us far afield
from the main point of our paper. The theoretical uncertainty
in the nuclear matrix elements \cite{Vogel:2006sq,0503063}
plays an essential role in the numerical analysis. However, as we show below,
the nuclear-model dependence of the angular coefficient $K$ is rather modest.

The fact that the dependence of $K$ on the nuclear matrix elements
is much weaker than the uncertainty in $T_{1/2}$ from this source
is illustrated in Table 8 for QRPA models
\cite{Pantis,Kortelainen:2007rh} for the assumed values of the
parameters: $|\mu_{V\mp
A}^{V-A}|=|\epsilon_{V+A}^{V+A}|=5\times10^{-7}$,
$|\epsilon_{V-A}^{V+A}|=5\times10^{-9}$. It is clear from Table 8
that measuring $K$ with 10\% accuracy (or better) produces useful
experimental data that could be sensitive to the new physics. We
note that for the parameters ${\mu _{V \mp A}^{V - A} }$ the
angular coefficient does not depend actually on the nuclear matrix
elements as it is seen from Tables 2, 3 (for $|\mu | = 0$) and
Eqs. (28), (31): $K = \left( {\chi _F  \mp 1} \right)^2 B_{01}
/\left[ {\left( {\chi _F  \mp 1} \right)^2 A_{01} } \right] =
B_{01} /A_{01}  \simeq 0.81$.
\vspace{3mm}
\begin{center}
\noindent Table 8: $T_{1/2}$ and $K$ for the fixed values of the
parameters $|\mu_{V\mp A}^{V-A}|, |\epsilon_{V\mp A}^{V+A}|$ for
decay of $^{76}\mbox{Ge}$ for the case of $|\langle m\rangle|=0$
in QRPA without (with) p-n pairing \cite{Pantis} [pnQRPA with
$g_{pp}$=1.02(1.06) \cite{Kortelainen:2007rh}]. \label{Tab8}

\nopagebreak

\vspace*{3mm}

\begin{tabular}{|c|c|c|c|c|}
  \hline
  & $|\mu_{V-A}^{V-A}|=5\times 10^{-7}$ & $|\mu_{V+A}^{V-A}|=5\times 10^{-7}$
  & $|\epsilon_{V-A}^{V+A}|= 5\times10^{-9}$ & $|\epsilon_{V+A}^{V+A}|=5\times 10^{-7}$ \\
  \hline
    $T_{1/2}/(10^{25}~\mbox{yr})$   & 1.6(8.4)[0.44(0.52)] & 1.8(2.7)[1.3(1.6)] & 1.5(11) &  4.0(39) \\
  \hline
  $K$   & 0.81(0.81)[0.81(0.81)] & 0.81(0.81)[0.81(0.81)] & $-0.73(-0.73)$ & $-0.79(-0.87)$ \\
  \hline
\end{tabular}
\end{center}
\vspace{3mm}

Using the numerical results given above,
 the current lower bound $T_{1/2}>1.6\times 10^{25}~\mbox{yr}$
for the $^{76}\mbox{Ge}$ nucleus~\cite{Aalseth} yields the upper
bounds on the parameters $|\mu_{V\mp A}^{V-A}|$ and
$|\epsilon_{V\mp A}^{V+A}|$ shown in Table 9. The bound on
$|\epsilon_{V-A}^{V+A}|$ is stronger than the others shown in this
table due to the relatively large values of the recoil and $P$-
wave matrix elements in this case. The bounds on $|\epsilon_{V\mp
A}^{V+A}|$ given in Table 9 are comparable with the bounds
$|\epsilon_{V-A}^{V+A}|<4\times10^{-9}$,
$|\epsilon_{V+A}^{V+A}|<6\times10^{-7}$ given in Ref.
\cite{0002109}.

\vspace{3mm}
\begin{center}
\noindent Table 9: Upper bounds on $|\mu_{V\mp A}^{V-A}|$,
$|\epsilon_{V\mp A}^{V+A}|$ for decays of $^{76}\mbox{Ge}$ for the
case of $|\langle m\rangle|=0$ in QRPA. \label{Tab9}

\nopagebreak

\vspace*{3mm}

\begin{tabular}{|c|c|c|c|c|}
  \hline
  Nuclear model & $|\mu_{V-A}^{V-A}|$ & $|\mu_{V+A}^{V-A}|$
  & $|\epsilon_{V-A}^{V+A}|$ & $|\epsilon_{V+A}^{V+A}|$ \\
  \hline
  pnQRPA with $g_{pp}$=1.02(1.06) \cite{Kortelainen:2007rh} & $2.6(2.9)\times10^{-7}$& $4.5(5.0)\times10^{-7}$ & --- & --- \\
  \hline
  QRPA without (with) p-n pairing \cite{Pantis} & $5.0(11)\times10^{-7}$& $5.4(6.5)\times10^{-7}$ & $4.8(13)\times10^{-9}$ & $7.9(25)\times10^{-7}$ \\
  \hline
\end{tabular}
\end{center}
\vspace{3mm}

To be definite, we use the QRPA model without p-n pairing \cite{Pantis} in the
following. The bounds given in Table 9 could be used
for deriving the bounds on the parameters of the particular models
(see section \ref{section_Lagrangian}). For example, using Eq.
(\ref{LQ1}) we have the following conservative constraints on the
couplings of the effective LQ-quark-lepton interactions:
\begin{eqnarray}
    |\alpha_I^{(L)}|\leq1.1\times10^{-9}\left(\frac{M_I}{100
    ~\mbox{GeV}}\right)^2, \quad
    |\alpha_I^{(R)}|\leq2.6\times10^{-7}\left(\frac{M_I}{100
    ~\mbox{GeV}}\right)^2, \quad I=S,V.
\end{eqnarray}

\begin{itemize}
\item Consider a more general case of $|\langle m\rangle|\neq 0$,
$\cos\psi_i=0$, where the index $i$ depends on $\alpha$, $\beta$
(as above, we take only one nonzero $\epsilon_\alpha^\beta$ at a
time). Using Tables 2 and 4 we have
\begin{eqnarray}
    {\mathcal A} = C_1|\mu|^2+4C_i|\mu_\alpha^\beta|^2, \nonumber\\
    K{\mathcal A} = D_1|\mu|^2+4D_i|\mu_\alpha^\beta|^2,
\end{eqnarray}
and
\begin{eqnarray}
    {\mathcal A} = C_1|\mu|^2+C_i|\epsilon_\alpha^\beta|^2, \nonumber\\
    K{\mathcal A} = D_1|\mu|^2+D_i|\epsilon_\alpha^\beta|^2.
\end{eqnarray}
\end{itemize}
Hence, using Eq. (\ref{T_1/2}) we obtain
\begin{eqnarray}\label{Eq:mu}
  &&  |\mu|^2 = (\lambda_1-\lambda_2K)/T_{1/2}, \nonumber\\
  &&
    |\epsilon_\alpha^\beta|^2 = (-\lambda_3+\lambda_4K)/T_{1/2}
    = 4|\mu_\alpha^\beta|^2,
\end{eqnarray}
with the coefficients
\begin{eqnarray}
  &&  \lambda_1=\frac{D_i}{|M_{\rm GT}|^2\Delta_i},  \quad
  \lambda_2=\frac{C_i}{|M_{\rm GT}|^2\Delta_i}, \nonumber\\
  &&  \lambda_3=\frac{D_1}{|M_{\rm GT}|^2\Delta_i},  \quad
  \lambda_4=\frac{C_1}{|M_{\rm GT}|^2\Delta_i}, \label{Eq:lambda}
\end{eqnarray}
where $\Delta_i=C_1D_i-D_1C_i$.

Using Eqs (\ref{Eq:mu})--(\ref{Eq:lambda}) we have for
$\epsilon_{V+A}^{V+A}\neq 0$
\begin{eqnarray}\label{Eq:V+A}
  |\mu|^2 = (7.9+10K)\times10^{12}/T_{1/2}, \quad
    |\epsilon_{V+A}^{V+A}|^2 = (5.1-6.3K)\times10^{12}/T_{1/2}
\end{eqnarray}
and for $\epsilon_{V-A}^{V+A}\neq 0$
\begin{eqnarray}\label{Eq:V-A}
  |\mu|^2 = (7.7+10K)\times10^{12}/T_{1/2}, \quad
    |\epsilon_{V-A}^{V+A}|^2 = (1.9-2.4K)\times10^{8}/T_{1/2},
\end{eqnarray}
with $T_{1/2}$ in years. Fig.~1 shows the correlation among
$|\langle m\rangle|$, $T_{1/2}$, $K$ ({\it left}) and the
correlation among $|\epsilon_{V+A}^{V+A}|$, $T_{1/2}$, $K$ ({\it
right}) for the choice of a nonzero $\epsilon_{V+A}^{V+A}$.
 Fig.~2 shows the same for the parameter
$\epsilon_{V-A}^{V+A}$. It is clear from Figs 1 and 2 that the
closer is $K$ to 1 for the fixed value of $T_{1/2}$, the weaker is
bounded $|\langle m\rangle|$ and stronger is bounded
$|\epsilon_{V\mp A}^{V+A}|$. The correlations among
$|\epsilon_{V\mp A}^{V+A}|$, $T_{1/2}$, $K$ will be used in the
next section in the analysis of left-right symmetric models.

Note that if several $\epsilon_\alpha^\beta$ are nonzero in the
considered model than the respective interference terms should be
taken into account.
\begin{itemize}
\item To extract $|\mu|$, $|\mu_\alpha^\beta|$,
$|\epsilon_\alpha^\beta|$, $c_i$ in the general case of $|\langle
m\rangle|\neq0$, $c_i\neq 0$ we need to analyze the data on at
least two decaying nuclei. This analysis will be presented for the
five nuclei already discussed in a forthcoming paper~\cite{Ali:2010zza}.
\end{itemize}

\section{Electron angular correlation in left-right symmetric
models}

The experimental bounds on the $\epsilon_\alpha^\beta$ are
connected with the masses of new particles, their mixing angles,
and other parameters specific to particular extensions of the SM
\cite{Doi,Shchep,SUSY1,SUSY,BL,LQ}. To illustrate the kind of
correlations that the measurements of  $T_{1/2}$ and the angular
correlation coefficient $K$ in the $0\nu2\beta$ decay would imply,
we work out the case of the left-right symmetric models \cite{LR}.
In the model $SU(2)_L\times SU(2)_R\times U(1)$ the parameters
$\eta$ and $\lambda$ (see Eq. (\ref{Doi})) are expressed through
the masses $m_{W_{L}}$ and $m_{W_{R}}$ of the left- and
right-handed $W$ bosons and their mixing angle $\zeta$ \cite{Doi}:
\begin{equation}\label{r1}
    \eta=-\tan\zeta, \quad \lambda  = \left( {m_{W_L } /m_{W_R } } \right)^2 ,
\end{equation}
under the condition
\begin{equation}\label{mL<<mR}
 m_{W_L}\ll m_{W_R}.
\end{equation}
Eqs. (\ref{Doi}) and (\ref{eps}) and the relation \cite{Doi}
\begin{equation}
   V_{ei}=V_{ei}^\prime
\end{equation}
of the $SU(2)_L\times SU(2)_R\times U(1)$ model yield
\begin{equation}\label{r2}
\epsilon^{V+A}_{V+A}=\lambda\frac{g^\prime_V}{g_V}U_{ei}V_{ei}.
,\quad \epsilon^{V+A}_{V-A}=\eta U_{ei}V_{ei}.
\end{equation}
\noindent To reduce the number of free parameters, we assume the
equality of the form factors of the left- and right-handed
hadronic currents:
\begin{equation}\label{gV}
g_V=g^\prime_V.
\end{equation}
The small masses of the observable $\nu$s are likely described by
the seesaw formula that in the simplest case gives
\begin{equation}
    m_i\sim m_D^2/M_R,~~~ M_R\gg m_D,
\end{equation}
with the Dirac mass scale $m_D$ (for the charged leptons and the
light quarks $m_D\sim 1$ MeV) and the mass scale $M_R$ of right
$\nu_M$s (in the majority of theories $M_R>1$ TeV). In the
left-right symmetric models these scales arise usually from the
two scales of the vacuum expectation values of Higgs multiplets
\cite{LR}. In the seesaw mechanism, the values of the mixing
parameters $V_{ei}$ (for $i$ numbering light mass states) have the
same order of magnitude as $m_D/M_R$. In our discussion we use two
rather conservative values (compare with Eq. (\ref{nsm}))
\begin{equation}\label{e}
\epsilon=10^{-6},~ 5\times10^{-7}
\end{equation}
for the mixing parameter
\begin{equation}\label{epsilon}
\epsilon=|U_{ei}V_{ei}|.
\end{equation}
We recall that here the summation index $i$ runs only over the
light neutrino mass eigenstates (the summation over the total mass
spectrum including also heavy states gives strictly zero due to
the orthogonality condition \cite{Doi}).

From Eqs. (\ref{r1}), (\ref{r2}), (\ref{gV}), and (\ref{epsilon})
we have
\begin{equation}
\label{WR} m_{W_R} =
m_{W_L}\left(\epsilon/\left|\epsilon^{V+A}_{V+A}\right|\right)^{1/2},
\quad
\zeta=-\arctan\left(\left|\epsilon^{V+A}_{V-A}\right|/\epsilon\right).
\end{equation}
Using Eq. (\ref{mL<<mR}) we note the approximate equality of
$m_{W_{L}}$ and the mass of the observed charged gauge boson $W_1$
($m_{W_1}$=80.4~GeV \cite{PDG}).

The correlation among $m_{W_R}$ ($\zeta$), $K$, and $T_{1/2}$ for
the case of $|\langle m\rangle|\neq0$, $\cos\psi_i=0$ (see section
3.2) is shown in Fig. 3 (4) for the two chosen values of
$\epsilon$. The numerical results for these figures have been obtained
using Eqs.~(\ref{Eq:V+A}) and (\ref{Eq:V-A}).
 It is clear from Fig. 3 (4) that
the closer is $K$ to 1 for the fixed value of $T_{1/2}$ the
stronger is the lower bound on $m_{W_R}$ (the upper bound on
$\zeta$). However this bound is weaker than the one $m_{W_R}
> 715~\mbox{GeV}$, obtained  from the electroweak fits \cite{PDG}.
There is still a more stringent bound $m_{W_R} > 1.2~\mbox{TeV}$,
obtained in Ref.~\cite{2004} for the $0\nu2\beta$ decay mediated
by heavy Majorana neutrinos using arguments based on the vacuum
stability \cite{Mohapatra86} and additional theory input. We
assume $m_{W_R}\ge 1~\mbox{TeV}$ in the next figure.

While experiments in the $0\nu2\beta$ decay would measure the
product of the quantities called $\lambda$ and the neutrino mixing
matrix elements $U_{ei}V_{ei}$ in Eq.~(\ref{r2}), collider
experiments at the Tevatron and the LHC can, in principle, measure
$\lambda$ by determining $m_{W_R}$. Assuming these logically
independent possibilities,
 we plot the differential width
(\ref{dG}) vs. $\cos\theta$ in Fig.~5 for a set of values of
$|\langle m\rangle|$ and $m_{W_R}$, taking $\epsilon_{V+A}^{V+A}$
at a time and assuming $\epsilon=10^{-6}$. In this figure, we
consider the values of $|\langle m\rangle|$, starting from
$|\langle m\rangle|\leq 0.03$~eV up to $|\langle m\rangle|=
5$~meV, covering two of three scenarios of neutrino mass
hierarchies and mixing angles: normal and inverted mass
hierarchies (see Ref.~\cite{Bil} for a recent discussion and
update). It is seen that the sensitivity of the electron angular
correlation to the right-handed $W$-boson mass $m_{W_R}$
increases with decreasing values of the effective Majorana
neutrino mass $|\langle m\rangle|$, as can be seen from Fig. 5
(right), where this correlation is shown for $|\langle
m\rangle|$=5~meV, 10~meV.

In conclusion, we have presented a detailed study of the electron
angular correlation for the long range mechanism of $0\nu2\beta$
decays in a general theoretical context. This information,
together with the ability of observing these decays in several
nuclei, would help greatly in identifying the dominant mechanism
underlying these decays.
 At present, no experiment is geared
to measuring the angular correlation in $0\nu2\beta$ decays, as
the main experimental thrust is on establishing a non-zero signal
unambiguously in the first place. We note that the running
 experiment NEMO3 has already measured the electron angular
 distribution for the two neutrino double beta decay, and is capable of
 measuring this correlation in the future for the $0\nu2\beta$ decay
as well, assuming that the experimental sensitivity is sufficiently good
to establish this decay \cite{NEMO3}.
The proposed experimental facilities that can measure the electron
angular correlation in the $0\nu2\beta$ decays are SuperNEMO
\cite{SuperNEMO}, MOON \cite{MOON}, and EXO \cite{EXO}. We have argued
in this paper that there is a strong case in  building at least one of them.

\section*{Acknowledgments}
We thank Alexander Barabash and Fedor \v{S}imkovic for helpful
discussions, and the referee of our paper for critical comments
which helped us in improving and correcting the earlier version of this
manuscript. One of us (DVZ) would like to thank DESY for the
hospitality in Hamburg where a good part of this work was done. DVZ is grateful to Aritro Azad for collaboration on checking and correcting the Appendix A.


\appendix
\section{$0\nu2\beta$ decay rate for scalar nonstandard terms}
The nucleon currents in the impulse approximation in the
nonrelativistic form are used in this paper~\cite{Tomoda,Ericson}. Keeping all
terms up to order $p/m_p$ in the nonrelativistic expansion we have
\begin{eqnarray}
 J_{S\mp P}^+({\bf x}) = \sum\limits_a \tau_+^a \delta({\bf x-r}_a)
\left( F_S^{(3)}\mp F_P^{(3)}B_a \right), \quad
B_a=\frac{{\bm\sigma}_a\cdot{\bf q}}{2m_p}, \\
 J_{V-A}^{\mu+}({\bf x}) = \sum\limits_a \tau_+^a \delta({\bf x-r}_a)
\left[ g^{\mu0}(g_VI_a-g_AC_a) + g^{\mu m}(g_A\sigma_{am} -
g_VD_a^m -g_AP_a^m) \right], \label{J_V-A} \\
 C_a=\frac{{\bm \sigma}_a\cdot{\bm Q}}{2m_p}-\frac{q^0
{\bm\sigma}_a\cdot{\bm q}}{{\bf q}^2+m_\pi^2}, \quad
D_a^m=\frac{Q^m}{2m_p}I_a-\left(1+
\frac{g_M}{g_V}\right)\frac{i[{\bm \sigma}_a\times{\bm
q}]^m}{2m_p}, \quad P_a^m=\frac{q^m{\bm\sigma}_a\cdot {\bf
q}}{{\bf q}^2+m_\pi^2}, \label{C_D_P}
\end{eqnarray}
where $q^\mu=p^\mu-p^{\prime\mu}$ is the 4-momentum transferred
from hadrons to leptons, $Q^\mu=p^\mu+p^{\prime\mu}$; $p^\mu$ and
$p^{\prime\mu}$ are the initial and final 4-momenta of a nucleon;
$m_p$ is proton mass and $m_\pi$ is pion mass.

We neglect the dipole dependence of the form factors
$F_{S}^{(3)}$, $F_{P}^{(3)}$, $g_{V}$, $g_{A}$, $g_{M}$ on the
momentum transfer and omit the zero argument of the form factors.
Note that $g_V(0)=1$.

Consider the pure SP case assuming $\langle m \rangle=0$. In terms
of the combinations of hadronic currents
\begin{eqnarray}
 J_{\mp L}^\mu=\langle F|\tilde J^+_\mp|N\rangle \langle N|\hat J_L^{\mu+}|I\rangle, \quad
 J_{L\mp}^\mu=\langle F|\hat J^{\mu+}_L|N\rangle \langle N|\tilde J_\mp^+|I\rangle, \\
 \tilde J^+_-=\epsilon_{S-P,i}^{S-P}J_{S-P}^+ + \epsilon_{S+P,i}^{S-P}J_{S+P}^+, \quad
 \tilde J^+_+=\epsilon_{S+P,i}^{S+P}J_{S+P}^+ + \epsilon_{S-P,i}^{S+P}J_{S-P}^+, \\
 \hat J^{\mu+}_L=U_{ei}J_{V-A}^{\mu+}~,
\end{eqnarray}
and the combinations
\begin{eqnarray}
 \ell_\mu^{L,R}=\frac{s_\mu^{L,R}(2{\bf y},1{\bf x})}{\omega+A_1} -
\frac{s_\mu^{L,R}(1{\bf y},2{\bf x})}{\omega+A_2}, \\
 \ell_{\lambda\mu}^L=\frac{s_{\lambda\mu}^L(2{\bf y},1{\bf x})}{\omega+A_1}
 - \frac{s_{\lambda\mu}^L(1{\bf y},2{\bf x})}{\omega+A_2}
\end{eqnarray}
of electron currents
\begin{eqnarray}
    s_\mu^{L,R}(2{\bf y},1{\bf x}) =
    \bar e_2({\bf y})\gamma_\mu(1\mp\gamma_5)e_1^c({\bf x}), \quad
    s_{\lambda\mu}^L(2{\bf y},1{\bf x}) =
    \bar e_2({\bf y})\gamma_\lambda(1-\gamma_5)\gamma_\mu e_1^c({\bf
    x}),
\end{eqnarray}
$e_i({\bf x})\equiv e_{{\bf p}_is_i}({\bf x})$, the matrix element
is expressed as
\begin{eqnarray}\label{R^SP}
 R_{0\nu}^{SP}= \frac{1}{\sqrt{2!}}\left(\frac{G_F|V_{ud}|}{\sqrt{2}}\right)^2
2\sum\limits_i \int d{\bf x}d{\bf y}\frac{d{\bf k}}{(2\pi)^3}
\frac{e^{i{\bf k\cdot r}}}{2\omega} \nonumber \\
\times\sum\limits_N\left[ m_i\left(J_{-L}^\mu\ell_\mu^R
-J_{L-}^\mu\ell_\mu^L\right) + k^\lambda
\left(J_{+L}^\mu\ell_{\lambda\mu}^L
-J_{L+}^\mu\ell_{\mu\lambda}^L\right) \right],
\end{eqnarray}
where ${\bf r} ={\bf y} -{\bf x}$.
By using the identities
\begin{eqnarray}
    s_\mu^{L,R}(1{\bf y},2{\bf x}) = s_\mu^{R,L}(2{\bf x},1{\bf y}), \quad
    s_{\lambda\mu}^L(1{\bf y},2{\bf x}) = -s_{\mu\lambda}^L(2{\bf x},1{\bf y}),
\end{eqnarray}
the algebraic formula
\begin{eqnarray}\label{algebraic_formula}
    2(am\pm bn) = (a+b)(m\pm n) + (a-b)(m\mp n),
\end{eqnarray}
the constant
\begin{equation}\label{Cconstant}
 C_{0\nu}=\frac{G_F^2|V_{ud}|
 ^2}{8\sqrt{2}\pi}\frac{2m_e}{R}
\end{equation}
and the neutrino potentials
\begin{eqnarray} \label{neutrino_potentials}
&& (H_j, H_{\omega j},H_{kj}^l)=4\pi\int\frac{d{\bf
k}}{(2\pi)^3}\frac{e^{i{\bf k\cdot
r}}}{\omega}\frac{(1,\omega,k^l)}{\omega+A_j}~,
\end{eqnarray}
the matrix element (\ref{R^SP}) is expressed as
\begin{equation}\label{R0nuSP}
 R_{0\nu}^{SP}= -C_{0\nu}\sum\limits_i\sum\limits_N\left( \frac{m_i}{m_e}M_{SP}^m+M_{SP}^k \right).
\end{equation}

Each part of this matrix element is expressed as a sum of
nonvanishing (indexed by $n$) and vanishing (indexed by $c$) terms, in
the closure approximation:
\begin{equation}
  M_{SP}^{m,k} = \{M_{SP}^{m,k}\}_n + \{M_{SP}^{m,k}\}_c,
\end{equation}
\begin{eqnarray}\label{SPm}
&& \left\{M_{SP}^m\right\}_n =\frac{R}{2}\int d{\bf x}d{\bf y}T_N
(H_1+H_2) \nonumber\\ && \times\left[ (A_1+A_{1R})F_{5+}^0 +
(A_3^i+\tilde A_{3R}^i) F_{5+}^i + B_{1R}F_-^0 + (B_3^i+\tilde
B_{3R}^i)F_-^i \right], \\ && \left\{M_{SP}^m\right\}_c
=\frac{R}{2}\int d{\bf x}d{\bf y}T_N (H_1-H_2) \nonumber\\ &&
\times\left[ (A_1+A_{1R})F_{5-}^0 + (A_3^i+\tilde
A_{3R}^i)F_{5-}^i + B_{1R}F_+^0 + (B_3^i+\tilde B_{3R}^i)F_+^i
\right],
\end{eqnarray}
\begin{eqnarray}\label{SPk}
\left\{M_{SP}^k\right\}_n =\frac{R}{2m_e}\int d{\bf x}d{\bf
	y}T_N\left\{\right. (H_{\omega1}-H_{\omega2})
\left[ -(A_4^i+ \tilde A_{4R}^i)E_+^i + B_{2R}E_- \right] \nonumber \\
+ (H_{k1}^l+H_{k2}^l) \left[ - (A_2+A_{2R})E_-^l + (A_5^{lk}+\tilde A_{5R}^{lk})E_-^k
	+ (B_4^l+\tilde B_{4R}^l)E_+ \right]\left.\right\}, \\
\left\{M_{SP}^k\right\}_c =\frac{R}{2m_e}\int d{\bf x}d{\bf y}T_N
\left\{\right. (H_{\omega1}+H_{\omega2}) \left[ - (A_4^i+ \tilde
A_{4R}^i)E_-^i + B_{2R}E_+ \right] \nonumber \\
+ (H_{k1}^l-H_{k2}^l) \left[  - (A_2+A_{2R})E_+^l + (A_5^{lk}+\tilde A_{5R}^{lk})E_+^k
	+ (B_4^l+\tilde B_{4R}^l)E_- \right]\left.\right\},
\end{eqnarray}
with \begin{eqnarray}
 T_N= g_A^2\langle F|\sum\limits_a\tau^a_+|N\rangle \langle N|\sum\limits_
b\tau_+^b|I\rangle \delta({\bf x-r}_a)\delta({\bf y-r}_b)~.
\end{eqnarray}
The electron currents are defined as:
\begin{eqnarray}
 & F_+ =\frac{1}{2} \left[u({\bf yx}) \pm u({\bf xy})\right],
 \quad & F_{5\pm} =\frac{1}{2} \left[u_5({\bf yx}) \pm
u_5({\bf xy})\right], \nonumber \\
 & F_+^\mu =\frac{1}{2} \left[u^\mu({\bf yx}) \pm u^\mu({\bf xy})\right],
 \quad & F_{5\pm}^\mu =\frac{1}{2} \left[u_5^\mu({\bf yx}) \pm
u_5^\mu({\bf xy})\right], \nonumber \\
  & F_+^{\mu\nu} =\frac{1}{2} \left[u^{\mu\nu}({\bf yx}) \pm
u^{\mu\nu}({\bf xy})\right],
 \quad & F_{5\pm}^{\mu\nu} =\frac{1}{2} \left[u_5^{\mu\nu}({\bf yx})
\pm u_5^{\mu\nu}({\bf xy})\right], \nonumber \\
 & E_\pm =F_\pm + F_{5\pm}, \quad & E_\pm^i =
F_\pm^{0i} + F_{5\pm}^{0i},
\end{eqnarray}
with
\begin{eqnarray}
 & u({\bf yx}) = \bar e_2({\bf y}) e_1^c({\bf x}), \quad
 & u_5({\bf yx}) = \bar e_2({\bf y})\gamma_5 e_1^c({\bf x}),
 \nonumber \\
 & u^\mu({\bf yx}) = \bar e_2({\bf y})\gamma^\mu e_1^c({\bf x}), \quad
 & u_5^\mu({\bf yx}) = \bar e_2({\bf y})\gamma_5\gamma^\mu e_1^c({\bf x}),
 \nonumber \\
 & u^{\mu\nu}({\bf yx}) = -i\bar e_2({\bf y})\sigma^{\mu\nu}
 e_1^c({\bf x}), \quad & u_5^{\mu\nu}({\bf yx}) = -i
\bar e_2({\bf y})\gamma_5\sigma^{\mu\nu} e_1^c({\bf x})~.
\end{eqnarray}
The nucleon operator matrix elements are defined as follows:
\begin{eqnarray}
 &\tilde A=A+A^P, \quad & \tilde B=B+B^P,
\end{eqnarray}
\begin{eqnarray}
 &A_1= 2G_V^0\varepsilon_S, \quad& A_{1R}= -G_A^0\varepsilon_SC_+
-G_V^0\varepsilon_PB_+, \nonumber \\
 &A_2= 2G_V^0\varepsilon_S^\prime, \quad& A_{2R}= -G_A^0\varepsilon_S^
\prime C_+
+G_V^0\varepsilon_P^\prime B_+, \nonumber \\
 &A_3^i= G_A^0\varepsilon_S\sigma_+^i, \quad& A_{3R}^i= -G_A^0\varepsilon_P
B_{\sigma+}^i -G_V^0\varepsilon_SD_+^i, \quad A_{3R}^{Pi}=-G_A^0
\varepsilon_SP_+^i, \nonumber \\
 &A_4^i= G_A^0\varepsilon_S^\prime\sigma_+^i, \quad& A_{4R}^i=
G_A^0\varepsilon_P^\prime B_{\sigma+}^i -G_V^0\varepsilon_S^\prime
D_+^i,
 \quad A_{4R}^{Pi}= -G_A^0\varepsilon_S^\prime P_+^i, \nonumber \\
 & A_5^{lk}=i\varepsilon_{ilk}A_4^i, \quad& \tilde A_{5R}^{lk}=i
\varepsilon_{ilk}\tilde A_{4R}^i,
\end{eqnarray}
\begin{eqnarray}
 &B_{1R}= -G_A^0\varepsilon_SC_- +G_V^0\varepsilon_PB_-,  \nonumber \\
 &B_{2R}= -G_A^0\varepsilon_S^\prime C_- -G_V^0\varepsilon_P^\prime
B_-,  \nonumber \\
 &B_3^i= G_A^0\varepsilon_S\sigma_-^i, \quad  B_{3R}^i= -G_A^0\varepsilon_P
B_{\sigma-}^i -G_V^0\varepsilon_SD_-^i, \quad  B_{3R}^{Pi}= -G_A^0
\varepsilon_SP_-^i, \nonumber \\
 &B_4^i= G_A^0\varepsilon_S^\prime\sigma_-^i, \quad  B_{4R}^i= G_A^0
\varepsilon_P^\prime B_{\sigma-}^i -G_V^0\varepsilon_S^\prime
D_-^i,
 \quad  B_{4R}^{Pi}=-G_A^0\varepsilon_S^\prime P_-^i,
\end{eqnarray}
with
\begin{eqnarray}
 && \sigma_\pm^i = \sigma_a^iI_b \pm I_a\sigma_b^i, \quad B_\pm = B_aI_b \pm I_aB_b, \quad B_{\sigma\pm}^{i}=\sigma_a^iB_b \pm B_a\sigma_b^j,\nonumber\\
 && C_\pm = C_aI_b \pm I_aC_b, \quad D_\pm^i = D_a^iI_b \pm I_aD_b^i, \quad P_\pm^i=P_a^iI_b\pm I_aP_b^i.
\end{eqnarray}

Under the exchange of running indices $a$ and $b$ (i.e. {\bm x}
$\leftrightarrow$
 {\bm y}), nuclear operators $A$, electron
currents $E_+$ and $F_+$ and neutrino potentials $H_i$ and
$H_{\omega i}$ are  even, while $B$, $E_-$, $F_-$, and ${\bf
H}_{ki}$ are odd.

The constants are defined as:
\begin{eqnarray}
 G_V=\frac{g_V}{g_A}\left[\left(U_{ei}+\epsilon_{V-A,i}^{V-A}\right)
+\epsilon_{V+A,i}^{V-A}\right], \quad &
G_A=\left(U_{ei}+\epsilon_{V-A,i}^{V-A}\right)-\epsilon_{V+A,i}^{V-A},
\nonumber \\   G^0=G(\epsilon=0), \quad & \ds
G_V^0=\frac{g_V}{g_A}U_{ei}, \quad G_A^0=U_{ei},
\end{eqnarray}
\begin{eqnarray}
 \varepsilon_S=\frac{F_S^{(3)}}{g_A}\left(\epsilon_{S-P,i}^{S-P}+\epsilon_{S+P,i}^{S-P}\right),
 \quad \varepsilon_P=\frac{F_P^{(3)}}{g_A}\left(\epsilon_{S-P,i}^{S-P}
-\epsilon_{S+P,i}^{S-P}\right), \nonumber \\
 \varepsilon_S^\prime=\frac{F_S^{(3)}}{g_A}\left(\epsilon_{S+P,i}^{S+P}
+\epsilon_{S-P,i}^{S+P}\right),
 \quad \varepsilon_P^\prime=\frac{F_P^{(3)}}{g_A}\left(\epsilon_{S+P,i}^{S+P}
-\epsilon_{S-P,i}^{S+P}\right).
\end{eqnarray}

Note that in the notations of Ref. \cite{Doi}:
\begin{eqnarray}
& t=u+u_5, \quad & t^l=u^{0l}+u_5^{0l}.
\end{eqnarray}

Since the nucleon recoil term ${\bf P}_a$ behaves as an even
parity operator while the neutrino momentum {\bm k} and the recoil
terms $B_a$, $C_a$, ${\bf D}_a$ as odd ones, each of the $A_j$,
${\bf k\cdot A}_j$, $B_j$, ${\bf k\cdot B}_j$ has a definite
parity. The operators
\begin{eqnarray}
 A_1,\ A_3^i,\ A_4^i,\ A_{3R}^{Pi},\ A_{4R}^{Pi}; \quad B_3^i,\ B_{3R}^{Pi};
 \nonumber \\
{\bf r\cdot B}_{4R},\ r^lA_{2R},\ r^lA_{5R}^{lk}~,
\end{eqnarray}
have even parity and the operators
\begin{eqnarray}
 A_{1R},\ A_{3R}^i,\ A_{4R}^i; \quad B_{1R},\ B_{2R},\ B_{3R}^i; \nonumber \\
{\bf r\cdot B}_4,\ {\bf r\cdot B}_{4R}^P,\ r^lA_2,\ r^lA_5^{lk},\
r^lA_{5R}^{Plk}~,
\end{eqnarray}
have odd parity. The odd-parity operators do not contribute to the
$0^+\to J^+$ transition in the case where both the electrons are in
the $S$-wave state (the $S-S$ case) with no de Broglie wave length
correction (no FBWC).

Using the definitions of neutrino potentials
\begin{eqnarray}\label{nu_potentials}
&& h_+=\frac{R}{2}(H_1+H_2), \quad
h_0=\frac{1}{\varepsilon_{21}}(H_1-H_2), \nonumber\\
&& h_{\omega+}=\frac{R^2}{2}(H_{\omega1}+H_{\omega2}), 
\quad h_{0\omega}=\frac{R}{\varepsilon_{21}}(H_{\omega1}-H_{\omega2}), \nonumber\\
&& h_+^\prime\hat r^l =
-i\frac{rR}{2}\left(H_{k1}^l+H_{k2}^l\right), \quad
 h_0^\prime\hat r^l = -i\frac{r}{\varepsilon_{21}}\left(H_{k1}^l-H_{k2}^l\right)~,
\end{eqnarray}
in the $S-S$ case with no FWBC, Eqs. (\ref{SPm}), (\ref{SPk}) are
reduced to
\begin{eqnarray}\label{SPmSS}
&& \left\{M_{SP}^m\right\}_{n,S-S} = \int d{\bf x}d{\bf y}T_Nh_+
\left[ A_1F_{5+}^0 + (A_3^i+A_{3R}^{Pi}) F_{5+}^i \right],
\\ &&\left\{M_{SP}^m\right\}_{c,S-S}
=\frac{\varepsilon_{21}R}{2} \int d{\bf x}d{\bf y}T_Nh_0
(B_3^i+B_{3R}^{Pi})F_+^i,
\end{eqnarray}
\begin{eqnarray}\label{SPkSS}
&&\left\{M_{SP}^k\right\}_{n,S-S}
=-\frac{1}{2}\frac{\varepsilon_{21}}{m_e}\int d{\bf x} d{\bf
	y}T_Nh_{0\omega}  (A_4^i+ A_{4R}^{Pi})E_+^i \nonumber \\ 
&& +\frac{2}{m_eR}\int d{\bf x}d{\bf y}T_N\frac{iR}{2r} h_+^\prime
	{\bf \hat r\cdot B}_{4R}E_+, \\ 
&&\left\{M_{SP}^k\right\}_{c,S-S} =
\frac{1}{2}\frac{\varepsilon_{21}}{m_e}\int d{\bf x}d{\bf
		y}T_N\frac{iR}{r} h_0^\prime \hat r^l \left( -A_{2R}E_+^l +
	A_{5R}^{lk}E_+^k \right),
\end{eqnarray}
where $E$, $F$ are taken for {\bf x}=0, {\bf y}=0.

For the $0^+\to 0^+$ transition we have
\begin{eqnarray}
  \sum\limits_i\frac{m_i}{m_e}\sum\limits_N\left\{M_{SP}^m\right\}_{S-S}
  &=&
g_A^2 C_1^AF_{5+}^0, \\
 \sum\limits_i\sum\limits_N\left\{M_{SP}^k\right\}_{S-S} &=& g_A^2
\frac{2}{m_eR}C_{4R}^B E_+,
\end{eqnarray}
with
\begin{eqnarray}
  C_1^A=\left\langle \frac{m_i}{m_e}h_+A_1 \right\rangle, \quad C_{4R}^B=\left\langle
\frac{iR}{2r}h_+^\prime{\bf\hat r\cdot B}_{4R} \right\rangle,
\end{eqnarray}
where $\hat {\bf r}={\bf r}/r$ and $\langle X \rangle =\sum
\limits_i\sum\limits_N\langle 0_f^+|| X ||0_I^+ \rangle$, with
$h=h(r, E_N)$.

In the $S-P_{1/2}$ case with no FBWC for the $0^+\to 0^+$
transition we have
\begin{eqnarray}
 \left\{M_{SP}^m\right\}_{n,S-P_{1/2}} = \int d{\bf x}d{\bf y}T_Nh_+
\left( A_{3R}^iF_{5+}^i + B_{3R}^iF_-^i \right),  \\
\left\{M_{SP}^m\right\}_{c,S-P_{1/2}} =\frac{\varepsilon_{21}R}{2}
\int d{\bf x}d{\bf y}T_Nh_0 \left( A_{3R}^iF_{5-}^i +
B_{3R}^iF_+^i \right),
\end{eqnarray}

\begin{eqnarray}
\left\{M_{SP}^k\right\}_{n,S-P_{1/2}}
=-\frac{1}{2}\frac{\varepsilon_{21}}{m_e}\int d{\bf x}
d{\bf y}T_Nh_{0\omega}  A_{4R}^iE_+^i \nonumber \\
 + \frac{2}{m_eR}\int d{\bf x}d{\bf y}T_N\frac{iR}{2r} h_+^\prime
	\hat r^l \left[ -A_2E_-^l + (A_5^{lk}+A_{5R}^{Plk})E_-^k \right], \\
\left\{M_{SP}^k\right\}_{c,S-P_{1/2}} = -\frac{1}{m_eR}\int d{\bf
	x}d{\bf y}T_Nh_{\omega+} A_{4R}^iE_-^i \nonumber \\ 
 + \frac{1}{2}\frac{\varepsilon_{21}}{m_e}\int d{\bf x}d{\bf y}T_N\frac{iR}{r}
	h_0^\prime \hat r^l  \left[ -A_2E_+^l +
	(A_5^{lk}+A_{5R}^{Plk})E_+^k \right].
\end{eqnarray}

The squared modulus of the matrix element (\ref{R0nuSP}), summed
over the polarizations $s_j$ of the electrons and multiplied by
the phase space element (\ref{ps}),  yields the differential decay
rate for the $0^+\to0^+$ transition
\begin{eqnarray}
 d\Gamma = \sum\limits_{s_1, s_2} |R^{SP}_{0\nu}|^2 \frac{m_e^5}{4\pi^3}d\Omega_{0\nu} =
\frac{a_{0\nu}}{(m_eR)^2} \left[ A_0^{SP} - {\hat {\bf p}_1\cdot
\hat {\bf p}_2}B_0^{SP} \right] d\Omega_{0\nu},
\end{eqnarray}
with $a_{0\nu}$ being defined in Eq. (\ref{a0nu}). Here the
coefficients are
\begin{eqnarray}
 && A_0^{SP} = \sum\limits_{i=1}^4|M_i|^2, \\
 && B_0^{SP} = \mbox{Re}(M_1M_2^*+M_1^*M_2+M_3M_4^*+M_3^*M_4),
\end{eqnarray}
with
\begin{eqnarray}
&& M_1 = \alpha_{-1-1}^*\left\{
\left[-C_1^A+\frac{2}{m_eR}C_{4R}^B\right] + \left[
\left(\frac{m_eR}{3}\left( \frac{\zeta}{m_eR}-2 \right)C_{3R}^A +
\frac{\varepsilon_{21}R}{3}\{C_{3R}^A\}_c \right) \frac{r}{2R} \right.\right.\nonumber\\
&&\left.\left.
+\frac{\varepsilon_{21}^2R}{6m_e}\left(\{C_2^A\}_c-\{C_5^A\}_c-\{C_{5R}^A\}_c-C_{4R}^A\right)\frac{r}{2R}
+
\frac{1}{6}\left(\frac{\zeta}{m_eR}-2\right)\left(C_2^A-C_5^A-C_{5R}^A-\{C_{4R}^A\}_c\right)
\right] \right. \nonumber\\
&&+\left.\left[ \frac{(\alpha Z)^2}{2m_eR}\left(\{C_4^A\}_c
+\{C_{4R}^A\}_c
-3C_{4RF}^B\right) \right] \right\}, \label{M_1} \\
&& M_2 = \alpha_{11}^*  \left\{
\left[C_1^A+\frac{2}{m_eR}C_{4R}^B\right] + \left[
\left(\frac{m_eR}{3}\left( \frac{\zeta}{m_eR}+2 \right)C_{3R}^A -
\frac{\varepsilon_{21}R}{3}\{C_{3R}^A\}_c \right) \frac{r}{2R} \right.\right. \nonumber\\
&&\left.\left.
+\frac{\varepsilon_{21}^2R}{6m_e}\left(\{C_2^A\}_c-\{C_5^A\}_c-\{C_{5R}^A\}_c-C_{4R}^A\right)\frac{r}{2R}
+
\frac{1}{6}\left(\frac{\zeta}{m_eR}+2\right)\left(C_2^A-C_5^A-C_{5R}^A-\{C_{4R}^A\}_c\right)
\right] \right. \nonumber\\
&&+\left.\left[\frac{(\alpha Z)^2}{2m_eR}\left(\{C_4^A\}_c
+\{C_{4R}^A\}_c
-3C_{4RF}^B\right) \right] \right\}, \\
&& M_3 = \alpha_{1-1}^* \left\{
\left[\frac{2}{m_eR}C_{4R}^B\right] +\left[
\frac{\varepsilon_{21}R}{6}\left(\frac{\varepsilon_{21}}{m_e}+2\right)
\left(\{C_2^A\}_c-\{C_5^A\}_c-\{C_{5R}^A\}_c-C_{4R}^A\right)\frac{r}{2R} \right.\right.\nonumber\\
&&+\left.\left.\frac{1}{6}\frac{\zeta}{m_eR}\left(C_2^A-C_5^A-C_{5R}^A-\{C_{4R}^A\}_c\right) \right] \right. \nonumber\\
&&+\left.\left[ \frac{(\alpha Z)^2}{2m_eR}\left(\{C_4^A\}_c
+\{C_{4R}^A\}_c
-3C_{4RF}^B\right) \right] \right\}, \\
&& M_4 = \alpha_{-11}^* \left\{
\left[\frac{2}{m_eR}C_{4R}^B\right] +\left[
\frac{\varepsilon_{21}R}{6}\left(\frac{\varepsilon_{21}}{m_e}-2\right)
\left(\{C_2^A\}_c-\{C_5^A\}_c-\{C_{5R}^A\}_c-C_{4R}^A\right)\frac{r}{2R} \right.\right.\nonumber\\
&&+\left.\left.\frac{1}{6}\frac{\zeta}{m_eR}\left(C_2^A-C_5^A-C_{5R}^A-\{C_{4R}^A\}_c\right) \right] \right. \nonumber\\
&&+\left.\left[ \frac{(\alpha Z)^2}{2m_eR}\left(\{C_4^A\}_c
+\{C_{4R}^A\}_c -3C_{4RF}^B\right) \right] \right\},
\label{M_4}
\end{eqnarray}
where $\alpha_{ij}=\tilde A_i(\varepsilon_2)\tilde
A_j(\varepsilon_1)$ and the nucleon matrix elements are
\begin{eqnarray}
  && C_{3R}^B=\left\langle \frac{m_i}{m_e}ih_+\hat{\bf r}\cdot{\bf B}_{3R} \right\rangle, \quad
  \{C_{3R}^B\}_c=\left\langle \frac{m_i}{m_e}ih_0\hat{\bf r}_+\cdot{\bf B}_{3R} \right\rangle, \nonumber \\
  && C_{3R}^A=\left\langle \frac{m_i}{m_e}ih_+\hat{\bf r}_+\cdot{\bf A}_{3R} \right\rangle, \quad
  \{C_{3R}^A\}_c=\left\langle \frac{m_i}{m_e}ih_0\hat{\bf r}\cdot{\bf A}_{3R} \right\rangle, \nonumber\\
  && C_{4R}^A=\left\langle ih_{0\omega}\hat{\bf r}_+\cdot{\bf A}_{4R} \right\rangle, \quad
  \{C_{4R}^A\}_c=\left\langle i\frac{r}{R}h_{\omega+}\hat{\bf r}\cdot{\bf A}_{4R} \right\rangle, \nonumber\\
  && \{C_2^A\}_c=\left\langle \frac{R}{r}h_0^\prime\hat{\bf r}\cdot\hat{\bf r}_+A_2 \right\rangle, \quad
  C_2^A=\left\langle h_+^\prime A_2 \right\rangle, \nonumber \\
  && \{C_{5(R)}^A\}_c=\left\langle \frac{R}{r}h_0^\prime \hat r^i\hat r_+^jA_{5}^{ij} \left(A_{5R}^{Pij}\right) \right\rangle, \quad
  C_{5(R)}^A=\left\langle h_+^\prime \hat r^i\hat r^jA_{5}^{ij} \left(A_{5R}^{Pij}\right) \right\rangle,\nonumber \\
 && C_{4RF}^B=\left\langle \frac{iR}{2r}\frac{r_a^2+r_b^2}{2R^2}h_+^\prime{\bf \hat r}\cdot{\bf B}_{4R}\right\rangle,
\end{eqnarray}
with ${\bf r}_+={\bf y+x}=r_+\hat{\bf r}_+$.


The terms in the first brackets in Eqs. (\ref{M_1})--(\ref{M_4})
come from the $S-S$ case, the terms in the second brackets come
from the $S-P_{1/2}$ case and in the third brackets there are the
most important terms due to the $P_{1/2}-P_{1/2}$ case and FBWC.

Assuming now $\langle m \rangle \neq 0$ for the dominant terms we
have
\begin{eqnarray}
&& M_1 = \alpha_{-1-1}^*\left\{ \left[Z_1^X-C_1^A+\frac{2}{m_eR}C_{4R}^B\right]\right.\nonumber\\
&&
+\left.\left[\frac{\varepsilon_{21}^2R}{6m_e}\left(\{C_2^A\}_c-\{C_5^A\}_c\right)\frac{r}{2R}
+
\frac{1}{6}\left(\frac{\zeta}{m_eR}-2\right)\left(C_2^A-C_5^A\right)\right] \right\},\\
&& M_2 = \alpha_{11}^*  \left\{ \left[Z_1^X+C_1^A+\frac{2}{m_eR}C_{4R}^B\right]\right.\nonumber\\
&& +\left.\left[
\frac{\varepsilon_{21}^2R}{6m_e}\left(\{C_2^A\}_c-\{C_5^A\}_c\right)\frac{r}{2R}
+
\frac{1}{6}\left(\frac{\zeta}{m_eR}+2\right)\left(C_2^A-C_5^A\right) \right] \right\}, \\
&& M_3 = \alpha_{1-1}^* \left\{ \left[Z_1^X+\frac{2}{m_eR}C_{4R}^B\right]\right.\nonumber\\
&& +\left.\left[
\frac{\varepsilon_{21}R}{6}\left(\frac{\varepsilon_{21}}{m_e}+2\right)
\left(\{C_2^A\}_c-\{C_5^A\}_c\right)\frac{r}{2R} +
\frac{1}{6}\frac{\zeta}{m_eR}\left(C_2^A-C_5^A\right) \right] \right\}, \\
&& M_4 = \alpha_{-11}^* \left\{ \left[Z_1^X+\frac{2}{m_eR}C_{4R}^B\right]\right.\nonumber\\
&& +\left.\left[
\frac{\varepsilon_{21}R}{6}\left(\frac{\varepsilon_{21}}{m_e}-2\right)
\left(\{C_2^A\}_c-\{C_5^A\}_c\right)\frac{r}{2R} +
\frac{1}{6}\frac{\zeta}{m_eR}\left(C_2^A-C_5^A\right)
\right] \right\}.
\end{eqnarray}

In the expressions for $M_1,..., M_4$, the terms with $\zeta$ are
due to the inclusion of the $P$-wave in the electron wave function
and those with $C_{4R}^B$ are from the inclusion of the nucleon
recoil effect. 
Note that some of the subdominant terms should be taken into account in
case of large cancellation among the dominant terms.

%
\section{$0\nu2\beta$ decay rate for vector nonstandard
terms} In this appendix we in general follow  the derivation of
Ref. \cite{Doi}. However in addition to Ref. \cite{Doi} we keep in
our calculations the terms associated with the parameters
$\epsilon_{V\mp A}^{V-A}$ and the pseudoscalar form factor.

The nucleon currents in the impulse approximation up to
order $p/m_p$ in the nonrelativistic expansion are
\cite{Tomoda,Ericson}:
\begin{eqnarray}
 J_{V\mp A}^{\mu+}({\bf x}) = \sum\limits_a \tau_+^a \delta({\bf x-r}_a)
\left[ g^{\mu0}(g_VI_a \mp g_AC_a) + g^{\mu m}(\pm g_A\sigma_{am}
- g_VD_a^m \mp g_AP_a^m) \right],
\end{eqnarray}
with $C_a$, $D_a^m$, $P_a^m$ given in Eq.~(\ref{C_D_P}).

In terms of $S_{L\mu\nu}$, $V_{\alpha\mu\nu}$,
$J_{\alpha\beta}^{\mu\nu}$ ($\alpha,\beta=L,R$) \cite{Doi} the
matrix element
\begin{equation}
 R_{0\nu}^{VA}=C_{0\nu}\sum\limits_i\sum\limits_N\frac{R}{2m_e} \int d{\bf x}d{\bf y}\ 4\pi
\frac{d{\bf k}}{(2\pi)^3}\frac{e^{i{\bf k}\cdot{\bf r}}}{\omega}
\left( m_iJ_{LL}^{\mu\nu}S_{L\mu\nu} +
 J_{LR}^{\mu\nu}V_{L\mu\nu} + J_{RL}^{\mu\nu}V_{R\mu\nu} \right),
\end{equation}
may be expressed as
\begin{equation}
 R_{0\nu}^{VA}=C_{0\nu}\sum\limits_i\sum\limits_N \left(
 \frac{m_i}{m_e}M_{VA}^m +M_{VA}^k\right), \\
 M_{VA}^{m,k} = \{M_{VA}^{m,k}\}_n + \{M_{VA}^{m,k}\}_c.
\end{equation}

The analogues of the Eqs. (C.2.11), (C.2.23), and (C.2.24) from
Ref. \cite{Doi} are as follows:
\begin{eqnarray}\label{M_VA}
&& \{M_{VA}^m\}_n \equiv \{M_{m_\nu}\}_n =\frac{R}{2}\int d{\bf
x}d{\bf y}T_N (H_1+H_2)\left[(X_1+\tilde
X_{1R})E_+ + (Y_1^i+\tilde Y_{1R}^i)E_-^i \right], \\
&& \{M_{VA}^m\}_c \equiv \{M_{m_\nu}\}_c =\frac{R}{2}\int d{\bf
x}d{\bf y}T_N (H_1-H_2)\left[(X_1+\tilde
X_{1R})E_- + (Y_1^i+\tilde Y_{1R}^i)E_+^i \right], \\
&& \{M_{VA}^k\}_n \equiv\{M_{V+A}(a)\}_n = \frac{R}{m_e}\int d{\bf
x}d{\bf y} T_N
\left\{\right. (H_{\omega1}-H_{\omega2})   \nonumber \\
&& \times \left[ (X_3+\tilde X_{5R})F_+^0 + Y_{3R}F_{5-}^0 +
(X_5^l + \tilde X_{4R}^l)F_+^l + (Y_4^l + \tilde Y_{6R}^l)F_{5-}^l
\right]
 + (H_{k1}^l+H_{k2}^l) \nonumber  \\
&& \times \left[ (X_5^l+\tilde X_{3R}^l)F_-^0 + (Y_3^l + \tilde
Y_{5R}^l)F_{5+}^0 + (X_4^{lk} + \tilde X_{6R}^{lk})F_-^k +
(Y_6^{lk} + \tilde Y_{4R}^{lk})F_{5+}^k
 \right]\left.\right\}, \label{V+An} \\
&& \{M_{VA}^k\}_c \equiv\{M_{V+A}(a)\}_c = \frac{R}{m_e}\int d{\bf
x}d{\bf y} T_N
\left\{\right. (H_{\omega1}+H_{\omega2})   \nonumber \\
&& \times \left[ (X_3+\tilde X_{5R})F_-^0 + Y_{3R}F_{5+}^0 +
(X_5^l + \tilde X_{4R}^l)F_-^l + (Y_4^l + \tilde Y_{6R}^l)F_{5+}^l
\right]
 + (H_{k1}^l-H_{k2}^l) \nonumber  \\
&& \times \left[ (X_5^l+\tilde X_{3R}^l)F_+^0 + (Y_3^l + \tilde
Y_{5R}^l)F_{5-}^0 + (X_4^{lk} + \tilde X_{6R}^{lk})F_+^k +
(Y_6^{lk} + \tilde Y_{4R}^{lk})F_{5-}^k
 \right]\left.\right\}, \label{V+Ac}
\end{eqnarray}
with $\tilde X = X + X^P$, $\tilde Y = Y + Y^P$. The operators $X$
and $Y$ are defined in \cite{Doi}, except for the operator
$Y_{6R}^l=-Y_{5R}^l$ which is defined to remove the minus sign from the Eqs.
(\ref{V+An}) and (\ref{V+Ac}); $X_1=X_{1S}$, $Y_1=Y_{1S}$.

The additional operators are
\begin{eqnarray}
&&X_{1R}^P=G_A^2P_{\sigma+}^{ii}, \quad X_{3R}^{Pl}=X_{4R}^{Pl}=
G_-P_+^l, \quad  X_{5R}^P = G_A\varepsilon_A P_{\sigma+}^{ii},
\nonumber \\ &&X_{6R}^{Plk} = -G_A\varepsilon_A \left[
\delta_{lk}P_{\sigma+}^{ii} - \left(
P_{\sigma+}^{lk}+P_{\sigma+}^{kl} \right) \right] + iG_+
\varepsilon_{ilk}P_+^i,\nonumber \\
&&Y_{1R}^{Pi}=G_VG_AP_-^i+G_A^2i\varepsilon_{ijk}P_{\sigma+}^{jk},
\quad
 Y_{4R}^{Plk} = -iG_-\varepsilon_{ilk}P_-^i, \nonumber \\
&&Y_{5R}^{Pl}=iG_A\varepsilon_A\varepsilon_{lij}P_{\sigma+}^{ij}-G_+P_-^l,
 \quad
Y_{6R}^{Pl} =
-iG_A\varepsilon_A\varepsilon_{lij}P_{\sigma+}^{ij}-G_+P_-^l,
\end{eqnarray}
with
\begin{equation}
 P_{\sigma+}^{ij}=\sigma_a^iP_b^j+P_a^i\sigma_b^j.
\end{equation}

Under the exchange of running indices $a$ and $b$, nuclear
operators $X$, electron currents $E_+$ and $F_+$ and neutrino
potentials $H_i$ and $H_{\omega i}$ are even, while $Y$, $E_-$,
$F_-$, and ${\bf H}_{ki}$ are odd.

New constants are defined as:
\begin{equation}
\varepsilon_V=\frac{g_V}{g_A}\left(\epsilon_{V+A,i}^{V+A}+
\epsilon_{V-A,i}^{V+A}\right), \quad
\varepsilon_A=\epsilon_{V+A,i}^{V+A}- \epsilon_{V-A,i}^{V+A}.
\end{equation}

The operators
\begin{eqnarray}
&& X_1, \ X_{1R}^P; \quad Y_1^i, \ Y_{1R}^{Pi}; \nonumber \\
&&X_3, \ X_5^l, X_{5R}^P, \ X_{4R}^{Pl}, \ {\bf r}\cdot{\bf
X}_{3R}, \ r^lX_{6R}^{lk}; \nonumber \\ && Y_4^l, \ Y_{6R}^{Pl}, \
{\bf r}\cdot{\bf Y}_{5R}, r^lY_{4R}^{lk}~,
\end{eqnarray}
have even parity and the operators
\begin{eqnarray}
 X_{1R}; \quad Y_{1R}^i;
X_{5R}, \ X_{4R}^l, \ {\bf r}\cdot{\bf X}_5, \ {\bf r}\cdot{\bf
X}_{3R}^P, \ r^lX_4^{lk}, \  r^lX_{6R}^{Plk}; \nonumber \\ Y_{3R},
\ Y_{6R}^l, \ {\bf r}\cdot{\bf Y}_3, \ {\bf r}\cdot{\bf Y}_{5R}^P,
\ r^lY_6^{lk}, \ r^lY_{4R}^{Plk}~,
\end{eqnarray}
have odd parity.

Using the definitions of the neutrino potentials from Eq.
(\ref{nu_potentials}) and
\begin{eqnarray}\label{nu_potential}
 h_\omega=\frac{R^2}{2}(H_{\omega1}+H_{\omega2})
\end{eqnarray}
in the $S-S$ case with no FBWC we have
\begin{eqnarray}\label{M_VA_SS}
&& \{M_{VA}^m\}_{n,S-S} =\int d{\bf x}d{\bf y}T_Nh_+(X_1+X_{1R}^P)E_+, \\
&& \{M_{VA}^m\}_{c,S-S} = \frac{\varepsilon_{21}R}{2} \int d{\bf
x}d{\bf y}T_Nh_0
(Y_1^i+Y_{1R}^{Pi})E_+^i, \\
&& \{M_{VA}^k\}_{n,S-S} = \frac{\varepsilon_{21}}{m_e}\int d{\bf
x}d{\bf y}T_Nh_{0\omega} \left[ (X_3+X_{5R}^P)F_+^0 + (X_5^l +
X_{4R}^{Pl})F_+
^l \right] \nonumber \\
&& + \frac{4}{m_eR}\int d{\bf x}d{\bf y}T_N\frac{iR}{2r}
h_+^\prime \hat r^l
 \left[ Y_{5R}^lF_{5+}^0 + Y_{4R}^{lk}F_{5+}^k
 \right], \label{V+AnSS} \\
&& \{M_{VA}^k\}_{c,S-S} = \frac{2}{m_eR}\int d{\bf x}d{\bf
y}T_Nh_\omega (Y_4^l + Y_{6R}^{Pl})F_{5+}^l
\nonumber\\
&& + \frac{\varepsilon_{21}}{m_e}\int d{\bf x}d{\bf
y}T_N\frac{iR}{r} h_0^\prime \hat r^l \left( X_{3R}^lF_+^0 + X
_{6R}^{lk}F_+^k \right), \label{V+AcSS}
\end{eqnarray}
where $E$ and $F$ are taken for ${\bf x} = {\bf y}$=0.

For the $0^+\to 0^+$ transition we have
\begin{eqnarray}
&&
\sum\limits_i\frac{m_i}{m_e}\sum\limits_N\left\{M_{VA}^m\right\}_{S-S}
=
g_A^2 (Z_1^X+Z_{1R}^{XP})E_+, \\
&& \sum\limits_i\sum\limits_N\left\{M_{VA}^k\right\}_{S-S} = g_A^2
\left[\frac{\varepsilon_{21}}{m_e}(Z_3^X+Z_{5R}^{XP}+\{Z_{3R}^X\}_c)F_+^0
+ \frac{4}{m_eR}Z_{4R}^YF_{5+}^0\right],
\end{eqnarray}
with
\begin{eqnarray}
  Z_1^X=\langle \frac{m_i}{m_e}h_+X_1 \rangle, &\quad  Z_{1R}^{XP}=\langle
\frac{m_i}{m_e}h_+X_{1R}^P \rangle, &\quad  Z_3^X=\langle
h_{0\omega}X_3 \rangle, \nonumber \\
  Z_{4R}^Y=\langle \frac{iR}{2r}h_+^\prime {\bf\hat r}\cdot{\bf  Y}_{5R} \rangle, &\quad
Z_{5R}^{XP}=\langle h_{0\omega}X_{5R}^P \rangle, &\quad
\left\{Z_{3R}^X\right\}_c=\langle \frac{iR}{r}h_0^\prime {\bf\hat
r}\cdot{\bf X}_{3R}\rangle.
\end{eqnarray}
In the $S-P_{1/2}$ case with no FBWC for the $0^+\to 0^+$
transition we have
\begin{eqnarray}
&& \{M_{VA}^m\}_{n,S-P_{1/2}} =\int d{\bf x}d{\bf y}T_Nh_+ Y_{1R}^iE_-^i,   \\
&& \{M_{VA}^m\}_{c,S-P_{1/2}} = \frac{\varepsilon_{21}R}{2} \int
d{\bf x}d{\bf y}T_Nh_0
   Y_{1R}^iE_+^i, \\
&& \{M_{VA}^k\}_{n,S-P_{1/2}} = \frac{\varepsilon_{21}}{m_e}\int
d{\bf x}d{\bf y}T_Nh_{0\omega}
  (X_{4R}^lF_+^l + Y_{6R}^lF_{5-}^l)  \nonumber \\
&& + \frac{4}{m_eR}\int d{\bf x}d{\bf y}T_N\frac{iR}{2r}
h_+^\prime \hat r^l \left[
  (X_4^{lk} + X_{6R}^{Plk})F_-^k + (Y_6^{lk} + Y_{4R}^{Plk})F_{5+}^k \right], \\
&& \{M_{VA}^k\}_{c,S-P_{1/2}} = \frac{2}{m_eR}\int d{\bf x}d{\bf
y}T_Nh_\omega
   (X_{4R}^lF_-^l + Y_{6R}^lF_{5+}^l)\nonumber\\
&& + \frac{\varepsilon_{21}}{m_e}\int d{\bf x}d{\bf
y}T_N\frac{iR}{r} h_0^\prime \hat r^l \left[
  (X_4^{lk} + X_{6R}^{Plk})F_+^k + (Y_6^{lk} + Y_{4R}^{Plk})F_{5-}^k \right].
\end{eqnarray}

The decay rate for the $0^+\to0^+$ transition takes the form
\begin{eqnarray}
 d\Gamma = \sum\limits_{s_1, s_2} |R_{0\nu}|^2 \frac{m_e^5}{4\pi^3}d\Omega_{0\nu} =
\frac{a_{0\nu}}{(m_eR)^2} \left[ A_0^{VA} - {\bf \hat
p}_1\cdot{\bf \hat p}_2 B_0^{VA} \right] d\Omega_{0\nu},
\end{eqnarray}
where the coefficients are
\begin{eqnarray}
&& A_0^{VA} = \sum\limits_{i=1}^4|N_i|^2, \\ && B_0^{VA} =
\mbox{Re}(N_1N_2^*+N_1^*N_2+N_3N_4^*+N_3^*N_4),
\end{eqnarray}
with
\begin{eqnarray}
 && N_1 = \alpha_{-1-1}^* \left\{ \left[Z_1^X+Z_{1R}^{XP}-\frac{4}{m_eR}Z_{4R}^Y\right]\right.
+ \left[
\frac{m_er}{6}\left(\left(\frac{\zeta}{m_eR}-2\right)Z_{1R}^Y
+\frac{\varepsilon_{21}^2R}{2m_e}\{Z_{1R}^Y\}_c\right) +\right.\nonumber\\
&& \left.\left.
\frac{2}{3}\left(\frac{\zeta}{m_eR}-2\right)\left(Z_6^Y+Z_{4R}^{YP}+
\{Z_{6R}^Y\}_c\right)\frac{r}{2R} +
\frac{1}{3}\frac{\varepsilon_{21}^2R}{m_e}\left(Z_{6R}^Y-
\frac{1}{2}(\{Z_6^Y\}_c+\{Z_{4R}^Y\}_c)\right) \right] \right.\nonumber\\
&& \left. +\left[ \frac{(\alpha Z)^2}{m_eR}(\{Z_5^X\}_c+3Z_{5RF}^Y) \right]\right\}, \label{N_1}\\
 && N_2 = \alpha_{11}^* \left\{ \left[Z_1^X+Z_{1R}^{XP}+\frac{4}{m_eR}Z_{4R}^Y\right]\right.
+ \left[
\frac{m_er}{6}\left(\left(\frac{\zeta}{m_eR}+2\right)Z_{1R}^Y
+\frac{\varepsilon_{21}^2R}{2m_e}\{Z_{1R}^Y\}_c\right) +\right.\nonumber\\
&&
\left.\left.-\frac{2}{3}\left(\frac{\zeta}{m_eR}+2\right)\left(Z_6^Y+Z_{4R}^{YP}+
\{Z_{6R}^Y\}_c\right)\frac{r}{2R} -
\frac{1}{3}\frac{\varepsilon_{21}^2R}{m_e}\left(Z_{6R}^Y-
\frac{1}{2}(\{Z_6^Y\}_c+\{Z_{4R}^Y\}_c)\right)\right] \right.\nonumber\\
&& \left. +\left[-\frac{(\alpha Z)^2}{m_eR}(\{Z_5^X\}_c+3Z_{5RF}^Y)\right] \right\}, \label{N_2}\\
&& N_3 = \alpha_{1-1}^* \left\{ \left[Z_1^X+Z_{1R}^{XP} -
\frac{\varepsilon_{21}}{m_e}
(Z_3^X+Z_{5R}^{XP}+\{Z_{3R}^X\}_c)\right]\right. +
\left[\frac{r}{6R}\left( \zeta Z_{1R}^Y+
\frac{1}{2}\varepsilon_{21}(\varepsilon_{21}+2m_e)R^2Z_{2R}^Y \right)\right. \nonumber \\
&& + \left.\left.
\frac{1}{3}\frac{\varepsilon_{21}}{m_e}\zeta\left(Z_{4R}^X
-\frac{1}{2}(\{Z_4^X\}_c+\{Z_{6R}^{XP}\}_c)\right)\frac{r}{2R}
- \frac{1}{3}\left(\frac{\varepsilon_{21}}{m_e}+2\right)(Z_4^X+Z_{6R}^{XP}-2Z_{4R}^X) \right] \right\}, \\
&& N_4 = \alpha_{-11}^* \left\{ \left[Z_1^X+Z_{1R}^{XP} +
\frac{\varepsilon_{21}}{m_e}
(Z_3^X+Z_{5R}^{XP}+\{Z_{3R}^X\}_c)\right]\right. +
\left[\frac{r}{6R}\left( \zeta Z_{1R}^Y+
\frac{1}{2}\varepsilon_{21}(\varepsilon_{21}-2m_e)R^2Z_{2R}^Y \right)\right. \nonumber \\
 && \left.\left.-\frac{1}{3}\frac{\varepsilon_{21}}{m_e}\zeta\left(Z_{4R}^X
-\frac{1}{2}(\{Z_4^X\}_c+\{Z_{6R}^{XP}\}_c)\right)\frac{r}{2R} +
\frac{1}{3}\left(\frac{\varepsilon_{21}}{m_e}-2\right)(Z_4^X+Z_{6R}^{XP}-2Z_{4R}^X)
\right] \right\}, \label{N_4}
\end{eqnarray}
where the terms in the first brackets in Eqs.
(\ref{N_1})--(\ref{N_4}) come from the $S-S$ case and the terms in
the second ones come from the $S-P_{1/2}$ case. The terms in the
third brackets in Eqs. (\ref{N_1})--(\ref{N_2}) are the most
important terms of those that come from the $P_{1/2}-P_{1/2}$ case
and from the $S-S$ case due to FBWC. The nuclear matrix elements
are
\begin{eqnarray}
 && Z_{1R}^Y=\langle \frac{m_i}{m_e} \frac{i}{2R}h_+{\bf r}\cdot{\bf Y}_{1R} \rangle, \quad
 \{Z_{1R}^Y\}_c=\langle \frac{m_i}{m_e}\frac{i}{2R}h_0{\bf r}_+\cdot{\bf Y}_{1R} \rangle,\quad  \nonumber \\
 && Z_6^Y=\langle -\frac{1}{2r}h_+^\prime \hat r^ir_+^jY_6^{ij} \rangle, \quad
 Z_{4R}^{YP}=\langle -\frac{1}{2r}h_+^\prime \hat r^ir_+^jY_{4R}^{Pij} \rangle, \quad
 \{Z_{6R}^Y\}_c=\langle \frac{i}{2R}h_\omega {\bf r}\cdot{\bf Y}_{6R} \rangle, \nonumber \\
 && Z_{6R}^Y=\langle \frac{i}{2R}h_{0\omega} {\bf r}\cdot{\bf Y}_{6R} \rangle, \quad
 \{Z_6^Y\}_c=\langle \frac{1}{r}h_0^\prime \hat r^ir^jY_6^{ij} \rangle, \quad
 \{Z_{4R}^Y\}_c=\langle \frac{1}{r}h_0^\prime \hat r^ir^jY_{4R}^{ij} \rangle, \nonumber \\
 && Z_{4R}^X=\langle \frac{i}{2R}h_{0\omega} {\bf r}_+\cdot{\bf X}_{4R} \rangle, \quad
 \{Z_4^X\}_c=\langle \frac{1}{r}h_0^\prime \hat r^ir_+^jX_4^{ij} \rangle, \quad
 \{Z_{6R}^{XP}\}_c=\langle \frac{1}{r}h_0^\prime \hat r^ir_+^jX_{6R}^{Pij} \rangle, \nonumber \\
 && Z_4^X=\langle \frac{1}{r}h_+^\prime \hat r^ir^jX_4^{ij} \rangle, \quad
 \{Z_{5}^{X}\}_c=\langle \frac{ir^2}{2R^2}h_\omega [\hat{\bf r}_a\times\hat{\bf r}_b]\cdot{\bf X}_5 \rangle,
 \quad Z_{5RF}^Y=\langle \frac{iR}{2r}\frac{r_a^2+r_b^2}{2R^2}h_+^\prime\hat{\bf r}\cdot{\bf
 Y}_{5R}\rangle.
\end{eqnarray}

The dominant terms give
\begin{eqnarray}
&&  N_1 = \alpha_{-1-1}^* \left\{
\left[Z_1^X-\frac{4}{m_eR}Z_{4R}^Y\right] +
\left[ \frac{2}{3}\left(\frac{\zeta}{m_eR}-2\right)Z_6^Y\frac{r}{2R} \right]
\right\}, \label{Eq238} \\
&&  N_2 = \alpha_{11}^* \left\{
\left[Z_1^X+\frac{4}{m_eR}Z_{4R}^Y\right] +
\left[-\frac{2}{3}\left(\frac{\zeta}{m_eR}+2\right)Z_6^Y\frac{r}{2R} \right]
\right\},  \label{Eq239}\\
&&  N_3 = \alpha_{1-1}^* \left\{ \left[Z_1^X -
\frac{\varepsilon_{21}}{m_e} Z_3^X\right] + \left[
- \frac{1}{3}\left(\frac{\varepsilon_{21}}{m_e}+2\right)Z_4^X \right]
\right\}, \label{Eq240} \\
&&  N_4 = \alpha_{-11}^* \left\{ \left[Z_1^X +
\frac{\varepsilon_{21}}{m_e} Z_3^X\right] + \left[
 \frac{1}{3}\left(\frac{\varepsilon_{21}}{m_e}-2\right)Z_4^X \right] \right\}, \label{Eq241}
\end{eqnarray}
that agrees with the Eq. (C.3.7) of Ref. \cite{Doi} taking into
account the correspondence with their notations:
\begin{eqnarray}
 Z_1^X = Z_1, &\quad Z_3^X = Z_3, &\quad Z_6^Y = Z_6, \nonumber\\
 Z_{4R}^Y = Z_{4R}, &\quad Z_{4R}^X = Z_{5R}, &\quad Z_4^X = Z_5,
\end{eqnarray}
and the fact that $Z_2$ is absent, as we have calculated only
the leading contribution of the parameters
$\epsilon_\alpha^\beta$. Recall that in Ref. \cite{Doi} the
pseudoscalar form factor is not taken into account. However the
terms associated with this form factor do not contribute to the
dominant terms (\ref{Eq238})--(\ref{Eq241}). Note that  in the
expressions for $N_1$ and $N_2$ given above, the terms with
 $\zeta$ are due to the inclusion of the $P$-wave in the electron wave function and the
ones with $Z_{4R}^Y$ are due to the nucleon recoil effect. We
remark that some of the subdominant terms, like those with
$Z_{4R}^X$, $\{Z_4^X\}_c$, $\{Z_{6R}^Y\}_c$, $\{Z_5^X\}_c$ and
$Z_{5RF}^Y$, should be taken into account in the case of large
cancellation among the dominant terms. The same is valid for the
contribution due to the pseudoscalar form factor $g_AP_a^i$ which
yields corrections at about 10 \% to the dominant terms.

\section{$0\nu2\beta$ decay rate for tensor nonstandard terms}
The nucleon currents in the impulse approximation up to order
$p/m_p$ in the nonrelativistic expansion are used
\cite{Tomoda,Ericson}, $J_{V-A}^{\mu+}$ from Eq. (\ref{J_V-A}) and
\begin{eqnarray}
 J_{T_{L,R}}^{\mu\nu+}({\bf x}) &=& T_1^{(3)}\sum\limits_a \tau_+^a
\delta({\bf x-r}_a) \left\{ (g^{\mu k}g^{\nu0}-g^{\mu0}g^{\nu
k})T_a^k+g^{\mu m}g^{\nu
n}\varepsilon_{kmn}\sigma_{ak} \right. \nonumber\\
 &\mp& \left.\frac{i}{2}\varepsilon^{\mu\nu\rho\sigma}\left[
(g_{\rho k}g_{\sigma0} -g_{\rho0}g_{\sigma k})T_{ak}+g_{\rho
r}g_{\sigma s}\varepsilon_{rsk}\sigma_{ak}
\right]\right\},\\
 T_a^k&=&\left[ i\left(T_1^{(3)}-2T_2^{(3)}\right)q^kI_a +
T_1^{(3)}[{\bm\sigma_a}\times{\bf Q}]^k \right]/(2T_1^{(3)}m_p),
\end{eqnarray}
where, as before, $q^\mu=p^\mu-p^{\prime\mu}$ is the 4-momentum transferred
from hadrons to leptons, $Q^\mu=p^\mu+p^{\prime\mu}$, $p^\mu$ and
$p^{\prime\mu}$ are the initial and final 4-momenta of a nucleon.
We neglect the dipole dependence of the form factors $T_1^{(3)}$
and  $T_2^{(3)}$ on the momentum transfer and omit the zero
argument of the form factors.

Consider the pure $T_{L,R}$ case assuming $\langle m \rangle=0$.
In terms of the hadronic currents
\begin{eqnarray}
&& J_{LT_{L,R}}^{\alpha\mu\nu}=\langle F|\hat
J^{\alpha+}_L|N\rangle \langle N|\tilde
J_{T_{L,R}}^{\mu\nu+}|I\rangle, \quad
 J_{T_{L,R}L}^{\mu\nu\alpha}=\langle F|\tilde J^{\mu\nu+}_{T_{L,R}}|N\rangle
\langle N|\hat J_L^{\alpha+}|I\rangle, \\
&& \tilde
J^{\mu\nu+}_{T_L}=\epsilon_{T_L,i}^{T_L}J_{T_L}^{\mu\nu+} +
\epsilon_{T_R,i}^{T_L}J_{T_R}^{\mu\nu+}, \quad \tilde
J^{\mu\nu+}_{T_R}=\epsilon_{T_R,i}^{T_R}J_{T_R}^{\mu\nu+}
+ \epsilon_{T_L,i}^{T_R}J_{T_L}^{\mu\nu+}, \\
&& \hat J^{\mu+}_L=U_{ei}J_{V-A}^{\mu+}~,
\end{eqnarray}
and the leptonic tensors
\begin{eqnarray}
&& \ell_{\alpha\mu\nu}^1=\frac{t_{\alpha\mu\nu}^1(2{\bf y},1{\bf
x})}{\omega+A_1} -
\frac{t_{\alpha\mu\nu}^1(1{\bf y},2{\bf x})}{\omega+A_2}, \\
&&
\ell_{\alpha\lambda\mu\nu}^1=\frac{t_{\alpha\lambda\mu\nu}^1(2{\bf
y},1{\bf x})}{\omega+A_1} - \frac{t_{\alpha\lambda\mu\nu}^1
(1{\bf y},2{\bf x})}{\omega+A_2}, \\
&& \ell_{\mu\nu\alpha}^2=\frac{t_{\mu\nu\alpha}^2(2{\bf y},1{\bf
x})}{\omega+A_1} -
\frac{t_{\mu\nu\alpha}^2(1{\bf y},2{\bf x})}{\omega+A_2}, \\
&&
\ell_{\mu\nu\lambda\alpha}^2=\frac{t_{\mu\nu\lambda\alpha}^2(2{\bf
y},1{\bf x})}{\omega+A_1} - \frac{t_{\mu\nu\lambda\alpha}^2(1{\bf
y},2{\bf x})}{\omega+A_2}~,
\end{eqnarray}
with the electron currents defined as
\begin{eqnarray}
&&    t_{\alpha\mu\nu}^1(2{\bf y},1{\bf x}) = \bar e_2({\bf
y})\gamma_\alpha(1-\gamma_5)\sigma_{\mu\nu}e_1^c({\bf x}),
\nonumber\\ &&    t_{\alpha\lambda\mu\nu}^1(2{\bf y},1{\bf x}) =
\bar e_2({\bf
y})\gamma_\alpha(1-\gamma_5)\gamma_\lambda\sigma_{\mu\nu}
e_1^c({\bf x}), \nonumber\\ &&    t_{\mu\nu\alpha}^2(2{\bf
y},1{\bf x}) = \bar e_2({\bf
y})\sigma_{\mu\nu}(1-\gamma_5)\gamma_\alpha e_1^c({\bf x}),
\nonumber\\ && t_{\mu\nu\lambda\alpha}^2(2{\bf y},1{\bf x}) = \bar
e_2({\bf y})\sigma_{\mu\nu}\gamma_\lambda(1-\gamma_5)\gamma_\alpha
e_1^c({\bf x})~,
\end{eqnarray}
the matrix element is expressed as
\begin{eqnarray}\label{R^T}
 R_{0\nu}^{T}= \frac{1}{\sqrt{2!}}\left(\frac{G_F|V_{ud}|}{\sqrt{2}}\right)^2
2\sum\limits_i \int d{\bf x}d{\bf y}\frac{d{\bf k}}{(2\pi)^3}
\frac{e^{i{\bf k}\cdot{\bf r}}}{2\omega} \nonumber \\
\times\sum\limits_N\left[
m_i\left(J_{LT_L}^{\alpha\mu\nu}\ell_{\alpha\mu\nu}^1
+J_{T_LL}^{\mu\nu\alpha}\ell_{\mu\nu\alpha}^2\right) + k^\lambda
\left(J_{LT_R}^{\alpha\mu\nu}
\ell_{\alpha\lambda\mu\nu}^1+J_{T_RL}^{\mu\nu\alpha}\ell_{\mu\nu\lambda\alpha}^2\right)
\right].
\end{eqnarray}

For the electron currents we have the identities
\begin{eqnarray}
&&    t_{\alpha\mu\nu}^1(1{\bf y},2{\bf x}) =
-t_{\mu\nu\alpha}^2(2{\bf y},1{\bf x}), \nonumber\\ &&
t_{\alpha\lambda\mu\nu}^1(1{\bf y},2{\bf x}) =
t_{\mu\nu\lambda\alpha}^2(2{\bf y},1{\bf x}).
\end{eqnarray}

Using Eqs (\ref{algebraic_formula}), (\ref{Cconstant}), and
(\ref{neutrino_potentials}), the matrix element (\ref{R^T}) is
expressed as
\begin{equation}
 R_{0\nu}^{T}=C_{0\nu}\sum\limits_i\sum\limits_N\left(
 \frac{m_i}{m_e}M_{T}^m+M_{T}^k \right),
\end{equation}
\begin{equation}
 M_T^{m,k} = \{M_T^{m,k}\}_n+\{M_T^{m,k}\}_c,
\end{equation}
with  nonvanishing ($n$) and vanishing ($c$) in the closure
approximation parts:
\begin{eqnarray}\label{Tm}
&& \left\{M_{T}^m\right\}_n =R\int d{\bf x}d{\bf y}T_N (H_1+H_2) \nonumber\\
&&\times\left[ (U_1+\tilde U_{1R})F_{5+}^0 + (U_3^i+\tilde
U_{3R}^i)F_{5+}^i
+ \tilde V_{1R}F_-^0 + (V_3^i+\tilde V_{3R}^i)F_-^i \right], \\
&&\left\{M_{T}^m\right\}_c =R\int d{\bf x}d{\bf y}T_N (H_1-H_2) \nonumber\\
&&\times\left[ (U_1+\tilde U_{1R})F_{5-}^0 + (U_3^i+\tilde
U_{3R}^i)F_{5-}^i + \tilde V_{1R}F_+^0 + (V_3^i+\tilde
V_{3R}^i)F_+^i \right],
\end{eqnarray}
\begin{eqnarray}\label{Tk}
&&\left\{M_{T}^k\right\}_n =\frac{R}{m_e}\int d{\bf x}d{\bf y}T_N
(H_{\omega1}-H_{\omega2}) \left[ \tilde V_{2R}E_- + (U_4^i+\tilde
U_{4R}^i)F_+^{0i}
+ (U_6^{ij}+\tilde U_{6R}^{ij})F_+^{ij} \right] \nonumber \\
&&+ (H_{k1}^i+H_{k2}^i) \left[ (V_4^i+\tilde V_{4R}^i)E_+ +
(U_2+\tilde U_{2R})F_-^{0i}
+ (U_5^{j}+\tilde U_{5R}^{j})F_-^{ij} + (U_7^{ij}+\tilde U_{7R}^{ij})F_-^{0j} \right. \nonumber \\
&&\left. + (U_8^{ijk}+\tilde U_{8R}^{ijk})F_-^{jk} \right], \\
&&\left\{M_{T}^k\right\}_c =\frac{R}{m_e}\int d{\bf x}d{\bf y}T_N
(H_{\omega1}+H_{\omega2}) \left[ \tilde V_{2R}E_+ + (U_4^i+\tilde
U_{4R}^i)F_-^{0i}
+ (U_6^{ij}+\tilde U_{6R}^{ij})F_-^{ij} \right] \nonumber\\
&& + (H_{k1}^i-H_{k2}^i)\left[ (V_4^i+\tilde V_{4R}^i)E_- +
(U_2+\tilde U_{2R})F_+^{0i}
+ (U_5^{j}+\tilde U_{5R}^{j})F_+^{ij} + (U_7^{ij}+\tilde U_{7R}^{ij})F_+^{0j}\right.  \nonumber \\
&& \left.+ (U_8^{ijk}+\tilde U_{8R}^{ijk})F_+^{jk} \right],
\label{Tkc}
\end{eqnarray}
where the nucleon operators are
\begin{eqnarray}
 &\tilde U=U+U^P, \quad & \tilde V=V+V^P,
\end{eqnarray}
\begin{eqnarray}
  U_1 &=& -2G_A^0(\varepsilon_{T_1}+\varepsilon_{T_2}){\bm\sigma}_a{\bm\sigma}_b,\quad
U_{1R}^P= G_A^0(\varepsilon_{T_1}+\varepsilon_{T_2})P_{\sigma+}^{ii}, \nonumber\\
  U_{1R} &=& G_V^0(\varepsilon_{T_1}+\varepsilon_{T_2})D_{\sigma+}^{ii}
- iG_A^0(\varepsilon_{T_1}+\varepsilon_{T_2})T_{\sigma+}^{ii}, \nonumber \\
  U_2 &=& 2iG_A^0(\varepsilon_{T_1}^\prime+\varepsilon_{T_2}^\prime){\bm\sigma}_a{\bm\sigma}_b, \quad
U_{2R}^P = -iG_A^0(\varepsilon_{T_1}^\prime+\varepsilon_{T_2}^\prime)P_{\sigma+}^{ii}, \nonumber\\
  U_{2R} &=&  -iG_V^0(\varepsilon_{T_1}^\prime+\varepsilon_{T_2}^\prime)D_{\sigma+}^{ii}
+ G_A^0(\varepsilon_{T_1}^\prime+\varepsilon_{T_2}^\prime)T_{\sigma+}^{ii}, \nonumber \\
  U_3^i &=& -G_V^0(\varepsilon_{T_1}+\varepsilon_{T_2})\sigma_+^i, \quad
U_{3R}^{Pi}= -iG_A^0(\varepsilon_{T_1}+\varepsilon_{T_2})\varepsilon_{ijk}P_{\sigma-}^{jk},\nonumber\\
  U_{3R}^i &=& G_A^0(\varepsilon_{T_1}+\varepsilon_{T_2})C_{\sigma+}^{i}
- iG_V^0(\varepsilon_{T_1}+\varepsilon_{T_2})\varepsilon_{ijk}D_{\sigma-}^{jk} \nonumber\\
 &-& iG_V^0(\varepsilon_{T_1}+\varepsilon_{T_2})T_{+}^{i}
- iG_A^0(\varepsilon_{T_1}+\varepsilon_{T_2})\varepsilon_{ijk}T_{\sigma-}^{jk}, \nonumber\\
  U_4^i &=& -iG_V^0(\varepsilon_{T_1}^\prime+\varepsilon_{T_2}^\prime)\sigma_+^i, \quad
U_{4R}^{Pi}=
-G_A^0(\varepsilon_{T_1}^\prime+\varepsilon_{T_2}^\prime)\varepsilon_{ijk}P_{\sigma-}^{jk},
\nonumber\\
  U_{4R}^i &=& iG_A^0(\varepsilon_{T_1}^\prime+\varepsilon_{T_2}^\prime)C_{\sigma+}^{i}
- G_V^0(\varepsilon_{T_1}^\prime+\varepsilon_{T_2}^\prime)\varepsilon_{ijk}D_{\sigma-}^{jk} \nonumber\\
 &-& G_V^0(\varepsilon_{T_1}^\prime+\varepsilon_{T_2}^\prime)T_{+}^{i}
+ iG_A^0(\varepsilon_{T_1}^\prime+\varepsilon_{T_2}^\prime)\varepsilon_{ijk}T_{\sigma-}^{jk}, \nonumber\\
  U_5^i &=& -iG_V^0(\varepsilon_{T_1}^\prime+\varepsilon_{T_2}^\prime)\sigma_+^i, \quad
U_{5R}^{Pi}= G_A^0\varepsilon_{T_1}^\prime\varepsilon_{ijk}P_{\sigma-}^{jk}, \nonumber\\
  U_{5R}^i &=& -iG_A^0(\varepsilon_{T_1}^\prime+\varepsilon_{T_2}^\prime)C_{\sigma+}^{i}
+ G_V^0\varepsilon_{T_1}^\prime\varepsilon_{ijk}D_{\sigma-}^{jk} \nonumber\\
 &-& G_V^0(\varepsilon_{T_1}^\prime+\varepsilon_{T_2}^\prime)T_{+}^{i}
+ iG_A^0\varepsilon_{T_2}^\prime\varepsilon_{ijk}T_{\sigma-}^{jk}, \nonumber\\
  U_6^{ij} &=& \frac{1}{2}G_V^0(\varepsilon_{T_1}^\prime+\varepsilon_{T_2}^\prime)\varepsilon_{ijk}\sigma_+^k,
\quad U_{6R}^{Pij} = iG_A^0(\varepsilon_{T_1}^\prime+\varepsilon_{T_2}^\prime)P_{\sigma_+}^{ij}, \nonumber\\
  U_{6R}^{ij} &=& -\frac{1}{2}G_A^0(\varepsilon_{T_1}^\prime+\varepsilon_{T_2}^\prime)
\varepsilon_{ijk}C_{\sigma+}^k
- \frac{i}{2}G_V^0(\varepsilon_{T_1}^\prime+\varepsilon_{T_2}^\prime)\varepsilon_{ijk}T_{+}^k \nonumber\\
 &+& iG_V^0(\varepsilon_{T_1}^\prime+\varepsilon_{T_2}^\prime)D_{\sigma+}^{ij}
- iG_A^0(\varepsilon_{T_1}^\prime+\varepsilon_{T_2}^\prime)T_{\sigma+}^{ij}, \nonumber\\
U_7^{ij} &=&
+G_V^0(\varepsilon_{T_1}^\prime+\varepsilon_{T_2}^\prime)\varepsilon_{ijk}\sigma_+^k
-2iG_A^0(\varepsilon_{T_1}^\prime+\varepsilon_{T_2}^\prime)
(\sigma_a^i\sigma_b^j+\sigma_a^j\sigma_b^i),
 \nonumber\\
U_{7R}^{ij} &=&
-G_A^0(\varepsilon_{T_1}^\prime+\varepsilon_{T_2}^\prime)\varepsilon_{ijk}C_{\sigma+}^k
- iG_V^0(\varepsilon_{T_1}^\prime+\varepsilon_{T_2}^\prime)\varepsilon_{ijk}T_+^k \nonumber\\
&+&
iG_V^0(\varepsilon_{T_1}^\prime+\varepsilon_{T_2}^\prime)(\tilde
D_{\sigma+}^{ij}+\tilde D_{\sigma+}^{ji}) -
G_A^0(\varepsilon_{T_1}^\prime+\varepsilon_{T_2}^\prime)
(\tilde T_{\sigma+}^{ij}+\tilde T_{\sigma+}^{ji}), \nonumber\\
U_{7R}^{Pij} &=&
+iG_A^0(\varepsilon_{T_1}^\prime+\varepsilon_{T_2}^\prime)
(\tilde P_{\sigma+}^{ij}+\tilde P_{\sigma+}^{ji}), \nonumber\\
U_8^{ijk} &=&
+\frac{1}{2}G_A^0(\varepsilon_{T_1}^\prime+\varepsilon_{T_2}^\prime)
[\varepsilon_{ljk}(\sigma_a^i\sigma_b^l+\sigma_a^l\sigma_b^i)
+ 2\varepsilon_{ilj}(\sigma_a^l\sigma_b^k+\sigma_a^k\sigma_b^l)], \nonumber\\
U_{8R}^{ijk} &=&
-\frac{1}{2}G_V^0(\varepsilon_{T_1}^\prime+\varepsilon_{T_2}^\prime)
\varepsilon_{ljk}\tilde D_{\sigma+}^{li} -
G_V^0\varepsilon_{ilj}(\varepsilon_{T_1}^\prime\tilde
D_{\sigma+}^{lk}
+ \varepsilon_{T_2}^\prime\tilde D_{\sigma+}^{kl}) \nonumber\\
&-&
\frac{i}{2}G_A^0(\varepsilon_{T_1}^\prime+\varepsilon_{T_2}^\prime)\varepsilon_{ljk}\tilde
T_{\sigma+}^{il} -
iG_A^0\varepsilon_{ilj}(\varepsilon_{T_1}^\prime\tilde
T_{\sigma+}^{lk}
+ \varepsilon_{T_2}^\prime\tilde T_{\sigma+}^{kl}), \nonumber\\
U_{8R}^{Pijk} &=&
-\frac{1}{2}G_A^0(\varepsilon_{T_1}^\prime+\varepsilon_{T_2}^\prime)
\varepsilon_{ljk}\tilde P_{\sigma+}^{li} -
G_A^0\varepsilon_{ilj}(\varepsilon_{T_1}^\prime\tilde
P_{\sigma+}^{lk} + \varepsilon_{T_2}^\prime\tilde
P_{\sigma+}^{kl}),
\end{eqnarray}
\begin{eqnarray}
 V_{1R} &=& -G_V^0(\varepsilon_{T1}+\varepsilon_{T2})D_{\sigma-}^{ii}
- iG_A^0(\varepsilon_{T1}+\varepsilon_{T2})T_{\sigma-}^{ii}, \quad
V_{1R}^P=-G_A^0(\varepsilon_{T1}+\varepsilon_{T_2})P_{\sigma-}^{ii},\nonumber \\
 V_{2R} &=& -G_V^0(\varepsilon_{T1}^\prime+\varepsilon_{T2}^\prime)D_{\sigma-}^{ii}
+
iG_A^0(\varepsilon_{T1}^\prime+\varepsilon_{T2}^\prime)T_{\sigma-}^{ii},
\quad
V_{2R}^P=-G_A^0(\varepsilon_{T1}^\prime+\varepsilon_{T_2}^\prime)P_{\sigma-}^{ii},\nonumber \\
 V_3^i &=& G_V^0(\varepsilon_{T_1}+\varepsilon_{T_2})\sigma_-^i
+ 2iG_A^0(\varepsilon_{T_1}+\varepsilon_{T_2})[{\bm\sigma}_a\times{\bm\sigma}_b]^i,\nonumber\\
 V_{3R}^i &=& -G_A^0(\varepsilon_{T1}+\varepsilon_{T2})C_{\sigma-}^{i}
+ iG_V^0(\varepsilon_{T_1}+\varepsilon_{T_2})\varepsilon_{ijk}D_{\sigma+}^{jk} \nonumber\\
 &+& iG_V^0(\varepsilon_{T_1}+\varepsilon_{T_2})T_{-}^{i}
- iG_A^0(\varepsilon_{T_1}+\varepsilon_{T_2})\varepsilon_{ijk}T_{\sigma+}^{jk}, \nonumber\\
 V_{3R}^{Pi} &=& iG_A^0(\varepsilon_{T_1}+\varepsilon_{T_2})\varepsilon_{ijk}P_{\sigma+}^{jk},\nonumber\\
 V_4^i &=& G_V^0(\varepsilon_{T_1}^\prime+\varepsilon_{T_2}^\prime)\sigma_-^i
- 2iG_A^0(\varepsilon_{T_1}^\prime+\varepsilon_{T_2}^\prime)[{\bm\sigma}_a\times{\bm\sigma}_b]^i,\nonumber\\
 V_{4R}^i &=& -G_A^0(\varepsilon_{T1}^\prime+\varepsilon_{T2}^\prime)C_{\sigma-}^{i}
+ iG_V^0(\varepsilon_{T_1}^\prime+\varepsilon_{T_2}^\prime)\varepsilon_{ijk}D_{\sigma+}^{jk} \nonumber\\
 &-& iG_V^0(\varepsilon_{T_1}^\prime+\varepsilon_{T_2}^\prime)T_{-}^{i}
- G_A^0(\varepsilon_{T_1}^\prime+\varepsilon_{T_2}^\prime)\varepsilon_{ijk}T_{\sigma+}^{jk}, \nonumber\\
 V_{4R}^{Pi} &=& iG_A^0(\varepsilon_{T_1}^\prime+\varepsilon_{T_2}^\prime)\varepsilon_{ijk}P_{\sigma+}^{jk},
\end{eqnarray}
with
\begin{equation}
  T_\pm^i=T_a^iI_b\pm I_aT_b^i, \quad
  T_{\sigma\pm}^{ij}=\sigma_a^iT_b^j \pm T_a^i\sigma_b^j, \\
  \tilde X_{\sigma\pm}^{ij}=\sigma_a^iX_b^j \pm X_a^j\sigma_b^i,
  \quad X=D,\,T,\,P.
\end{equation}

Under the exchange of indices $a$ and $b$, nuclear operators $U$,
electron currents $F_+$ and neutrino potentials $H_i$ and
$H_{\omega i}$ are even, while $V$, $F_-$, and ${\bf H}_{ki}$ are
odd.

The new constants are defined as:
\begin{eqnarray}
 \varepsilon_{T_1}=\frac{T_1^{(3)}}{g_A}\left(\epsilon_{T_L,i}^{T_L}
 +\epsilon_{T_R,i}^{T_L}\right),
 \quad \varepsilon_{T_2}=\frac{T_1^{(3)}}{g_A}\left(\epsilon_{T_L,i}^{T_L}
-\epsilon_{T_R,i}^{T_L}\right), \nonumber \\
 \varepsilon_{T_1}^\prime=\frac{T_1^{(3)}}{g_A}\left(\epsilon_{T_R,i}^{T_R}
+\epsilon_{T_L,i}^{T_R}\right),
 \quad \varepsilon_{T_2}^\prime=\frac{T_1^{(3)}}{g_A}\left(\epsilon_{T_R,i}^{T_R}
-\epsilon_{T_L,i}^{T_R}\right).
\end{eqnarray}

The even parity  operators are
\begin{eqnarray}
 && U_1, \ U_{1R}^P, \ k^iU_{2R}, \ U_3^i,\ U_{3R}^{Pi}, \ U_4^i, \ U_{4R}^{Pi}, \ k^iU_{5R}^j,
\ U_6^{ij}, \ U_{6R}^{Pij}, \ k^iU_{7R}^{ij}, \ k^iU_{8R}^{ijk}; \nonumber\\
 && V_{1R}^P, \ V_{2R}^P, \ V_3^i, \ V_{3R}^{Pi}, \ {\bf k}\cdot{\bf V}_{4R};
\end{eqnarray}
and the odd parity  operators are
\begin{eqnarray}
 && U_{1R},\ k^iU_2, \ k^iU_{2R}^P, \ U_{3R}^i, \ U_{4R}^i, \ k^iU_5^j, \ k^iU_{5R}^{Pj},
\ U_{6R}^{ij}, \ k^iU_7^{ij}, \ k^iU_{7R}^{Pij}, \nonumber\\
 && k^iU_8^{ijk}, \ k^iU_{8R}^{Pijk}; \quad V_{1R}, \ V_{2R}, \ V_{3R}^i, \ {\bf k}\cdot{\bf  V}_4,
\ {\bf k}\cdot{\bf V}_{4R}^P.
\end{eqnarray}

Using the definitions of the neutrino potentials from Eqs.
(\ref{nu_potentials}) and (\ref{nu_potential}), in the $S-S$ case
with no FBWC we have
\begin{eqnarray}\label{M_T_SS}
&& \left\{M_{T}^m\right\}_{n,S-S} =2\int d{\bf x}d{\bf y}T_Nh_+
\left[ (U_1+U_{1R}^P)F_{5+}^0
+ (U_3^i+U_{3R}^{Pi})F_{5+}^i \right], \\
&&\left\{M_{T}^m\right\}_{c,S-S} =\varepsilon_{21}R\int d{\bf
x}d{\bf y}T_Nh_0 \left[ V_{1R}^PF_+^0 + (V_3^i+V_{3R}^{Pi})F_+^i
\right],
\end{eqnarray}
\begin{eqnarray}
&&\left\{M_{T}^k\right\}_{n,S-S} =\frac{\varepsilon_{21}}{m_e}\int
d{\bf x}d{\bf y}T_N h_{0\omega} \left[ (U_4^i+U_{4R}^{Pi})F_+^{0i}
+ (U_6^{ij}+U_{6R}^{Pij})F_+^{ij} \right] \nonumber \\ &&+
\frac{4}{m_eR}\int d{\bf x}d{\bf y}T_N \frac{iR}{2r}h_+^\prime
{\bf\hat r}\cdot{\bf V}_{4R}E_+, \\
&&\left\{M_{T}^k\right\}_{c,S-S} =\frac{2}{m_eR}\int d{\bf x}d{\bf
y}T_N h_{\omega}
 V_{2R}^PE_+ \nonumber\\
&& + \frac{\varepsilon_{21}}{m_e}\int d{\bf x}d{\bf y}T_N
\frac{iR}{r}h_0^\prime\hat r^i \left[ U_{2R}F_+^{0i} +
U_{5R}^{j}F_+^{ij} + U_{7R}^{ij}F_+^{0j} + U_{8R}^{ijk}F_+^{jk}
\right],
 \label{M_TcSS}
\end{eqnarray}
where $E$ and $F$ are taken for ${\bf x}={\bf y}=0$.

For the $0^+\to 0^+$ transition we have
\begin{eqnarray}
 \sum\limits_i\frac{m_i}{m_e}\sum\limits_N\left\{M_T^m\right\}_{S-S} &=&
g_A^2 \left[ 2(W_1^U+W_{1R}^{UP})F_{5+}^0 + \varepsilon_{21}R\{W_{1R}^{VP}\}_cF_+^0 \right], \\
 \sum\limits_i\sum\limits_N\left\{M_{T}^k\right\}_{S-S} &=&
 \frac{2g_A^2}{m_eR}(2W_{4R}^V+\{W_{2R}^{VP}\}_c)E_+,
\end{eqnarray}
with
\begin{eqnarray}
 && W_1^U=\langle \frac{m_i}{m_e}h_+U_1 \rangle, \quad  W_{1R}^{UP}=\langle
\frac{m_i}{m_e}h_0U_{1R}^P \rangle, \quad
\{W_{1R}^{VP}\}_c=\langle \frac{m_i}{m_e}h_0V_{1R}^P \rangle,
\nonumber\\
 && W_{4R}^V=\langle \frac{iR}{2r}h_+^\prime {\bf\hat r}\cdot{\bf  V}_{4R} \rangle, \quad
\left\{W_{2R}^{VP}\right\}_c=\langle h_\omega V_{2R}^P \rangle.
\end{eqnarray}

In the $S-P_{1/2}$ case with no FBWC for the $0^+\to 0^+$
transition we have
\begin{eqnarray}
&& \{M_{T}^m\}_{n,S-P_{1/2}} = 2\int d{\bf x}d{\bf y}T_Nh_+\left( U_{3R}^iF_{5+}^i + V_{3R}^iF_-^i \right), \\
&& \{M_{T}^m\}_{c,S-P_{1/2}} = \varepsilon_{21}R\int d{\bf x}d{\bf
y}T_Nh_0
\left( U_{3R}^iF_{5-}^i + V_{3R}^iF_+^i \right), \\
&& \{M_{T}^k\}_{n,S-P_{1/2}} = \frac{\varepsilon_{21}}{m_e}\int
d{\bf x}d{\bf y}T_N h_{0\omega} U_{4R}^iF_+^{0i} \nonumber\\ &&+
\frac{4}{m_eR}\int d{\bf x}d{\bf y}T_N \frac{iR}{2r} h_+^\prime
\hat r^i
\left[ (U_2+U_{2R}^P)F_-^{0i} + (U_7^{ij}+U_{7R}^{Pij})F_-^{0j} \right], \\
&& \{M_{T}^k\}_{c,S-P_{1/2}} = \frac{2}{m_eR}\int d{\bf x}d{\bf y}T_N h_{\omega} U_{4R}^iF_-^{0i} \nonumber\\
&&+ \frac{\varepsilon_{21}}{m_e}\int d{\bf x}d{\bf y}T_N
\frac{iR}{r} h_0^\prime \hat r^i \left[ (U_2+U_{2R}^P)F_+^{0i} +
(U_7^{ij}+U_{7R}^{Pij})F_+^{0j} \right].
\end{eqnarray}

The decay rate for the $0^+\to0^+$ transition takes the form
\begin{eqnarray}
 d\Gamma = \sum\limits_{s_1, s_2} |R_{0\nu}|^2 \frac{m_e^5}{4\pi^3}d\Omega_{0\nu} =
\frac{a_{0\nu}}{(m_eR)^2} \left[ A_0^{T} - {\bf \hat p}_1\cdot{\bf
\hat p}_2 B_0^{T} \right] d\Omega_{0\nu},
\end{eqnarray}
where the coefficients are
\begin{eqnarray}
&& A_0^T = \sum\limits_{i=1}^4|O_i|^2, \\ && B_0^T =
\mbox{Re}(O_1O_2^*+O_1^*O_2+O_3O_4^*+O_3^*O_4),
\end{eqnarray}
with
\begin{eqnarray}
 && O_1 = \alpha_{-1-1}^* \left\{ \left[ -2(W_1^U+W_{1R}^{UP}) + \frac{2}{m_eR}
(W_{4R}^V+\{W_{2R}^{VP}\}_c) \right] \right.\nonumber\\
&&\left.+ \left[
\frac{m_er}{3}\left(\frac{\zeta}{m_eR}-2\right)W_{3R}^U
+\frac{\varepsilon_{21}^2rR}{6}\{W_{3R}^U\}_c \right] + \left[
-\frac{3(\alpha
Z)^2}{m_eR}\left(W_{4RF}^V+\frac{1}{2}\{W_{2RF}^{VP}\}_c\right)
\right] \right\}, \label{O_1}\\
 && O_2 = \alpha_{11}^* \left\{ \left[ 2(W_1^U+W_{1R}^{UP}) + \frac{2}{m_eR}
(W_{4R}^V+\{W_{2R}^{VP}\}_c) \right] + \right.\nonumber\\
&&\left.+ \left[
-\frac{m_er}{3}\left(\frac{\zeta}{m_eR}+2\right)W_{3R}^U
+\frac{\varepsilon_{21}^2rR}{6}\{W_{3R}^U\}_c \right] + \left[
-\frac{3(\alpha
Z)^2}{m_eR}\left(W_{4RF}^V+\frac{1}{2}\{W_{2RF}^{VP}\}_c\right)
\right] \right\}, \label{O_2}\\
 && O_3 = \alpha_{1-1}^* \left\{ \left[ -\varepsilon_{21}R\{W_{1R}^{VP}\}_c + \frac{2}{m_eR}
(W_{4R}^V+\{W_{2R}^{VP}\}_c) \right] \right.\nonumber\\
&& \left.+ \left[
\frac{m_er}{3}\left(\frac{\varepsilon_{21}}{m_e}+2\right)W_{3R}^V
+\zeta\frac{\varepsilon_{21}r}{6}\{W_{3R}^V\}_c \right.\right.\nonumber\\
&& \left.\left.+ \frac{\varepsilon_{21}R}{3}
\left(\frac{\varepsilon_{21}}{m_e}+2\right)
(W_{4R}^U-\{W_2^U\}_c-\{W_7^U\}_c-\{W_{2R}^{UP}\}_c-\{W_{7R}^{UP}\}_c)\right.\right.\nonumber\\
&& \left.\left.-\frac{4}{3}\frac{\zeta}{m_eR}
(W_2^U+W_7^U+W_{2R}^{UP}+W_{7R}^{UP}-\frac{1}{2}\{W_{4R}^U\}_c)
\right] + \left[ -\frac{3(\alpha
Z)^2}{m_eR}\left(W_{4RF}^V+\frac{1}{2}\{W_{2RF}^{VP}\}_c\right)
\right] \right\}, \\
 && O_4 = \alpha_{-11}^* \left\{ \left[ \varepsilon_{21}R\{W_{1R}^{VP}\}_c + \frac{2}{m_eR}
(W_{4R}^V+\{W_{2R}^{VP}\}_c) \right] \right.\nonumber\\
&& \left.+ \left[
-\frac{m_er}{3}\left(\frac{\varepsilon_{21}}{m_e}-2\right)W_{3R}^V
-\zeta\frac{\varepsilon_{21}r}{6}\{W_{3R}^V\}_c \right.\right.\nonumber\\
&& \left.\left.+
\frac{\varepsilon_{21}R}{3}\left(\frac{\varepsilon_{21}}{m_e}-2\right)
(W_{4R}^U-\{W_2^U\}_c-\{W_7^U\}_c-\{W_{2R}^{UP}\}_c-\{W_{7R}^{UP}\}_c)\right.\right.\nonumber\\
&& \left.\left.-\frac{4}{3}\frac{\zeta}{m_eR}
(W_2^U+W_7^U+W_{2R}^{UP}+W_{7R}^{UP}-\frac{1}{2}\{W_{4R}^U\}_c)
\right] + \left[ -\frac{3(\alpha
Z)^2}{m_eR}\left(W_{4RF}^V+\frac{1}{2}\{W_{2RF}^{VP}\}_c\right)
\right] \right\}, \label{O_4}
\end{eqnarray}
where the terms in the first brackets in Eqs.
(\ref{O_1})--(\ref{O_4}) come from the $S-S$ case and the terms in
the second ones come from the $S-P_{1/2}$ case. The terms in the
third brackets in Eqs. (\ref{O_1})-(\ref{O_2}) are the most
important terms of those that come from the $S-S$ case due to
FBWC. Note that in the $S-S$ case there is the contribution to
Eqs. (\ref{O_1}) and (\ref{O_2}) from the
$(H_{\omega1}+H_{\omega2})$ combination in Eq. (\ref{Tkc}).
Therefore the contribution from the $P_{1/2}-P_{1/2}$ case should
not be taken into account.

The nuclear matrix elements are
\begin{eqnarray}
&& W_{3R}^U=\left\langle \frac{m_i}{m_e} ih_+\hat{\bf
r}_+\cdot{\bf U}_{3R} \right\rangle, \quad
 \{W_{3R}^U\}_c=\left\langle \frac{m_i}{m_e}ih_0\hat{\bf r}\cdot{\bf U}_{3R} \right\rangle, \nonumber \\
&& W_{3R}^V=\left\langle \frac{m_i}{m_e} ih_+\hat{\bf r}\cdot{\bf
V}_{3R} \right\rangle, \quad
 \{W_{3R}^V\}_c=\left\langle \frac{m_i}{m_e}ih_0\hat{\bf r}_+\cdot{\bf V}_{3R} \right\rangle,  \nonumber\\
&& W_{4R}^U=\left\langle ih_{0\omega}\hat{\bf r}_+\cdot{\bf
U}_{4R} \right\rangle,  \nonumber\\ &&
\{W_{2}^U\}_c=\left\langle\frac{R}{r}h_0^\prime\hat{\bf r}\cdot\hat{\bf r}_+U_{2} \right\rangle,\quad
\{W_{2R}^{UP}\}_c=\left\langle\frac{R}{r}h_0^\prime\hat{\bf
r}\cdot\hat{\bf r}_+U_{2R}^P \right\rangle,\quad   \nonumber\\ 
&&
\{W_{7}^U\}_c=\left\langle\frac{R}{r}h_0^\prime\hat r^i\hat
r_+^jU_{7}^{ij} \right\rangle,\quad
\{W_{7R}^{UP}\}_c=\left\langle\frac{R}{r}h_0^\prime\hat r^i\hat
r_+^jU_{7R}^{Pij} \right\rangle.
\end{eqnarray}

Assuming now $\langle m \rangle\neq0$ for the dominant terms we
have
\begin{eqnarray}
 && O_1 = \alpha_{-1-1}^* \left\{ \left[Z_1^X -2W_1^U + \frac{2}{m_eR}
(W_{4R}^V+\{W_{2R}^{VP}\}_c) \right] \right\}, \label{OO_1}\\
 && O_2 = \alpha_{11}^* \left\{ \left[Z_1^X+ 2W_1^U + \frac{2}{m_eR}
(W_{4R}^V+\{W_{2R}^{VP}\}_c) \right] \right\}, \label{OO_2}\\
 && O_3 = \alpha_{1-1}^* \left\{ \left[Z_1^X + \frac{2}{m_eR}
(W_{4R}^V+\{W_{2R}^{VP}\}_c) \right] \right.\nonumber\\
&& \left.+ \left[
\frac{\varepsilon_{21}R}{3}\left(\frac{\varepsilon_{21}}{m_e}+2\right)
(W_{4R}^U-\{W_2^U\}_c-\{W_7^U\}_c) -\frac{4}{3}\frac{\zeta}{m_eR}(W_2^U+W_7^U) \right] \right\}, \\
 && O_4 = \alpha_{-11}^* \left\{ \left[Z_1^X + \frac{2}{m_eR}
(W_{4R}^V+\{W_{2R}^{VP}\}_c) \right] \right.\nonumber\\ && \left.+
\left[
\frac{\varepsilon_{21}R}{3}\left(\frac{\varepsilon_{21}}{m_e}-2\right)
(W_{4R}^U-\{W_2^U\}_c-\{W_7^U\}_c) -\frac{4}{3}\frac{\zeta}{m_eR}
(W_2^U+W_7^U) \right] \right\}. \label{OO_4}
\end{eqnarray}

Again, in the above expressions, the terms with $\zeta$ are due to
the inclusion of the $P$-wave in the electron wave function and
the ones with $W_{2R}^{VP}$ and  $W_{4R}^X~(X=U,V)$ are due to the
nucleon recoil effect.

\newpage



\newpage

\begin{figure}
\vspace{-1.6cm} \hspace{-1.5cm}
\begin{minipage}{0.49\textwidth}
\includegraphics[scale=0.85]{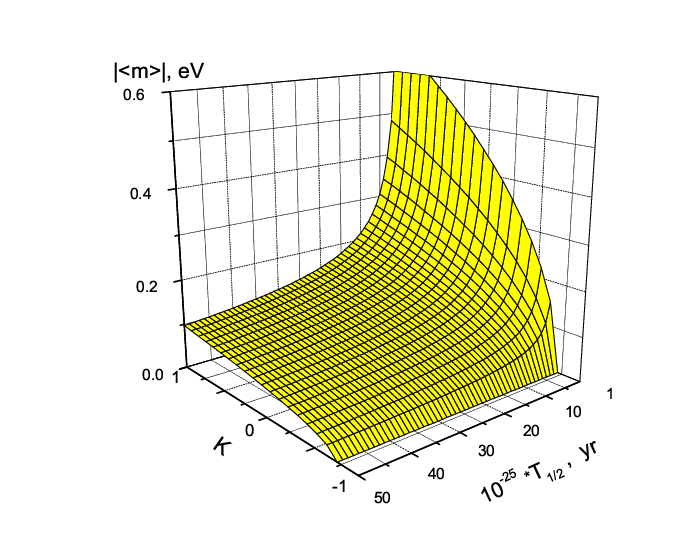}
\end{minipage}
\begin{minipage}{0.49\textwidth}
\includegraphics[scale=0.85]{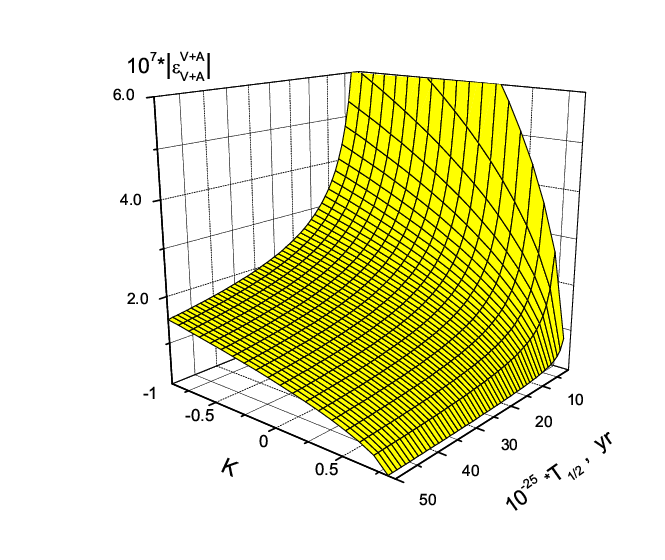}
\end{minipage}
\vspace{-0.7cm} \caption{Correlation between the neutrino
effective mass $|\langle m\rangle|$ ({\it left}) [
$|\epsilon_{V+A}^{V+A}|$ ({\it right}], the angular correlation
coefficient $K$, and the half-life $T_{1/2}$ for the $0\nu2\beta$
decay of $^{76}\mbox{Ge}$ for the case $\cos\psi_1=0$.}
\label{Fig1}
\end{figure}


\begin{figure}
\vspace{0.6cm} \hspace{-1.5cm}
\begin{minipage}{0.49\textwidth}
\includegraphics[scale=0.85]{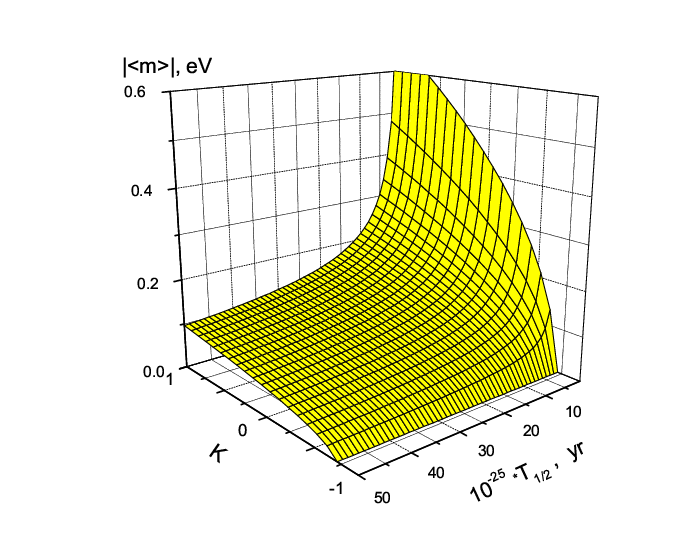}
\end{minipage}
\begin{minipage}{0.49\textwidth}
\includegraphics[scale=0.85]{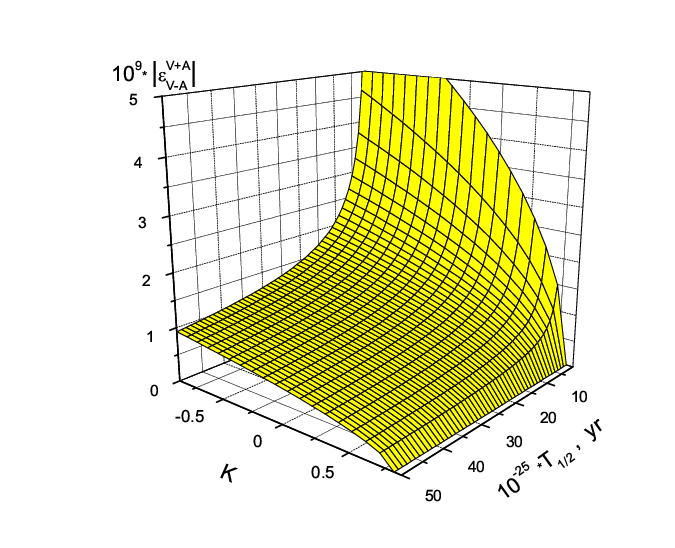}
\end{minipage}
\vspace{-0.7cm} \caption{Correlation between the neutrino
effective mass $|\langle m\rangle|$ ({\it left})
[$|\epsilon_{V-A}^{V+A}|$ ({\it right})], the angular correlation
coefficient $K$, and the half-life $T_{1/2}$ for the $0\nu2\beta$
decay of $^{76}\mbox{Ge}$ for the case $\cos\psi_1=0$.}
\label{Fig2}
\end{figure}


\begin{figure}
\vspace{-1.6cm} \hspace{-1.5cm}
\begin{minipage}{0.49\textwidth}
\includegraphics[scale=0.85]{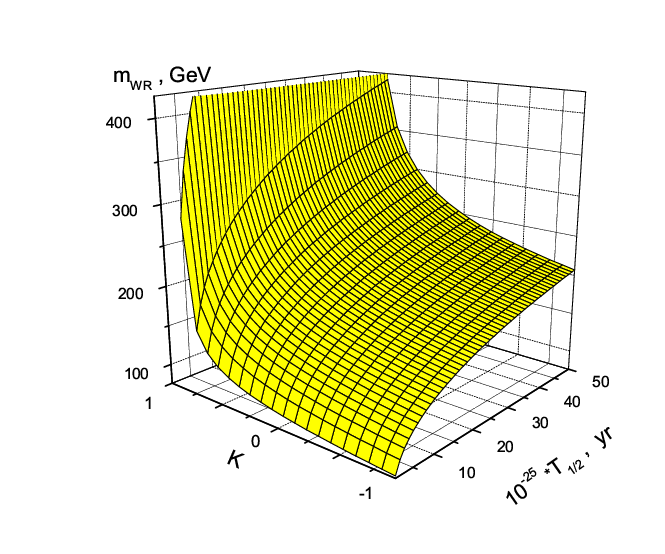}
\end{minipage}
\begin{minipage}{0.49\textwidth}
\includegraphics[scale=0.85]{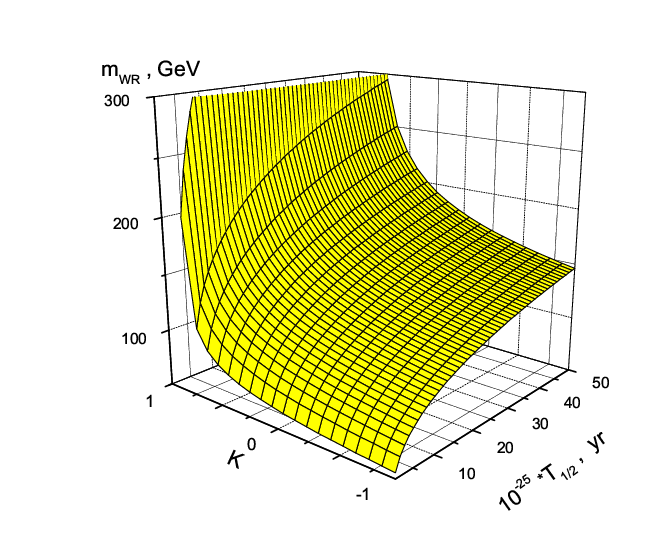}
\end{minipage}
\vspace{-0.7cm} \caption{Correlation between the right-handed
$W$-boson mass $m_{W_R}$, the angular correlation coefficient $K$,
and the half-life $T_{1/2}$ for the $0\nu2\beta$ decay of
$^{76}\mbox{Ge}$ for the case $\cos\psi_1=0$ and
$\epsilon=10^{-6}$ ({\it left}) and for $\epsilon=5\times10^{-7}$
({\it right}).} \label{Fig3}
\end{figure}


\begin{figure}
\vspace{0.6cm}
\hspace{-1.5cm}
\begin{minipage}{0.49\textwidth}
\centering
\includegraphics[scale=0.85]{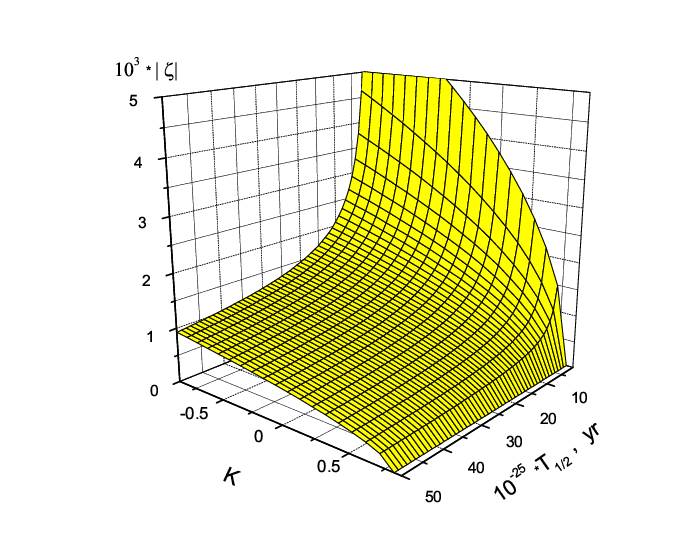}
\end{minipage}
\begin{minipage}{0.49\textwidth}
\centering
\includegraphics[scale=0.85]{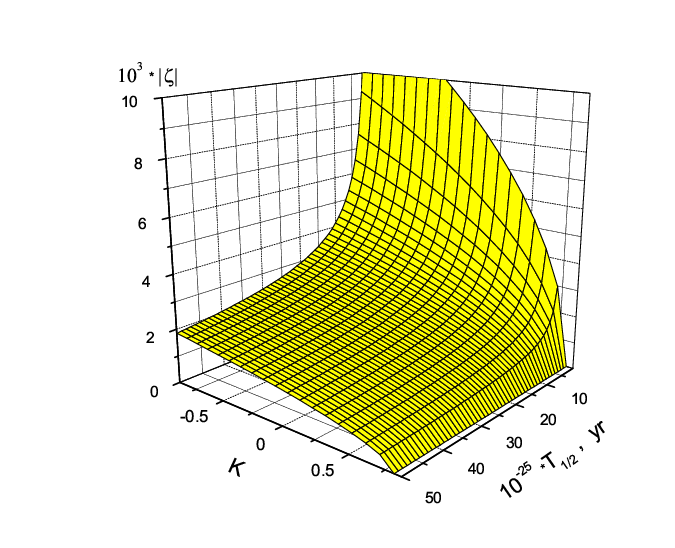}
\end{minipage}
\vspace{-0.7cm} \caption{Correlation between the mixing parameter
$\zeta$, the angular correlation coefficient $K$, and the
half-life $T_{1/2}$ for the $0\nu2\beta$ decay of $^{76}\mbox{Ge}$
for the case $\cos\psi_1=0$ and $\epsilon=10^{-6}$ ({\it
left}) and for $\epsilon=5\times10^{-7}$ ({\it right}).} \label{Fig4}
\end{figure}

\begin{figure}
\hspace{-0.8cm}
\begin{minipage}{0.49\textwidth}
\includegraphics[scale=0.7]{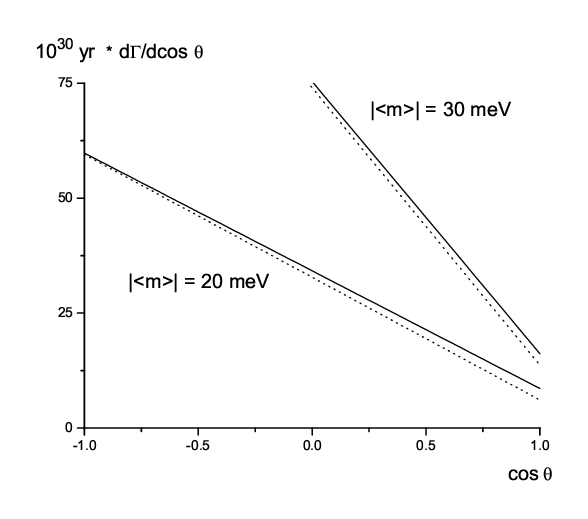}
\end{minipage}
\begin{minipage}{0.49\textwidth}
\includegraphics[scale=0.7]{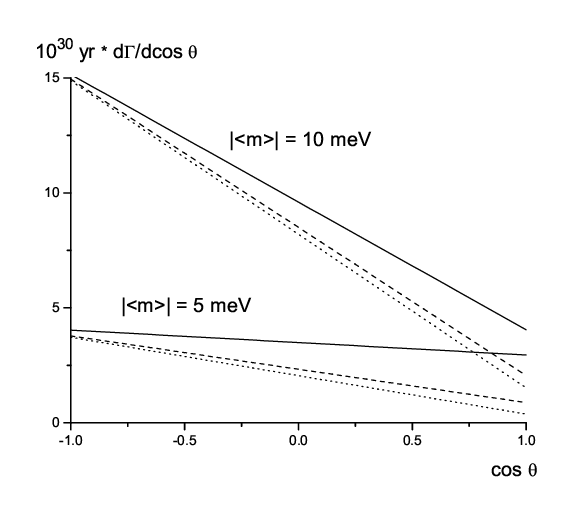}
\end{minipage}
\caption{{\it Left}:~Differential width in $\cos\theta$ for the
$0\nu2\beta$ decay of $^{76}\mbox{Ge}$  for a fixed
value of $\epsilon=10^{-6}$ and $|\langle m\rangle|=20, 30
~\mbox{meV}$. The straight and dotted lines correspond to
$m_{W_R}=1~\mbox{TeV}, \infty$, respectively (the latter is the
conventional case of the light Majorana neutrino exchange
mechanism). {\it Right}:~The same as the left figure but for
smaller values of $|\langle m\rangle|= 5, 10~\mbox{meV}$. In
addition, the dashed lines correspond to $m_{W_R}=
1.5~\mbox{TeV}$.} \label{Fig5}
\end{figure}

\end{document}